\documentclass[preprint]{aastex}

\begin{document}

\title{The Distance to SN 1999em from the Expanding Photosphere Method \altaffilmark{1}}
\author{Mario Hamuy\altaffilmark{2} \altaffilmark{3} \altaffilmark{4} }
\affil{Steward Observatory, The University of Arizona, Tucson, AZ 85721} 
\email{mhamuy@as.arizona.edu}
\author{Philip A. Pinto}
\affil{Steward Observatory, The University of Arizona, Tucson, AZ 85721}
\email{ppinto@as.arizona.edu}
\author{Jos\'{e} Maza\altaffilmark{2} \altaffilmark{3} }
\affil{Departamento de Astronom\'{i}a, Universidad de Chile, Casilla 36-D,
Santiago, Chile}
\email{jose@das.uchile.cl}
\author{Nicholas B. Suntzeff}
\affil{National Optical Astronomy Observatories\altaffilmark{5}, Cerro
Tololo Inter-American Observatory, Casilla 603, La Serena, Chile}
\email{nsuntzeff@noao.edu}
\author{M. M. Phillips}
\affil{Carnegie Institution of Washington, Las Campanas Observatory,
Casilla 601, La Serena, Chile} 
\email{mmp@lco.cl}
\author{Ronald G. Eastman}
\affil{Lawrence Livermore National Laboratory, Livermore, CA 94550}
\email{reastman@llnl.gov}

\author{R. C. Smith}
\affil{National Optical Astronomy Observatories\altaffilmark{5}, Cerro
Tololo Inter-American Observatory, Casilla 603, La Serena, Chile}
\email{csmith@noao.edu}
\author{C. J. Corbally}
\affil{Vatican Observatory, The University of Arizona, Tucson, AZ 85721}
\email{ccorbally@as.arizona.edu}
\author{D. Burstein}
\affil{Department of Physics \& Astronomy, Arizona State University, Tempe, AZ 85287-1504}
\email{burstein@samuri.la.asu.edu}
\author{Yong Li}     
\affil{Department of Physics \& Astronomy, Arizona State University, Tempe, AZ 85287-1504}
\email{li@samuri.la.asu.edu}
\author{Valentin Ivanov}
\affil{Steward Observatory, The University of Arizona, Tucson, AZ 85721}
\email{vivanov@as.arizona.edu}
\author{Amaya Moro-Martin}
\affil{Steward Observatory, The University of Arizona, Tucson, AZ 85721}
\email{amaya@as.arizona.edu}
\author{L. G. Strolger}
\affil{Department of Astronomy, University of Michigan, Ann Arbor, MI 48109-1090}
\email{lou47@astro.lsa.umich.edu}
\author{R. E. de Souza\altaffilmark{6}}
\affil{Steward Observatory, The University of Arizona, Tucson, AZ 85721}
\email{ronaldo@iagusp.usp.br}
\author{S. dos Anjos\altaffilmark{6}}
\affil{Steward Observatory, The University of Arizona, Tucson, AZ 85721}
\email{sandra@iagusp.usp.br}  
\author{Elizabeth M. Green}
\affil{Steward Observatory, The University of Arizona, Tucson, AZ 85721}
\email{bgreen@as.arizona.edu}
\author{T. E. Pickering}
\affil{Steward Observatory, The University of Arizona, Tucson, AZ 85721}
\email{tim@as.arizona.edu}
\author{Luis Gonz\'{a}lez\altaffilmark{2}}
\affil{Departamento de Astronom\'{i}a, Universidad de Chile, Casilla 36-D,
Santiago, Chile}
\email{lgonzale@das.uchile.cl}
\author{Roberto Antezana\altaffilmark{2}}
\affil{Departamento de Astronom\'{i}a, Universidad de Chile, Casilla 36-D,
Santiago, Chile}
\email{rantezan@das.uchile.cl}
\author{Marina Wischnjewsky\altaffilmark{2}}
\affil{Departamento de Astronom\'{i}a, Universidad de Chile, Casilla 36-D,
Santiago, Chile}
\email{marina@das.uchile.cl}
\author{G. Galaz}
\affil{Carnegie Institution of Washington, Las Campanas Observatory,
Casilla 601, La Serena, Chile} 
\email{gaspar@azul.lco.cl}
\author{M. Roth}
\affil{Carnegie Institution of Washington, Las Campanas Observatory,
Casilla 601, La Serena, Chile} 
\email{mroth@lco.cl}
\author{S. E. Persson}
\affil{Observatories of the Carnegie Institution of Washington, 813 Santa Barbara Street, Pasadena, CA 91101}
\email{persson@ociw.edu}
\author{W. L. Freedman}
\affil{Observatories of the Carnegie Institution of Washington, 813 Santa Barbara Street, Pasadena, CA 91101}
\email{wendy@rmdm.ociw.edu}
\author{R. A. Schommer}
\affil{National Optical Astronomy Observatories\altaffilmark{5}, Cerro Tololo Inter-American Observatory, Casilla 603, La Serena, Chile}
\email{rschommer@noao.edu}

\altaffiltext{1}{Based on observations collected at the European Southern Observatory, Chile (program ESO 164.H-0376).} 
\altaffiltext{2}{Visiting Astronomer, Cerro Tololo Inter-American Observatory.
CTIO is operated by AURA, Inc. under contract to the National Science
Foundation.}
\altaffiltext{3}{Visiting Astronomer, European Southern Observatory.}
\altaffiltext{4}{Visiting Astronomer, Las Campanas Observatory.}
\altaffiltext{5}{Cerro Tololo Inter-American Observatory, Kitt Peak
National Observatory, National Optical Astronomy Observatories,
operated by the Association of Universities for Research in Astronomy,
Inc., (AURA), under cooperative agreement with the National Science
Foundation.}
\altaffiltext{6}{Permanent address: Astronomy Department, University of Sao Paulo, C.Postal 9638, SP 01065-970, Brazil}

\begin{abstract}
We present optical and infrared spectroscopy of the first two months of
evolution of the Type II SN~1999em. We combine these data with high-quality
optical/infrared photometry beginning only three days after shock breakout,
in order to study the performance of the ``Expanding Photosphere Method'' (EPM) in the
determination of distances. With this purpose
we develop a technique to measure accurate
photospheric velocities by cross-correlating observed and model spectra.
The application of this technique to SN~1999em
shows that we can reach an average uncertainty of 11\% in
velocity from an individual spectrum. Our analysis
shows that EPM is quite robust to the effects of dust. In particular,
the distances derived from the $VI$ filters
change by only 7\% when the adopted visual extinction in
the host galaxy is varied by 0.45 mag. 
The superb time sampling of the $BVIZJHK$ light-curves
of SN~1999em permits us to study the internal consistency of EPM and test
the dilution factors computed from atmosphere models for Type II plateau supernovae.
We find that, in the first week since explosion, the EPM distances are up to 50\%
lower than the average, possibly due the presence of circumstellar material.
Over the following 65 days, on the other hand, our tests lend strong credence to the
atmosphere models, and confirm previous claims that EPM can produce
consistent distances without having to craft specific models to each supernova.
This is  particularly true for the $VI$ filters which yield distances with an
internal consistency of 4\%. From the whole set of $BVIZJHK$ photometry,
we obtain an average distance of 7.5$\pm$0.5 Mpc, where
the quoted uncertainty (7\%) is a conservative estimate of the internal precision
of the method obtained from the analysis of the first 70 days of the supernova evolution.
\end{abstract}


\section{INTRODUCTION}

The last ten years have witnessed an enormous progress in our
knowledge of the optical properties of supernovae (SNe) of all types.
However, comparatively little is still known about these objects in
infrared (IR) wavelengths.
Light-curves in the $JHK$ bands have been obtained only for a handful events
since the pioneering work of Elias et al. (1981, 1985) on Type Ia SNe. 
Aside from SN~1987A (e.g. Suntzeff \& Bouchet 1990), SN~1990E (Schmidt et al. 1993),
and SN~1980K (Dwek et al. 1983), virtually nothing is known of the $JHK$ light curves of Type II SNe.
The spectroscopic studies have been mostly limited to optical wavelengths
(3000-10000 \AA) also, and few spectra have been obtained beyond this range \cite{bowers97}.

Given the rapid technological development of IR light detection
over recent years, the next logical step is to expand the
SN observations to the broadest possible spectral range.
With this idea in mind, in 1999 we started a program to
obtain optical and IR photometry and spectroscopy of nearby SNe ($z$$<$0.08),
in order to better understanding 1) the nature of SNe, 2) the explosion mechanisms,
3) the relation of SN properties to their stellar environments, and
4) the use of SNe as distance indicators.
The ``Supernova Optical and Infrared Survey'' (SOIRS)
uses telescopes at Cerro Tololo Inter-American Observatory (CTIO),
the Carnegie Institution of Washington at Las Campanas Observatory (LCO),
the European Southern Observatory (ESO) at La Silla and Cerro Paranal,
the Steward Observatory (SO) of The University of Arizona,
and the Cerro El Roble observatory of the University of Chile.
To ensure sufficient targets for the observing runs,
we schedule SN search runs with the Maksutov camera
at Cerro El Roble \cite{maza81}, prior to the follow-up runs.
With photographic film this camera permits us to sample
a wide field of view (5$^\circ$ x 5$^\circ$) down to a limiting magnitude of 18.
With this setup the photographic survey finds SNe up to $z$=0.08, which
are ideally suited for the follow-up program.
We also coordinate the follow-up runs with the Nearby Galaxies Supernova Survey (NGSS)
carried out with the Kitt Peak 0.9-m telescope and the wide field mosaic CCD array,
which discovers SNe down to a limiting magnitude of $R\sim$21 ($z$$<$0.15) \cite{strolger99}.
We include in our list of targets SNe found at other observatories and
reported to the IAU Circulars.

In the course of 1999-2000 the SOIRS program has gathered high-quality observations for
$\sim$20 SNe. In this paper we report observations of one of the best-observed
objects included in our program, the Type II SN~1999em
discovered on 1999 October 29 (JD 2451480.94) in the course of the Lick Observatory
Supernova Search \citep{li99}. The supernova occurred
in NGC~1637, a spiral galaxy with a heliocentric
radial velocity of 717 km~s$^{-1}$ \cite{haynes98}.
A pre-discovery image of NGC~1637 taken on October 20.45 showed
nothing at the position of SN~1999em (with a limiting magnitude of $\sim$19.0),
which indicated that the SN had been caught at an early stage.
An optical spectrum taken one day later (October 30.34) showed that SN~1999em had the
H line P-Cygni profiles characteristic of a Type II event.  The blue
continuum and the presence of the He I $\lambda$5876 line confirmed
that SN~1999em had been found at an early epoch \citep{jha99}.
Given its proximity and early evolutionary stage,
SN~1999em proved to be an excellent target to
test the ``Expanding Photosphere Method'' (EPM, hereafter) in detail, and
determine the distance to the host galaxy.

The discovery of SN~1999em occurred right in the middle of a SOIRS
observing campaign previously scheduled for 1999 October-December.
As soon as the discovery of SN~1999em was reported, we decided to
initiate a follow-up program in order to obtain high-quality data for this event.
During this observing run we obtained superb optical and IR
sampling of the first 180 days of the $UBVRIZJHK$ light-curves, 
as well as optical/IR spectroscopy for the first 60 days
of the evolution of SN~1999em. Except for the sub-luminous/peculiar SN~1987A, this
object is the best observed event of the Type II class to date.

All of the photometry for SN~1999em, except for the $Z$-band,
was reduced and analyzed by Suntzeff et al. and will be published separately.
In Sec. 2.1 we give a brief summary of these observations and describe the main features
of the light-curves. Also, we report the $Z$-band photometry of SN~1999em.
In Sec. 2.2 we summarize our spectroscopic observations and
present the resulting spectra.
In Sec. 3 we use the ``Expanding Photosphere Method'' (EPM)
to compute the distance to the host galaxy, after which (Sec. 4)
we discuss our results.
Finally (Sec. 5), we present the main conclusions of this study.
In Appendix A we summarize the basic ideas behind EPM
and our implementation to compute distances to SNe~II.
Appendix B describes the details of computing
the synthetic magnitudes required by EPM. Having defined
our synthetic photometric system, in Appendix C we proceed 
to compute the distance correction factors
from the SN~II atmosphere models published by Eastman, Schmidt, \& Kirshner 
(1996, E96 hereafter). In Appendix D we define the $Z$ photometric
system and we list magnitudes for the standards that
we used in our observations.

\section{OBSERVATIONS AND REDUCTIONS}

\subsection{Photometry}

\begin{figure}
\plotone{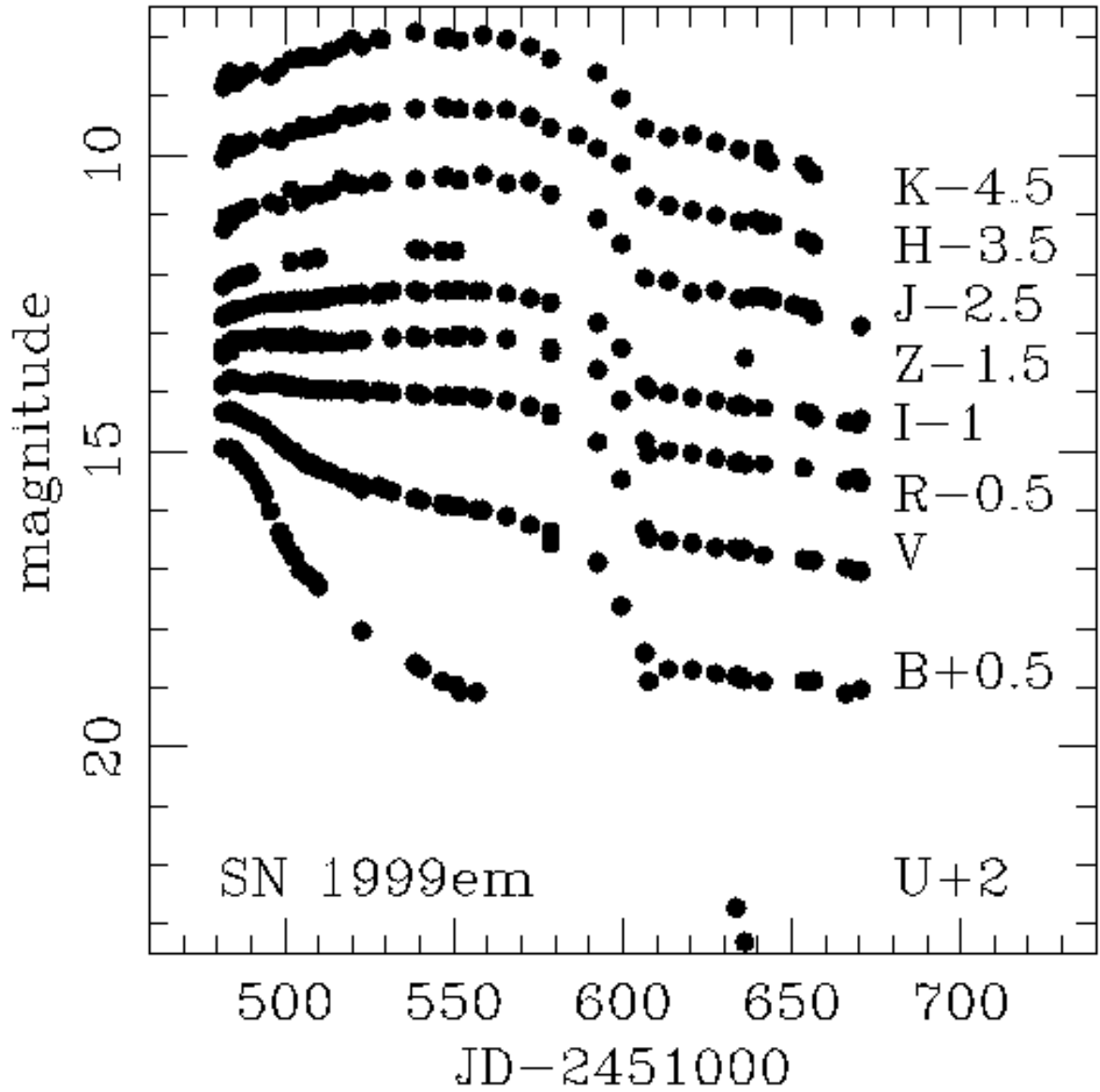}
\figcaption[figure1.ps]{$UBVRIZJHK$ light-curves of SN~1999em. \label{ubvrizjhk.fig}}
\end{figure}

We obtained extensive optical and IR photometric follow-up of SN~1999em
covering 180 days from  discovery until the SN went behind the Sun. 
The data will be presented in a separate paper by Suntzeff et al. (2001).
Table 1 lists additional $UBVRI$ observations taken with
the CTIO~1.5-m, SO~1.5-m, SO~2.3-m, and ESO NTT telescopes. We include
in this table $Z$-band photometry gathered with the CTIO~0.9-m and reduced
relative to a photometric sequence properly calibrated with respect to the
standards listed in Appendix D.  In what follows we adopt
a minimum photometric error of 0.015 mag in order to account for
photometric uncertainties beyond the photon statistics quoted by
Suntzeff et al.  Figure \ref{ubvrizjhk.fig} presents the light-curves which
reveal the exceptional sampling obtained. 
The $U$ light-curve shows that maximum light occurred 
just after discovery, on JD 2451482.8 (October 31), followed by
a phase of rapid decline during which the SN dimmed by $\sim$4 mag
in 70 days. The faintness of the SN made further observations
through the $U$ filter difficult.  The $B$ light-curve shows
that maximum occurred two days later than in $U$, a rapid
decline for $\sim$30 days during which the SN dimmed by $\sim$1 mag,
a phase of 70 days of slowly-decreasing luminosity (plateau)
during which the flux decreased by one additional magnitude,
a fast drop in flux by 2.5 mag in only 30 days, and
a linear decay in magnitude as of JD 2451610 that signaled
the onset of the nebular phase. The $V$ light-curve was characterized
by a plateau of nearly-constant brightness that lasted $\sim$100 days
(until JD 2451590), followed by a drop of 2 mag in $\sim$30 days, and a
linear decline at the slow pace of $\sim$0.01 mag day$^{-1}$.
The $R$, $I$, $Z$, $J$, $H$, and $K$ light-curves
had  the same basic features of the $V$ light-curve, except
that the SN gradually increased its luminosity during the plateau.
The brightening increased with wavelength, and reached
0.5 mag in the $K$ band.

Barbon, Ciatti, \& Rosino (1979) divided Type II SNe into two main subclasses
according to their photometric behavior in blue light. 
The observations of SN~1999em clearly show that this object
belongs to the ``plateau'' (SNe~II-P) group, the most frequent type of SNe~II,
as opposed to the  ``linear'' (SNe~II-L) class which is characterized
by a rapid post-maximum decline in brightness.

\begin{figure}
\epsscale{1.0}
\plotone{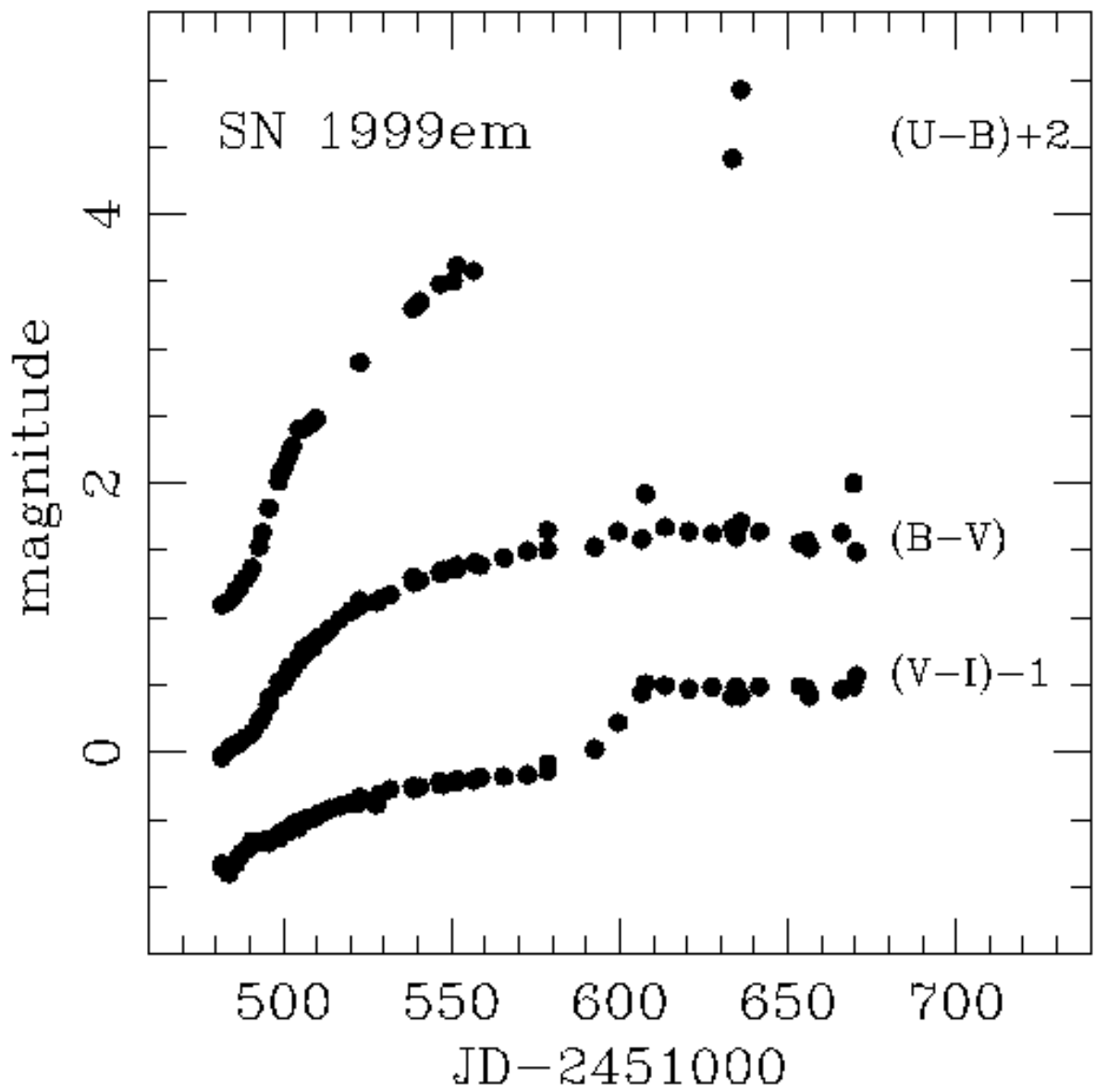}
\figcaption[figure2.ps]{Color curves of SN~1999em. \label{colors.fig}}
\end{figure}

Figure \ref{colors.fig} shows some of the color curves of SN~1999em.
They all reveal the reddening of the SN due to the cooling
of its atmosphere. In $U-B$ and $B-V$ the reddening was more pronounced due
to the many metal lines which depressed the SN flux at the wavelengths sampled by the $U$ and $B$ bands.
The $V-I$ color, on the other hand,
was less affected by line blanketing and better sampled 
the continuum emission. The evolution of SN~1999em in this color showed rapid cooling during
the first 50 days, after which $V-I$ remained nearly constant for
another $\sim$30 days. During this phase, the photosphere was near the hydrogen
recombination surface and was then of nearly constant temperature (E96). 
The end of the plateau phase at JD 2451590 coincided with a sudden reddening
of the SN, while the exponential tail was characterized by  a nearly 
constant $V-I$ color.

\subsection{Spectroscopy}

We obtained optical and IR spectra of SN~1999em with the ESO NTT/EMMI
at La Silla and VLT/ISAAC at Cerro Paranal between 1999 November 2
and November 28, and additional optical spectra with the CTIO~1.5-m telescope
on October 30 (one day after discovery), and the SO Bok 2.3-m telescope
and Boller \& Chivens spectrograph on December 16 and December 31. Table 2 presents a
journal of the observations.

\subsubsection{Optical Spectroscopy}

The NTT observations employed three different setups.
We used the blue channel of EMMI equipped 
with a Tek CCD (1024x1024) and grating 5 (158 lines mm$^{-1}$) which,
in first order, delivered spectra with a dispersion of 3.5~\AA~pix$^{-1}$ and a useful
wavelength range between 3300 and 5250 \AA. With the red channel,
CCD Tek 2048, and grating 13 (150 lines mm$^{-1}$) the dispersion was 2.7~\AA~pix$^{-1}$
and the spectral coverage ranged from 4700 through 11000 \AA~in first order.
Since this setup had potential second-order contamination beyond $\sim$6600 \AA~we
decided to take one spectrum with the OG530 filter and a second observation
without the filter, in order to provide an overlap with the blue spectrum.
Thus, a single-epoch observation usually consisted of three spectra. On two occasions
(November 3 and November 14) however, we were unable to obtain the observation
with the OG530 filter, so the red end of the spectra were most likely contaminated
by second-order blue light  (more below).

The observations with EMMI started with calibrations during day time 
(bias and dome flat-field exposures). The night began
with the observation of a spectrophotometric standard [from
the list of Hamuy et al. (1994)] through a wide slit of 10 arcsec, after which
we observed the SN with a slit of 1 arcsec. We took
two exposures per spectral setup, each of the same length  
(typically 120-180 sec) and always along the parallactic
angle. Immediately following this observation we observed
a He-Ar lamp, at the same position of the SN and before
changing the optical setup in order to ensure an accurate wavelength
calibration.  At the end of the night we observed a second flux standard.

We carried out all reductions using IRAF\footnote[7] 
{IRAF is distributed by the National Optical Astronomy Observatories,
which are operated by the Association of Universities for Research
in Astronomy, Inc., under cooperative agreement with the National
Science Foundation.}.
They consisted in subtracting the overscan and bias
from every frame. Next, we constructed a flat-field
from the quartz-lamp image, duly normalized along
the dispersion axis. We proceeded by flat-fielding all of
the object frames and  extracting 1-D spectra from the
2-D images. We followed the same procedure for the He-Ar
frames which we used to derive the wavelength calibration
for the SN. We then derived a
response curve from the two flux standards, which we applied
to the SN spectra in order to get flux calibrated spectra.
During this process we also corrected for atmospheric extinction
using an average continuum opacity curve scaled for the airmass 
at which we observed the object, but we did not attempt
to remove telluric lines.
From the pair of flux-calibrated spectra that we secured for each
spectral setup we removed cosmic rays and bad pixels, and
obtained a clean spectrum of the SN. The last step consisted
in merging the three spectra that sampled different wavelength ranges.
To avoid discontinuities  in the
combined spectrum we grey-shifted the three spectra
relative to each other. Finally, we computed the synthetic $V$-band
magnitude from the resulting spectrum (following the precepts described
in Appendix B) and grey-shifted it so that the
flux level matched our observed $V$ magnitude. We checked the
spectrophotometric quality of the spectra by computing synthetic
magnitudes for the $BRIZ$ bands. This test showed 
differences between the synthetic and observed magnitudes
of up to 0.03 mag in the $B$ and $R$ bands, which 
implies that the relative spectrophotometry at these wavelengths was very good.
The $I$ and $Z$ synthetic magnitudes, on the other hand,
disagreed with the observed magnitudes by up to 0.1 mag,
particularly when the blocking filter could not be used;
second-order contamination was as large as 10\%
at those wavelengths.

We obtained a spectrum of the SN one day after discovery with the
CTIO~1.5-m telescope and the Cassegrain spectrograph,
a 1200x800 LORAL CCD, grating 16 (527 lines mm$^{-1}$) and 2 arcsec slit,
in first order.
The resulting spectrum had a dispersion of 5.7~\AA~pix$^{-1}$ and
useful wavelength coverage of 3300-9700 \AA. Second-order
contamination was expected beyond 6600 \AA~due to first-order
3300 \AA~light since we did not use a blocking filter.
On two nights we employed the SO Bok telescope with the Boller \& Chivens
spectrograph, a 1200x800 LORAL CCD, and a 300 lines mm$^{-1}$ grating which, in first
order, produced spectra with a dispersion of 3.6~\AA~pix$^{-1}$. The wavelength
coverage was 4900-9300 \AA~on December 16 and 3500-7100 \AA~on December 31. We did not include a blocking filter in the
optical path, so it could well be that these spectra were affected by
second-order contamination beyond 6600 \AA.
The observing and reduction procedures for the CTIO
and SO spectra were the same as those described above for the NTT data.

\begin{figure}
\epsscale{1.0}
\plotone{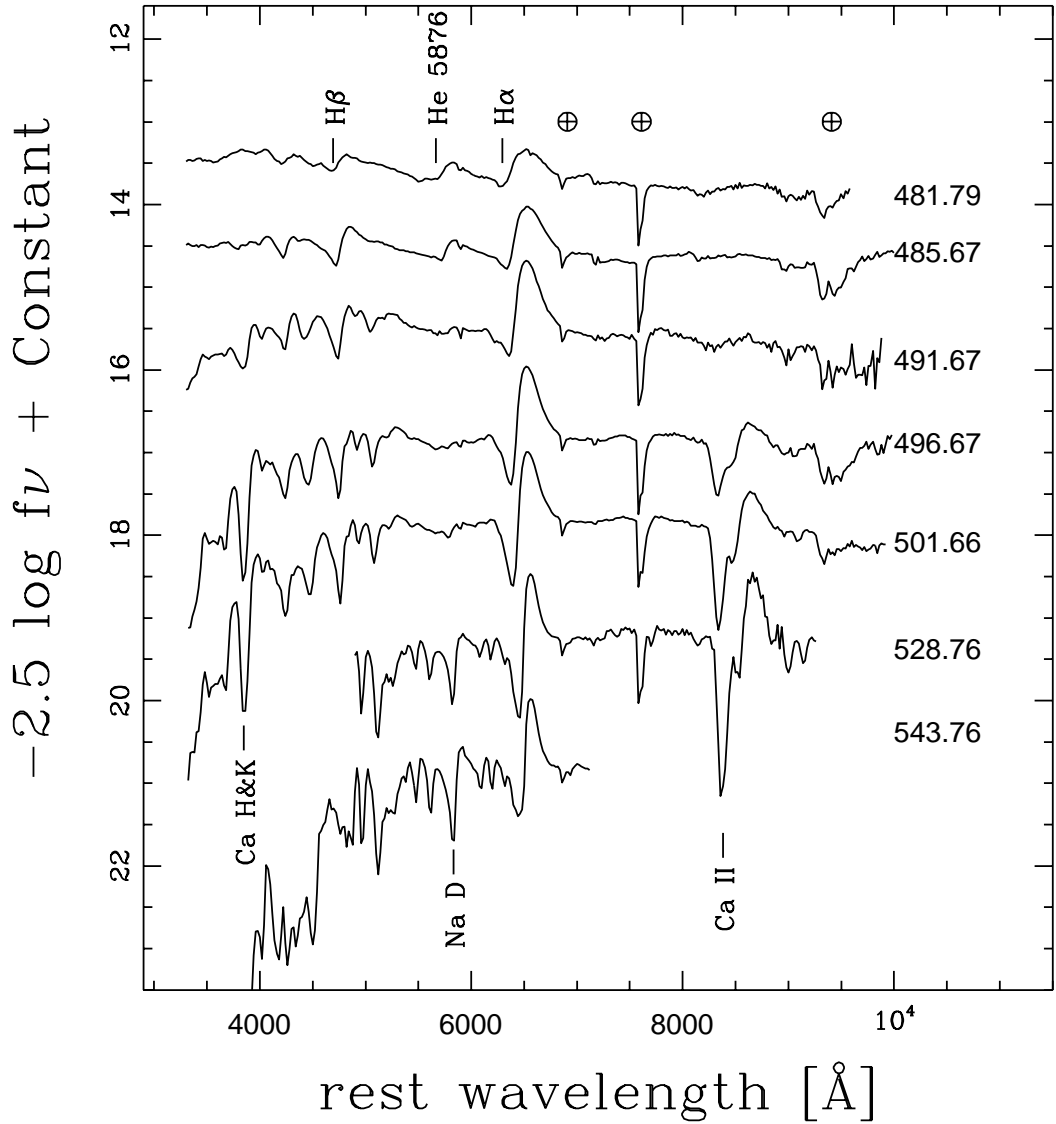}
\figcaption[figure3.ps]{Optical spectroscopic evolution of SN~1999em
in AB magnitudes.  Julian Day (-2451000) is indicated for each spectrum. Some of the
strongest lines are labeled. The $\oplus$ symbols show the
main telluric features.  \label{sn99em.opt.fig}}
\end{figure}

Figure \ref{sn99em.opt.fig} displays the optical spectra in the
rest-frame of the SN, after correcting the observed spectra for the 717 km~s$^{-1}$ recession velocity
of the host galaxy. 
The strongest SN lines are indicated along with the telluric lines (by the $\oplus$ symbol).
The first spectrum, taken on JD 2451481.79 (1 day after discovery), showed a blue continuum
with a $BVI$ color temperature of 15,600 K, P-Cygni profiles for the H Balmer lines, and the
He I $\lambda$5876 line which is characteristic of SNe~II during their hottest phases. The expansion
velocity from the minimum of the absorption features was $\sim$10,000 km~s$^{-1}$
which is typical of SNe~II during the initial phases. 
The presence of the interstellar Na I D lines $\lambda\lambda$5890, 5896
with an equivalent width of $\sim$2~\AA~ suggested substantial interstellar absorption
in the host galaxy.  However, there is additional evidence that the SN did not
suffer significant extinction (see Sec. 3.3) so we did not attempt to 
correct for dust the spectra of Figure \ref{sn99em.opt.fig}.

As the SN evolved the atmospheric temperature dropped,
the He I $\lambda$5876 line disappeared, and several new lines became evident,
namely, the Ca II H\&K $\lambda\lambda$3934,3968 blend, the Ca II triplet
$\lambda\lambda$8498,8542,8662, the Na I D blend, and several lines attributed to
Fe~II, Sc~II, and Ba~II. Table 3 summarizes the line identifications, their rest wavelengths
[taken from the list of Jeffery \& Branch (1990)], and their wavelengths measured from the absorption minimum.
We include several lines with unknown identifications in this table.
By the time of our last spectrum on JD 2451543.76 (62 days after discovery)
the $BVI$ color temperature was only 5,000 K, approximately the
recombination temperature of H. 

\subsubsection{Infrared Spectroscopy}

We obtained three IR spectra with the VLT/Antu telescope at Cerro Paranal,
between November 2-28. We employed the IR spectro-imager ISAAC \cite{moorwood97}
in low resolution mode (R$\sim$500), with four different gratings that
permitted us to obtain spectra in the $Z$, $J$, $H$, and $K$ bands. We used these gratings
in 5$^{th}$, 4$^{th}$, 3$^{rd}$ and 2$^{nd}$
order, respectively, which yielded useful data in the spectral ranges
9840-11360, 11090-13550, 14150-18180, and 18460-25600 \AA.
The detector was a Hawaii-Rockwell 1024x1024 array. 

A typical IR observation started during daytime by taking calibrations.
We began   taking flat-field images
using an internal source of continuum light. 
We secured multiple on- and off-image pairs with
the same slit used during the night (0.6 arcsec).
We then took Xe-Ar lamp images (and off-lamp images) with 
a narrow slit (0.3 arcsec) in order to map geometric distortions. 
The observations of SN~1999em 
consisted in a O-S-O-S-O cycle, where O is an on-source image
and S is a sky  (off-source) frame.
Given that the angular size of the host galaxy was comparable to the
slit length (2 arcmin), it was not possible to use the classical
technique of nodding the source along the slit and it proved
necessary to offset the telescope by several minutes of arc to
obtain the sky images. At each position we exposed for 200 sec,
conveniently split into two 100-sec images in order to
remove cosmic rays and bad pixels from the final spectra.
After completing the O-S-O-S-O cycle we immediately obtained
a pair of on-off arc lamp exposures 
without moving the telescope or changing optical elements to
ensure an accurate wavelength calibration. We then switched
to the next grating and repeated the above object-arc procedure
until completing the observations with the four setups.
For flux calibration we observed a bright solar-analog star,
close in the sky to the SN in order to minimize variations
in the atmospheric absorptions (Maiolino, Rieke, \& Rieke 1996). The selected star was
Hip 21488, of spectral type G2V, $V$=8.5, $B-V$=0.6,
and located only 12$^\circ$ from the SN (SIMBAD Astronomical Database). In this case we
nodded the object between two positions (A and B) along the slit
and we took two AB pairs for each grating. To avoid saturating
the detector, we took the shortest possible exposures (1.77 sec)
allowed by the electronics that controlled the detector.
Since the minimum time required before offsetting the telescope
was $\sim$60 sec, we took ten exposures at each
position which provided an exceedingly good signal-to-noise ratio (S/N)
for the flux standard.

The reductions of the IR data began by subtracting the off-lamp
images from the on-lamp flat-field frames, median filtering the multiple
flat images, and normalizing the resulting frame along the dispersion axis.
Then we divided all of the object images by the normalized flat-field.
The next step consisted in mapping the geometric distortions using
the arc lamp taken with the narrow slit. The emission lines displayed
a curvature of a few pixels across the spatial direction of the
image. From the narrow-slit
arc exposure we obtained sharp emission lines which allowed us to
form a 2-D map of the distortions. We then fit
a low-order 2-D polynomial to the line tracings, after which
we applied a geometric correction to all of the images obtained during
the night. Since the sky background is so large in the IR, a small
flat-fielding residual or a patchy sky can introduce large
background fluctuations. This made it necessary to subtract the sky
from the 2-D images before attempting to extract the
object spectrum. Given that the IR sky changes on short time-scales,
we always used the sky image taken immediately before or after the
SN frame. We followed the same procedure for the flux standard.
We then extracted 1-D spectra of the SN and the spectrophotometric
standard from the sky-subtracted frames, making sure to subtract residual sky from 
a window adjacent to the object. We also extracted a spectrum of
the Xe-Ar frame taken at the same position of the SN, in order
to derive a wavelength calibration.
We then combined the multiple spectra of the SN and the standard
with a `minmax' rejection algorithm that removed deviant pixels from the average
at each wavelength.

Flux calibration in the IR is in general quite involved because there are
no flux standards at these wavelengths. To get around this problem we 
adopted the technique described by Maiolino et al. (1996), which consists
in dividing the spectrum of interest by a solar-type star to remove the strong
telluric IR features, and multiplying the resulting spectrum by the
solar spectrum to eliminate the intrinsic features (pseudonoise)
introduced by the solar-type star. In its original version this
method used a solar spectrum normalized by its continuum slope,
so the object's spectrum was not properly flux calibrated. 
More recently (http://www.arcetri.astro.it/$\sim$maiolino/solar/solar.html)
this technique was modified to incorporate a flux-calibrated semi-empirical spectrum of the Sun, 
so that the object's final spectrum is in flux units.
The adopted spectrum is a combination of the observed solar line spectrum in the $JHK$ bands
\cite{livingston91} and the Kurucz solar model with parameters
$T_{eff}$=5,777 K, log $g$=4.4377, [Fe/H]=0, $V_{microturb}$=1.5 km s$^{-1}$ \cite{kurucz95}.
In the Vega magnitude system (described in Appendix B), the solar spectrum has
a visual magnitude of -26.752. The remaining broad-band magnitudes are listed in Table 12.

Before using the solar spectrum we convolved it with a 
kernel function in order to reproduce the spectral resolution of the solar-analog
standard Hip 21488, and we scaled it down to the equivalent of
$V$=8.5 which corresponds the the observed magnitude of Hip 21488.
From the ratio of this solar spectrum and the observed spectrum of Hip 21488
we derived an  instrumental response curve which included both
the telluric absorption lines and the instrumental sensitivity of ISAAC. 
Finally, we multiplied the SN spectrum by the response function to obtain
the flux-calibrated SN spectrum.
This technique worked very well to remove telluric lines. On the other hand,
it introduced a small systematic error in the flux calibration of the SN due
to departures between the solar spectrum and the actual spectral energy distribution
of the solar-analog standard. According to atmosphere models the difference
in continuum flux for stars with $T_{eff}$=5,500 and 6,000 K (which correspond to
spectral types G8V-F9V, Gray \& Corbally 1994) is smaller than 10\%
in the NIR region. A G2V star like Hip 21488 has the same spectral type of
the Sun, so its effective temperature must be close (within $\pm$100 K)
to that of the adopted solar model. Hence, the flux difference between the
solar-analog standard and the adopted spectrum should be less than 10\%.
The $B-V$=0.6 color of Hip 21488 suggests little or no reddening so 
the SN spectra fluxes are probably accurate to 5\% or better.

The result of these operations are four
spectra covering the $Z$, $J$, $H$, and $K$ bands, which we
combined into one final spectrum. Given the significant overlap of
the $Z$ and $J$ band spectra, we were able to grey-shift the $Z$
spectrum relative to the $J$ spectrum. Since none of the
other spectra overlapped (due to the strong absorption 
of telluric lines between the $J$, $H$, and $K$ bands), we shifted them individually
by computing synthetic magnitudes (see Appendix B) and bringing them into
agreement with the observed photometry.

\begin{figure}
\epsscale{1.0}
\plotone{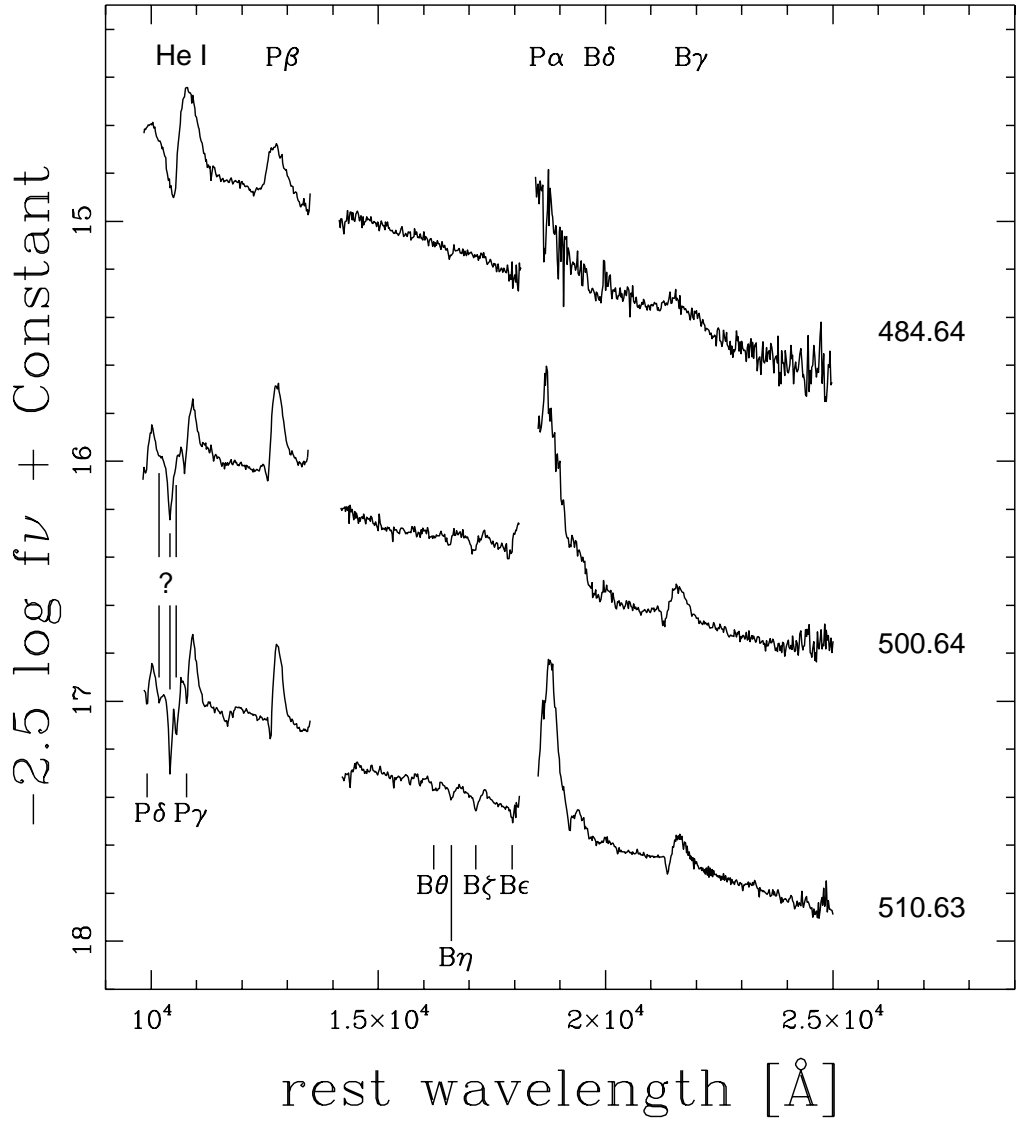}
\figcaption[figure4.ps]{IR spectroscopic evolution of SN~1999em
in AB magnitudes. The most prominent features are labeled.
Julian Day (-2451000) is indicated for each spectrum. \label{sn99em.ir.fig}}
\end{figure}

Figure \ref{sn99em.ir.fig} shows the resulting rest-frame spectra of SN~1999em, 
revealing the exquisite spectral resolution and the superb S/N
delivered by ISAAC. The first spectrum, taken four days after discovery,
showed a few prominent lines, namely He I $\lambda$10830, 
P$\alpha$, P$\beta$, B$\gamma$, and B$\delta$, on top of a
blue continuum. The second spectrum (20 days after discovery)
showed prominent P$\gamma$ and P$\delta$ lines in lieu  of the strong He I
feature.  A few faint lines appeared around 17000 \AA,
from higher transitions in the Brackett series.
The third spectrum, taken 30 days after discovery, confirmed
the presence of these faint lines. Three
lines could be seen at $\sim$10500 \AA~,
between P$\gamma$ and P$\delta$. It is possible that the
feature at 10180 \AA~was due to the 10327 \AA~line of the multiplet 2 of Sr II,
which was also observed in SN~1987A \cite{elias88}. 
The other two lines of the multiplet at 10037 and 10915 \AA~ 
were probably blended with P$\delta$ and P$\gamma$, respectively.
We cannot provide identifications with confidence for the other two
features at 10418 and 10549 \AA~. Meikle et al. (1989) identified
the C I $\lambda$10695 multiplet 1 line in an early-time spectrum
of SN~1987A, which might be responsible for the 10549 \AA~ feature
in the SN~1999em spectrum. They also reported the presence of the
He I $\lambda$10830 absorption line in late-time spectra of SN~1987A.
It is possible then that the 10418 \AA~feature in SN~1999em could be due to He I.
Table 4 summarizes the line identifications, their rest wavelengths
and the values measured from the absorption minima.

\begin{figure}
\epsscale{1.0}
\plotone{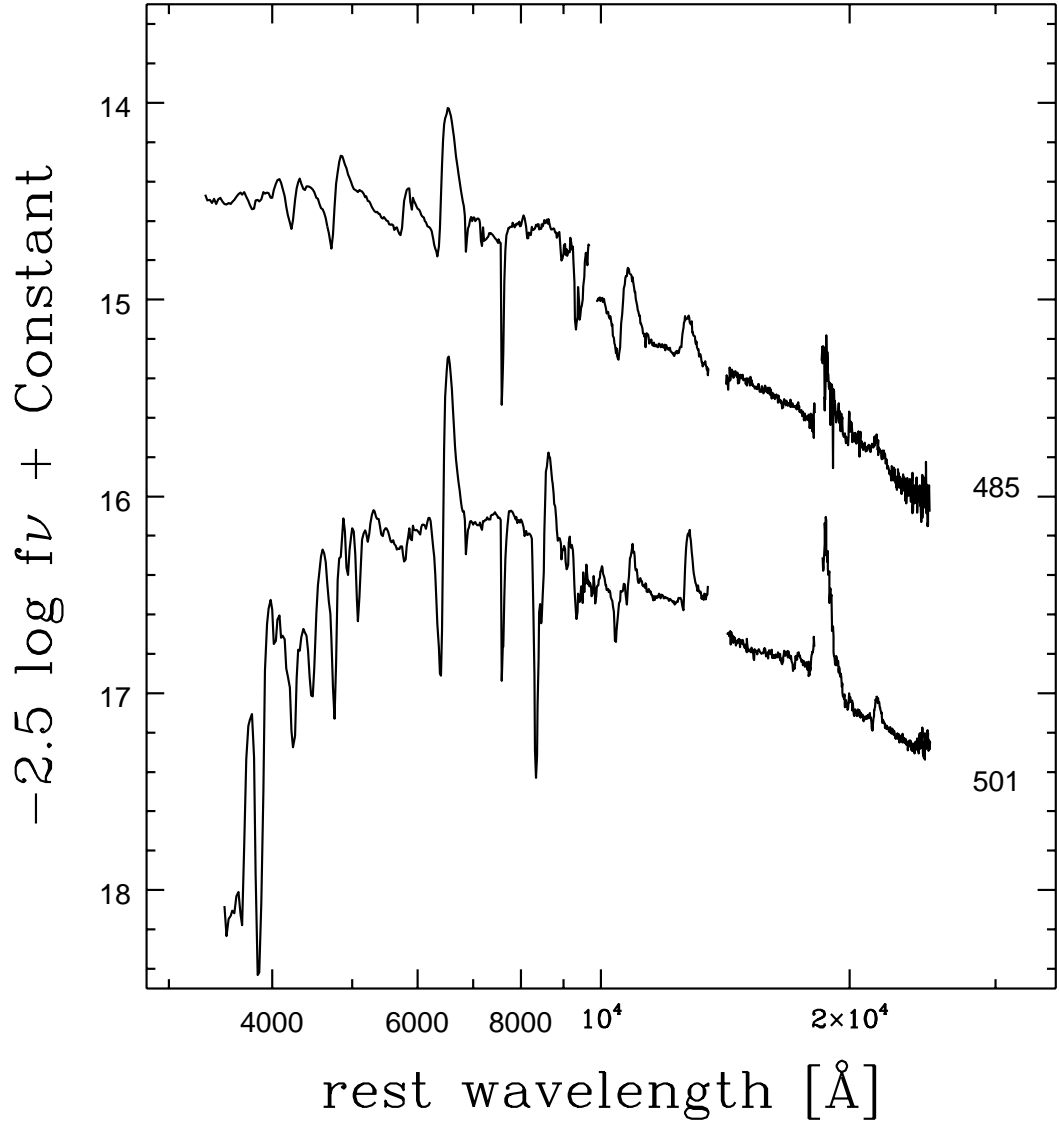}
\figcaption[figure5.ps]{Combined optical and IR spectra of SN~1999em
in AB magnitudes. The IR spectra spectra were obtained one day before
the optical observation and the mean Julian Day (-2451000) is indicated next
to each spectrum. The flux excess between 7500-10000 \AA~in
the first-epoch spectrum is due to second-order blue light contamination. \label{sn99em.optir.fig}}
\end{figure}

Given that two of the IR spectra were obtained one
day apart from the optical spectra, we were able to combine
these observations in Figure \ref{sn99em.optir.fig}. This exercise
revealed the excellent agreement between the optical and IR fluxes,
a result made possible by synthesizing broad-band magnitudes
and adjusting the flux scales according to the observed
magnitudes. Second-order contamination can be clearly seen in
the first-epoch spectrum as a flux excess between 7500-10000 \AA.

\section{THE APPLICATION OF THE EXPANDING PHOTOSPHERE METHOD TO SN~1999em}

EPM involves measuring a photometric angular radius and a spectroscopic
physical radius, from which the SN distance can be derived.
In this section we apply the method to SN~1999em following such order.
In Appendix A we summarize the basic ideas behind EPM.

\subsection{The angular radius of SN~1999em}

The angular radius, $\theta$, of the SN can be determined from equation \ref{angdist}
by fitting Planck curves [$B_\lambda(T_S)$] to the observed magnitudes.
With two wavelengths the solution is exact (two equations and two unknowns).
For three or more wavelengths we use the method of least-squares at each epoch to find
the color temperature $T_S$ and the parameter $\theta\zeta_S$ that minimize the quantity

\begin{equation}
\chi^2=\sum_{\overline{\lambda}~\epsilon~S} \frac {[m_{\overline{\lambda}}~+5~log~(\theta\zeta_S)~-~b_{\overline{\lambda}}(T_S)]^2}{\sigma_m^2}.
\label{aseqn}
\end{equation}

\noindent In this equation $m_{\overline{\lambda}}$ is the SN's apparent magnitude in a photometric band
with central wavelength ${\overline{\lambda}}$, $\sigma_m$ is the corresponding photometric error,
$b_{\overline{\lambda}}(T_S)$ is the synthetic magnitude of $\pi B_\lambda(T_S)10^{-0.4A(\lambda)}$
(computed with the precepts described in Appendix B), and $A({\lambda})$ is the dust extinction affecting the SN.
Once $\theta\zeta_S$ is determined from the $\chi^2$ fit, we solve for $\theta$ using
the dilution factor $\zeta_S$ corresponding to filter subset $S$, which we compute
from the color temperature using equation \ref{zetaeqn}. 

\begin{figure}
\epsscale{1.0}
\plotone{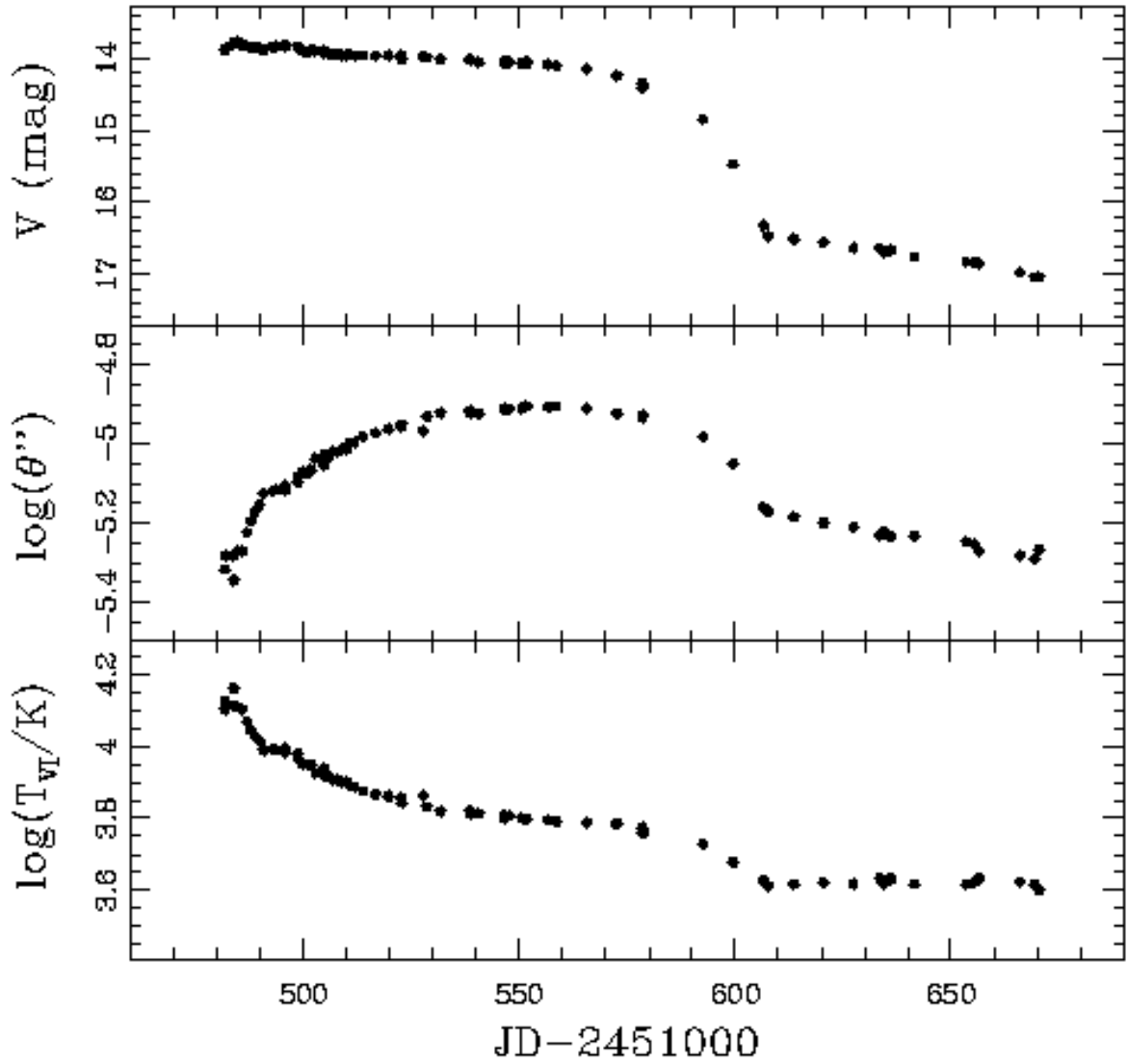}
\figcaption[figure6.ps]{$V$ magnitude, photospheric angular radius, and color temperature
of SN~1999em vs. Julian Day. \label{ltv.fig}}
\end{figure}

Using the $VI$ photometry and adopting a foreground extinction of $A_{Gal}(V)$=0.13 
(Schlegel, Finkbeiner, \& Davis 1998) and $A_{Host}(V)$=0.18 for dust absorption in the host galaxy \cite{baron00},
we compute $T_{VI}$ and the corresponding angular radius, which are plotted
in Figure \ref{ltv.fig}. This figure shows that
the initial color temperature was $\sim$13,000 K and that the SN 
cooled down for nearly 60 days until reaching a value near 6,500 K.
The photosphere remained at this temperature for 30 days more until the 
end of the plateau phase. In the rapid fall off the plateau, the atmosphere
cooled  to $\sim$4,000 K after which the temperature remained nearly constant.
At this stage the SN began the nebular phase and the spectrum was dominated
by recombination lines so the color temperature cannot be associated with a thermal 
process. During the initial cooling period
of 60 days the photospheric radius steadily increased as the photosphere
was swept outward by the expansion of the ejecta, and then decreased
as the wave of recombination moved into the flow with ever-increasing speed.
It is this balance between increasing radius and steady cooling which explains
the long plateau of nearly constant luminosity (top panel).
Following  the plateau phase, both the radius and the temperature
decreased, leading inevitably to the luminosity drop  around day JD 2451580.

\begin{figure}
\epsscale{1.0}
\plotone{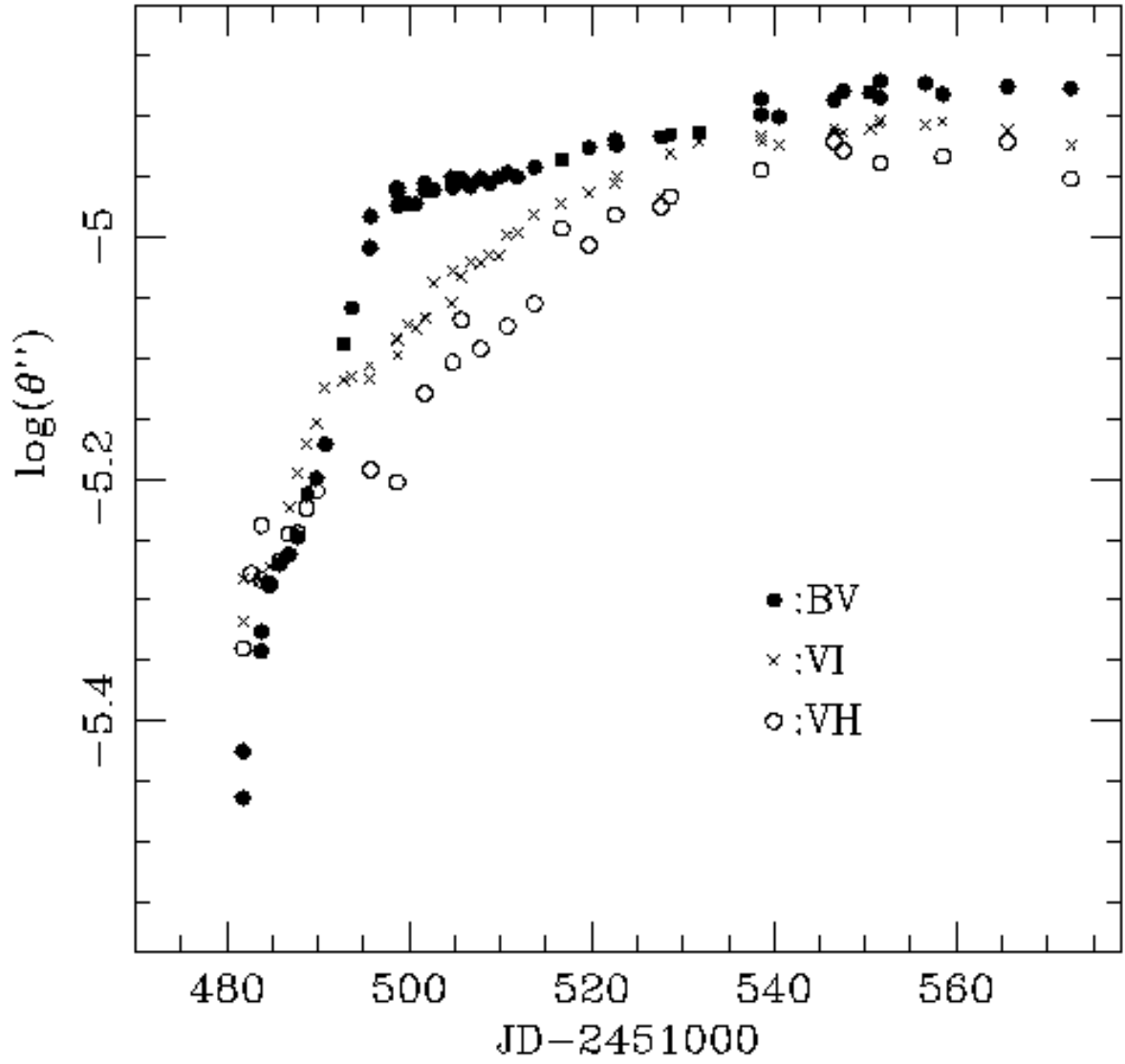}
\figcaption[figure7.ps]{Photospheric angular radius of SN~1999em derived from filter subsets $\{BV, VI, VH\}$,
as a function of time. \label{ast.fig}}
\end{figure}

Since the radiation field forms at a depth which varies with wavelength 
(depending on the value of the absorptive opacity), the angular radii derived
before corrections for the dilution factors are expected to vary significantly
depending on the filters employed in the blackbody fits.
By contrast, the photosphere is defined as the region of total optical depth
$\tau$=2/3, the last scattering surface. Since continuum opacity in the
optical and NIR is dominated by electron scattering, the opacity is grey,
and the photospheric angular radius is not expected to change with wavelength (E96).
We can check this prediction in Figure \ref{ast.fig} by comparing the photospheric angular radii
derived from filter subsets $\{BV, VI, VH\}$ for the entire plateau phase of SN~1999em.
The relative agreement is excellent ($<$5\%) over the first 10 days of SN
evolution, after which there is a period of $\sim$30 days in which
the photospheric radii  derived from the three filter subsets disagree
by up to 25\%. Over the following 60 days the agreement is much better, at 
the level of 10\%. This test reveals the overall good agreement
of the photospheric radii  determined from different filters
and the good performance of the dilution factors.

\subsection{The physical radius of SN~1999em}

\begin{figure}
\epsscale{1.0}
\plotone{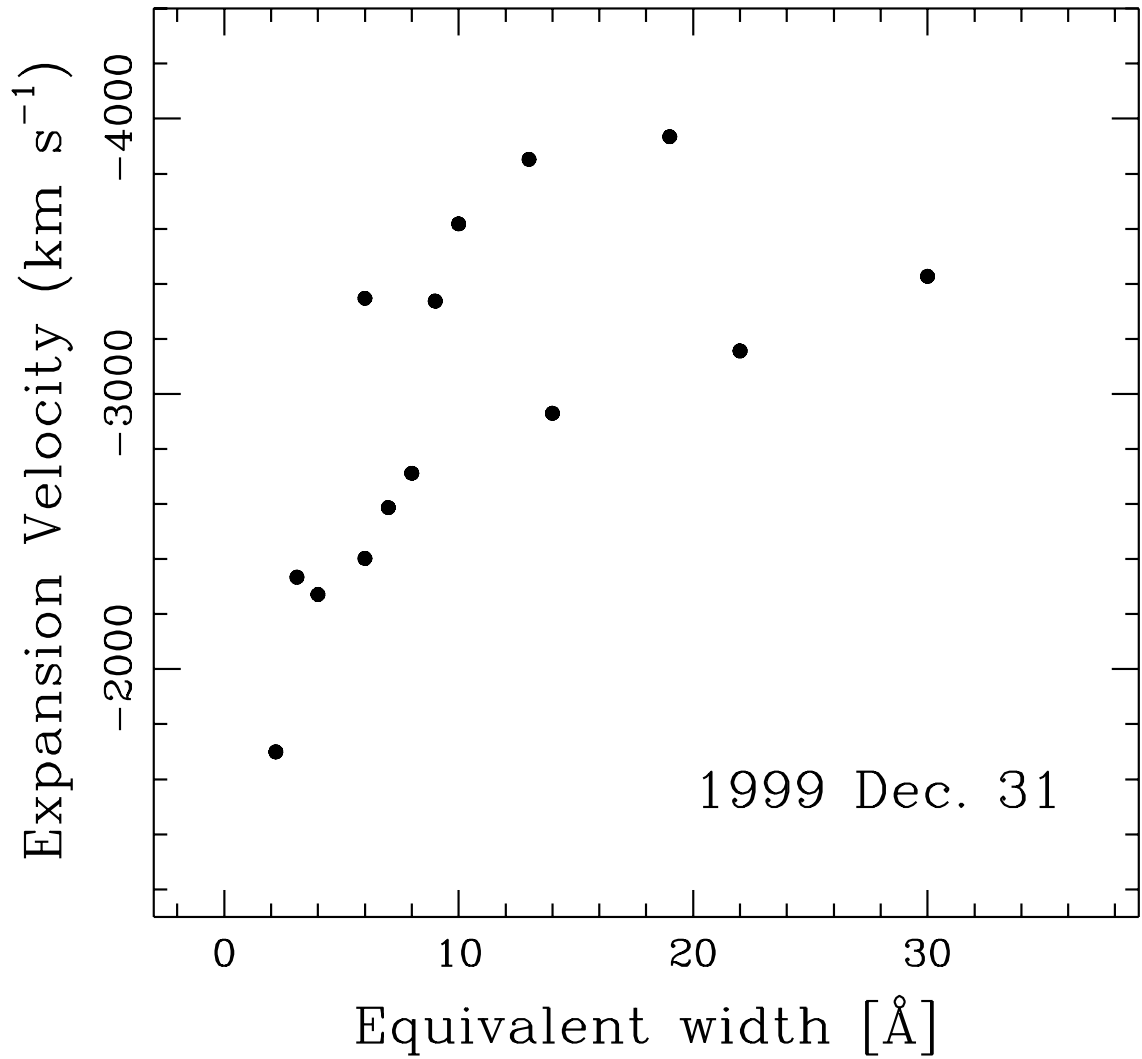}
\figcaption[figure8.ps]{Expansion velocity derived from the minimum of
spectral metal absorptions in the December 31 spectrum vs. the equivalent width of
the absorption. \label{lines_ew.fig}}
\end{figure}

The next step in deriving an EPM distance involves the determination of
photospheric velocities from the SN spectra.
To date the photospheric velocities have been estimated from the
minimum of spectral absorption features (Schmidt et al. 1992, 1994). There are several
problems with this approach, however.
First, the location of the line minimum
shifts toward bluer wavelengths (higher velocity) as the optical depth of the line 
and the scattering probability increase.
This is a prediction of line profile models in homologously
expanding scattering atmospheres [see Fig. 3 of Jeffery \& Branch (1990), for example],
and is also an observational fact as Figure \ref{lines_ew.fig} demonstrates.
This plot shows the expansion velocities derived from the metal (Sr, Ca, Fe, Na, Si)
lines of the spectrum taken on 1999 December 31, as a function of the equivalent width
of the line. Even though it proves difficult to measure 
line strengths due to the lack of a well-defined continuum,
the trend is quite evident in the sense
that stronger features yield higher velocities. It is well known
that hydrogen lines (which are not shown in this plot) yield
much higher expansion velocities than the metal lines in SN~II spectra due to their high optical depths, but no attention has
yet been paid to this effect among the weakest absorptions. This plot shows that it is
possible to incur significant errors if we measure expansion
velocities from the minima of absorption features, even from the weakest lines.
Second, even if we could extrapolate the observed velocities to zero strength,
the inferred velocity would correspond to that of the thermalization
surface (where the radiation field forms) and not to the photosphere
(the last scattering surface). Since the dilution factors computed
by E96 correspond to the ratio of the luminosity of the SN model to that
of a blackbody with the {\em photospheric} radius of such model (see Appendix C), the
use of the velocity of the thermalization surface is inappropriate,
even though it has been common practice in the past.
Third, velocities derived 
from absorption lines are affected by line blending or possible line misidentifications,
both of which can lead to an erroneous estimate of the photospheric velocity.

\begin{figure}
\epsscale{1.0}
\plotone{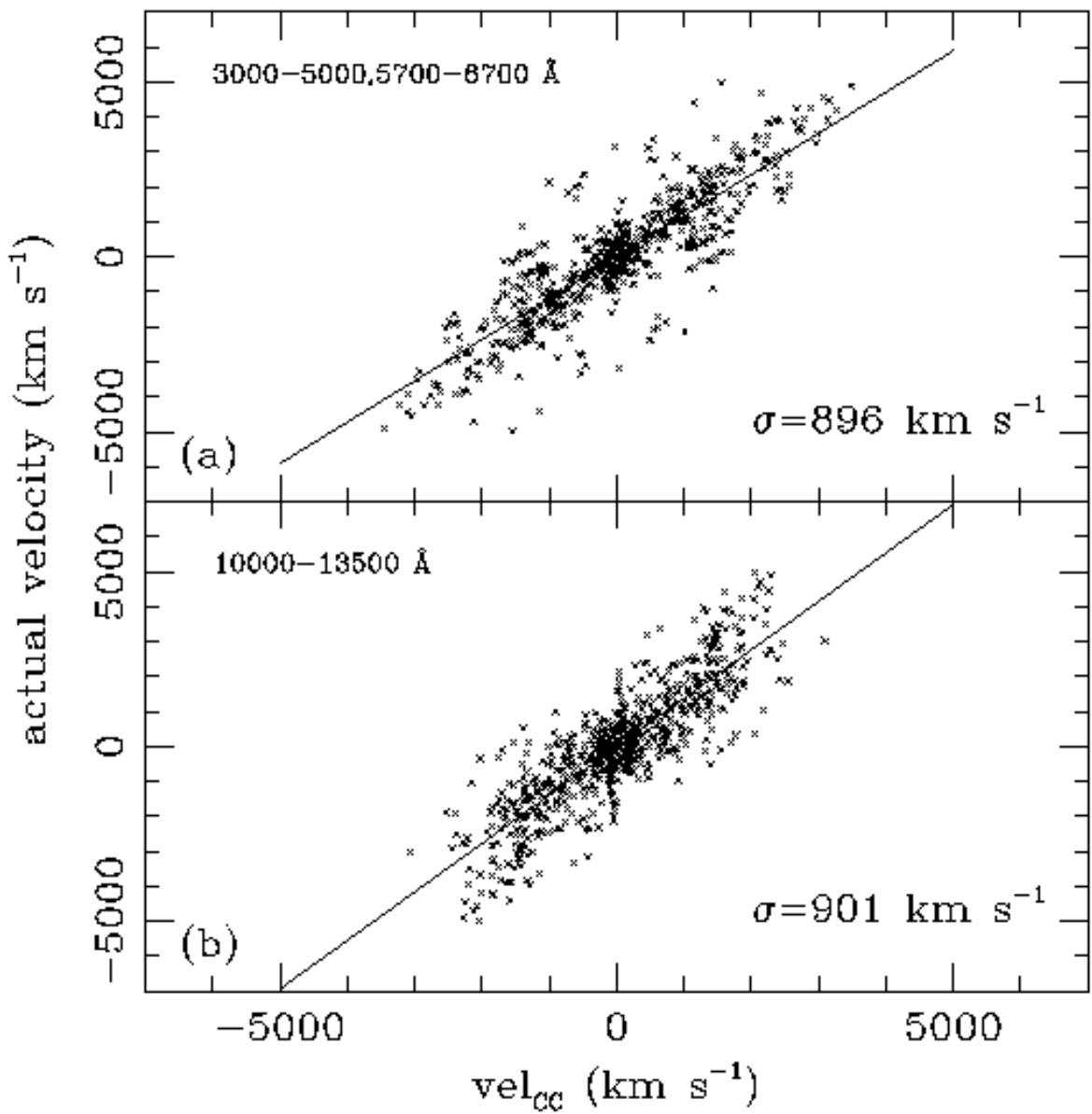}
\figcaption[figure9.ps]{(a) Relative expansion velocities derived from the cross-correlation
technique for pairs of model spectra with similar color temperatures, 
versus the actual relative velocities. The cross-correlation was
restricted to the range 3000-5000, 5700-6700 \AA. (b) Same as (a)
but for the range 10000-13500 \AA. \label{cc.fig}}
\end{figure}

To get around these problems we adopt an approach based on
cross-correlating the SN spectrum with the models of E96
(with known photospheric velocities) using the IRAF ``fxcor'' task.
Before applying this technique to the observed spectra, we must
test it with the model spectra of E96. In doing so we
cross-correlate models with other models having
$BVI$ color temperatures within $\pm$1,000 K of each other.
For each pair of models we end up with a relative velocity derived
from the cross-correlation (CC, hereafter) technique which can then
be compared with the actual value. The hope is that the whole set of
pairs can be used to derive a relationship between the CC relative
velocity and the actual value. We carry out tests separately
in the optical and the IR in order to apply this method 
to spectra observed in different spectral windows.
In the optical we select two wavelength ranges (3000-5000, 5700-6700 \AA)
and a NIR a window between 10000-13500 \AA. We end up with these ranges after
numerous experiments which show that beyond 13500 \AA~there are too few
spectral lines to help us constraining the expansion velocity. 
In the optical these tests suggest us to eliminate the red wings of the
strong H$\alpha$ and H$\beta$, which have the potential to bias
the derivation of expansion velocities from the CC procedure.
Figure \ref{cc.fig} compares the CC relative velocities and the actual values,
both in the optical and NIR, from the whole set of models (except for s15.5.1 and s25.5.1
which are not appropriate for SNe~II-P, E96). In both cases we get a reasonable correlation
which permits us to convert the velocity offsets measured from cross-correlation into
a photospheric velocity.  The scatter in these relationships is 900 km s$^{-1}$
which provides an estimate of the precision in the derivation of a photospheric velocity
from a single model. The data can be adequately modeled with straight lines, both
in the optical and IR. Least-squares fits yield slopes of 1.18 in the optical and
1.38 in the NIR, and zero-points of -3 km s$^{-1}$ in both cases. This implies that
the magnitude of the relative velocity is {\em smaller} than the actual value.
Overall, this is an encouraging result, although it must be mentioned that the
points with the largest scatter in Figure \ref{cc.fig} correspond to pairs
with the largest temperatures. This means that the precision of the CC technique
drops to $\sim$2000 km s$^{-1}$ when $T_{BVI}$$>$8,000 K.

\begin{figure}
\epsscale{1.0}
\plotone{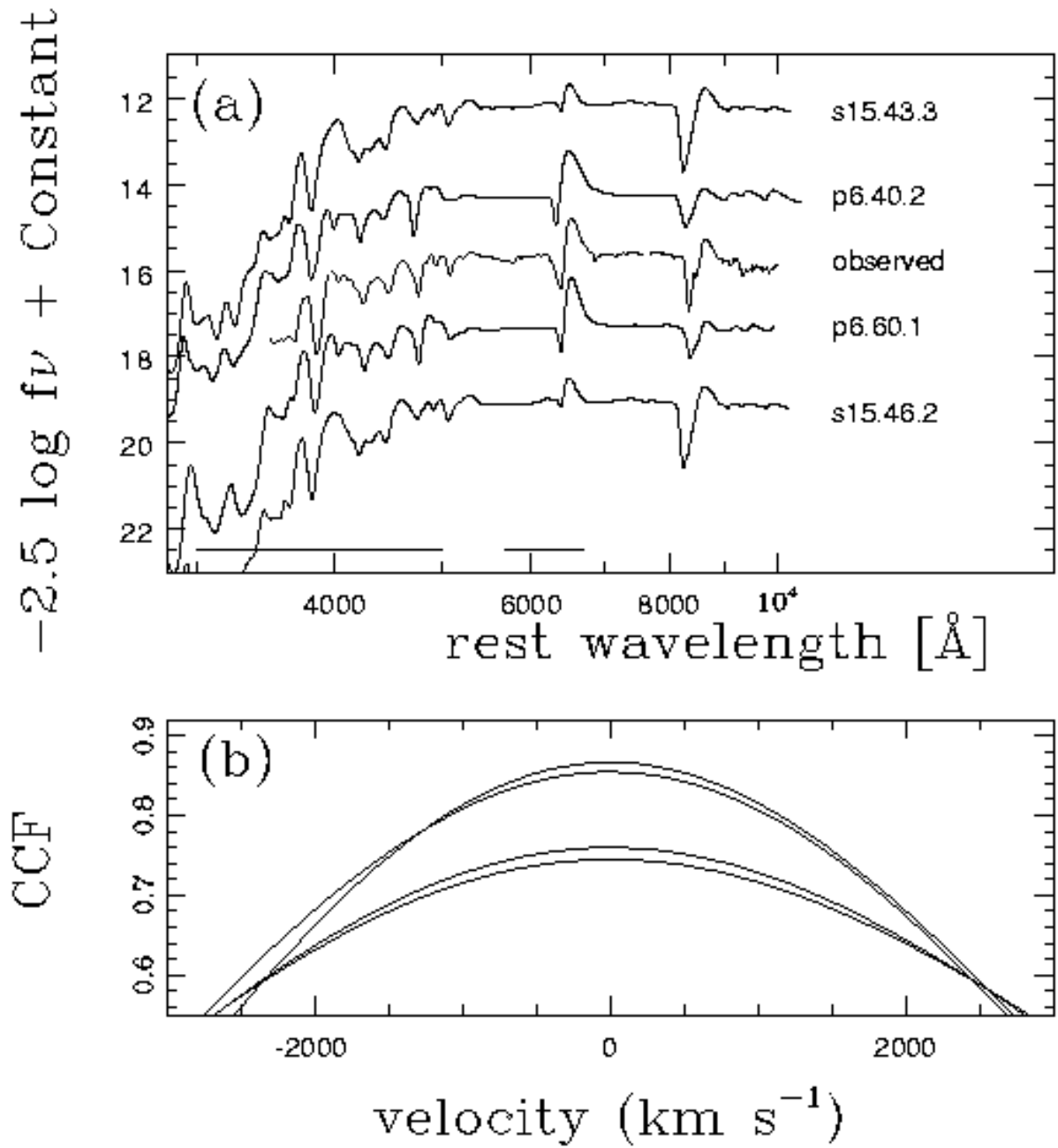}
\figcaption[figure10.ps]{(a) Optical spectrum taken on JD 2451501.66
(thick line) compared to four models with similar color temperature (thin lines).
The horizontal bars show the wavelength ranges (3000-5000, 5700-6700 \AA)
used in the derivation of relative velocities from the cross correlation technique.
(b) Cross correlation function between the observed spectrum and the four models
shown above. The two curves with the highest peaks correspond to models
p6.60.1 and p6.40.2, both of which match well the observed spectrum. 
The two lower curves correspond to models s15.43.3 and s15.46.2 which
provide a poorer match to the observed spectrum. \label{spec_comp.fig}}
\end{figure}

Having come up with a method to estimate photospheric expansion velocities we can
proceed to derive velocities for SN~1999em.
We start by selecting atmosphere models with $BVI$ color temperatures within $\pm$1,000 K
of each observed spectrum, after which we cross-correlate the SN spectrum and the subsample of
models in the aforementioned wavelength windows. 
The outcome of this operation is a cross-correlation function (CCF)
with a well-defined peak whose location in velocity space gives the
relative velocity of the observed and reference spectrum.
The height of the CCF is a measure of how well
the spectral features of the two spectra match each other.
Figure \ref{spec_comp.fig} shows examples of the CC technique. Panel (a) compares the 
optical spectrum obtained on JD 2451501.66
(thick line) with four models of similar color temperature (thin lines), and
panel (b) shows the corresponding CCFs. The two models
that best match the observed spectrum, p6.60.1 and p6.40.2,
are the ones that give the highest CCF peaks. 
Models s15.43.3 and s15.46.2, on the other hand, provide poorer
matches to the observed spectrum and, consequently, the lowest CCF peaks.
For each cross-correlation we get a relative velocity which we
proceed to correct using the calibrations shown in Figure \ref{cc.fig}, in order
to get the actual relative velocity.
Since the photospheric velocities of the models are known, we can
derive independent photospheric velocities for each SN spectrum.
Table 5 summarizes the velocities derived with this method for all
of the spectra of SN~1999em. In this table $v_{CC}$ is the relative
velocity determined from the CC method, $v_{model}-v_{SN}$ is
the actual relative velocity derived from the calibration
provided by the straight lines in Figure \ref{cc.fig}, and
$v_{SN}$ is the photospheric expansion velocity of the SN.
Note that, for a given epoch, the expansion velocities
derived from different models agree quite well.
Table 6 presents the average velocity obtained for each epoch
along with the rms value derived from the different models.
Note the great agreement in the photospheric
velocities determined from optical and IR wavelengths. While the lines used in the
different spectral regions form at different depths, the CC
technique yields photospheric velocities that do not depend on
the wavelength region employed in the cross-correlation, which
is the expected behavior for the photosphere. 
Inspection of Table 6 shows that the scatter in velocity yielded
by multiple cross-correlations varies between 5-21\%. 
The errors in the average velocities are probably smaller
than the tabulated values because we make use of
various models to compute such averages.

\begin{figure}
\epsscale{1.0}
\plotone{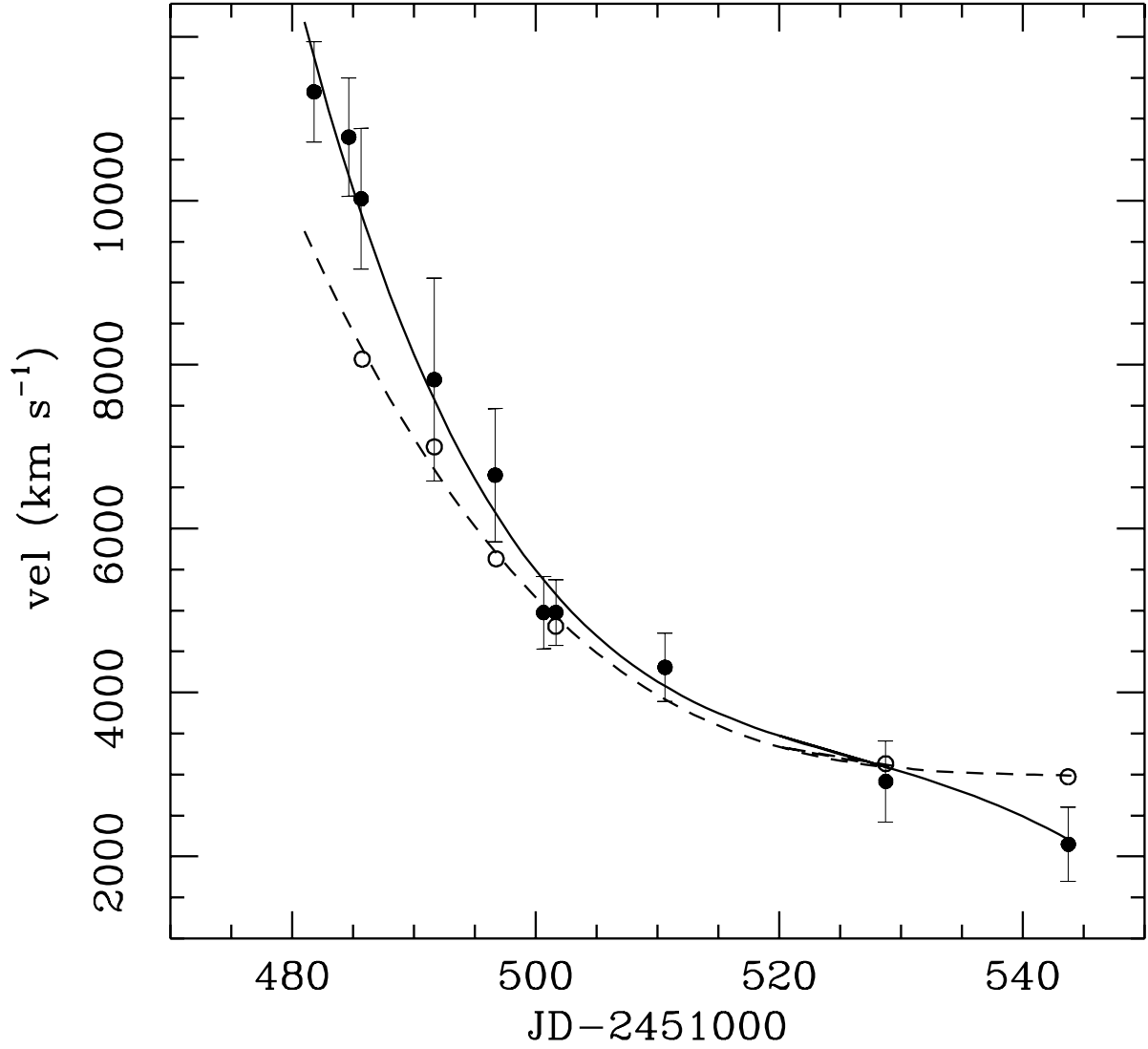}
\figcaption[figure11.ps]{Expansion velocity of SN~1999em vs. Julian Day,
derived from the cross-correlation technique (solid dots) and the minimum
of the spectral absorptions (open dots). The solid and dashed lines represent
polynomial fits to the solid and open points, respectively. \label{vel.fig}}
\end{figure}

For comparison we include in Table 6 velocities determined from the conventional
method of measuring the wavelength of weak absorption lines. Following the
approach of Schmidt et al. (1994), we list velocities measured from
Fe II $\lambda$5018 and Fe II $\lambda$5169 ($v_{Fe}$), and from
Sc II $\lambda$5526 and Sc II $\lambda$5658 ($v_{Sc}$). 
We include also estimates from
all metal lines but the strong calcium features ($v_{lines}$). 
This comparison shows that $v_{lines}$$\sim$$v_{Fe}$$>$$v_{Sc}$.
While $v_{Fe}$ and $v_{Sc}$ have low internal errors when considered
individually, the average derived from multiple weak lines
displays considerable scatter. This demonstrates that 
the technique of using a few pre-selected weak absorption lines has the
potential to produce very different results, so that the internal
precision of the Fe or Sc method must reflect the velocity scatter
yielded by all lines, which is $\sim$20\%.
Figure \ref{vel.fig} shows a comparison between the velocities derived
from measuring the weak metal lines and the CC method.
During the first 15 days, line velocities are $\sim$20\% lower than
the CC velocities, after which this difference becomes negligible.
In our last-epoch spectrum, on the other hand, there is a hint that
the velocity derived from absorption lines is higher than that obtained from
the CC method, although the difference amounts to only $\sim$1$\sigma$.

\subsection{The distance to SN~1999em}

In this section we use EPM to derive the distance to SN~1999em.
In doing so we adopt the foreground extinction $A_{Gal}(V)$=0.13 
measured by Schlegel et al. (1998). The presence of interstellar Na I D
in the SN spectrum at the wavelength corresponding to the velocity of the host galaxy with an equivalent
width of $\sim$2 \AA~suggests that additional reddening affected the SN. According to the
correlation between equivalent width and reddening of Munari \& Zwitter (1997),
SN~1999em was reddened by $E(B-V)$$_{host}$$\sim$1.0$\pm$0.15.
This estimate is highly inconsistent with the independent
estimate of $E(B-V)$$_{host}$$\approx$0.01-0.06 and $E(B-V)$$_{host}$$<$0.11
from theoretical modeling of the spectra of SN~1999em (Baron et al. 2000)
\footnote[8]{That paper provides constraints to $E(B-V)$$_{total}$, which
includes 0.04 mag of foreground reddening.}.
This disagreement suggests that the light of SN~1999em
was likely absorbed by a gas cloud with relatively large gas to dust ratio, perhaps
ejected by the supernova progenitor in an episode of mass loss.
This example demonstrates the difficulty of using
the equivalent width of interstellar lines to estimate
dust extinction due to the uncertainty in the calibration.
Another example that illustrates the problem of applying the Galactic
calibration of Munari \& Zwitter to SNe is 
the highly reddened Type Ia SN~1986G. The spectra of this SN revealed
interstellar Na I D absorption with an equivalent width of 4.1 \AA~\cite{phillips87} 
which implies $E(B-V)$$\sim$2, yet the reddening yielded by
color considerations  was $E(B-V)$=0.62 \cite{phillips99}.

\begin{figure}
\epsscale{1.0}
\plotone{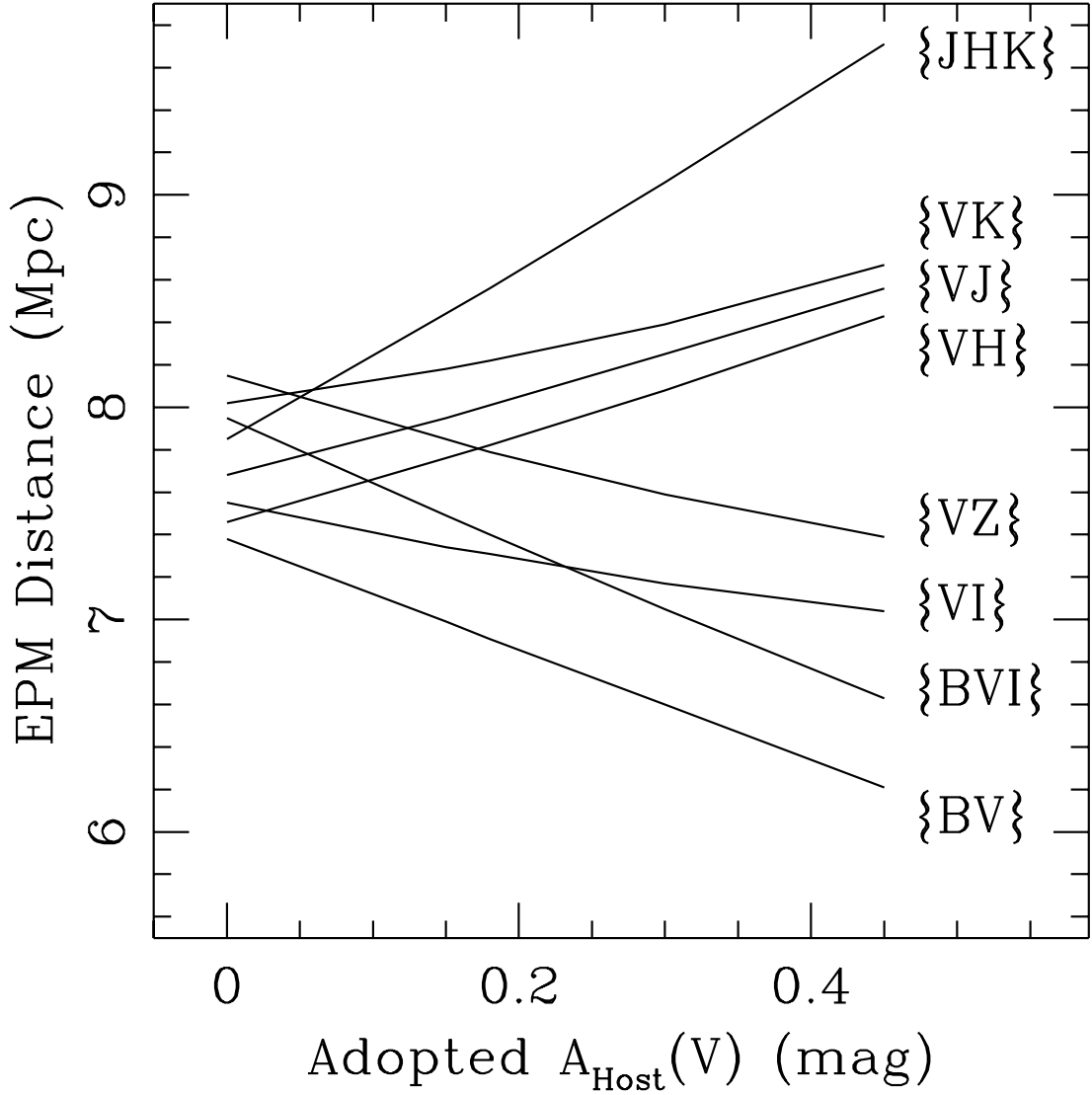}
\figcaption[figure12.ps]{EPM distance derived from different filter subsets,
as a function of the adopted visual extinction in the host galaxy. \label{dist_Av.fig}}
\end{figure}

In what follows we examine the sensitivity of EPM to dust
for which we adopt a   wide range of extinction values
between $A_{Host}(V)$=0-0.45 mag.
Figure \ref{dist_Av.fig} shows the EPM distances
derived for eight filter subsets 
(from $B$ through $K$). This plot reveals that the distance
from subset \{$BV$\} decreases from 7.4 to 6.2 Mpc 
when the adopted extinction increases from 0 to 0.45 mag.
A similar behavior can be appreciated from filters \{$BVI$\}.
From \{$VI$\} we find that the distance changes only from 7.6 to 7.0 Mpc, i.e.,
EPM is very insensitive to our choice of $A_{Host}(V)$ when observing with these filters.
This trend reverses with subsets including the IR filters,
i.e., the distance gets bigger as the  adopted extinction increases.
This plot shows empirically that the EPM distances are quite
robust to the effects of extinction.
The other interesting feature of this plot is that
these curves show a convergence toward small values of $A_{Host}(V)$,
which allows us to constrain the value of the extinction in the host galaxy.
In what follows we adopt $A_{Host}(V)$=0.18 which is 
the most likely value derived by Baron et al. (2000) and, as
Figure \ref{dist_Av.fig} shows, is also consistent with the EPM analysis.

Tables 7-10 summarize the EPM quantities derived for SN~1999em
using eight filter subsets. 
The photospheric velocities come from the polynomial fit
to the velocities obtained from the CC technique
(the solid line in Figure \ref{vel.fig}), to which we
assign a statistical uncertainty of 5\%. 
The color temperature $T_S$ and the quantity $\theta\zeta_S$ 
correspond to parameters obtained from the $\chi^2$
minimizing procedure described in Sec. 3.1 (equation \ref{aseqn}),
which allows us to estimate statistical uncertainties in the derived
parameters from the photometric errors. Also listed in these tables
is the dilution factor    $\zeta_S$ required for the derivation of the
$(\theta/v)_S$ quantity needed for EPM. 
In the models of E96, this factor is primarily determined by temperature 
and, for a given temperature, $\zeta_S$ changes only by 5-10\% over
a large range of other parameters. This is a remarkable result, permitting us
to compute $\zeta_S$ for SN~1999em without having to craft specific
atmospheric models. Using this approximation and the derived color
temperature, we compute $\zeta_S$ from the polynomial fits to
the dilution factors for our photometric system (Appendix C; equation \ref{zetaeqn}).
With this approach we expect a component of
systematic error ($\sim$5-10\%) in $(\theta/v)_S$. Since this
is not a statistical uncertainty, we do not include it
in the error quoted for $(\theta/v)_S$ in Tables 7-10.

\begin{figure}
\epsscale{1.0}
\plotone{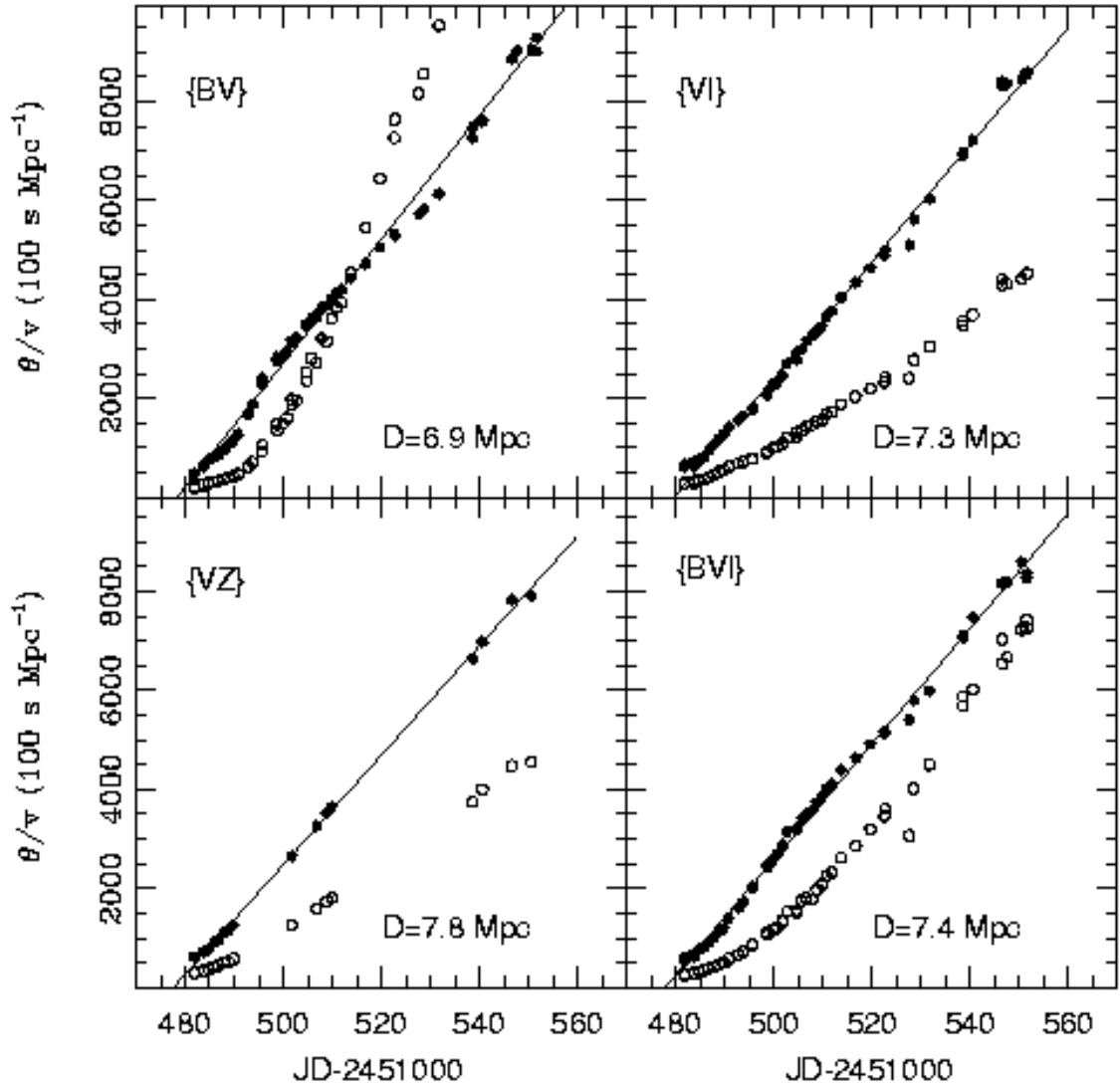}
\figcaption[figure13.ps]{$\theta$/$v$ as a function of time for filter subsets
$\{BV, VI, VZ, BVI\}$. Open dots show $\theta$$/v$ uncorrected for
dilution factor while filled dots show the
parameter corrected with the factors computed
by E96. In theory, this quantity should increase linearly with time 
and the slope of the relation gives the distance (Appendix A).
The small departures of these points from the ridge lines demonstrate
the good performance of the dilution factors at 
different times over a broad wavelength range. \label{sn99em.epm1.fig}}
\end{figure}

Figure \ref{sn99em.epm1.fig} shows $\theta$$/v$ as a function of time,
for four subsets including filters $B$ through $Z$. 
The open dots show $\theta/v$ uncorrected for dilution factors
while the filled dots show the parameter after correction.
In theory, $\theta/v$ should increase linearly with time 
(except for the very first days after explosion) and the slope
of the relation gives the distance (equation \ref{diseqn}).
Overall, this figure reveals that the dilution factors produce distances
with reasonable consistency during a period of 70 days of SN evolution in which the
photospheric temperature dropped from 15,000 to 5,000 K, although it
is evident that there are systematic departures from linearity.
In particular, the earliest points provide evidence that
the angular radius is too large relative to the physical size 
implied by homologous expansion of the photosphere from a point. This
suggests that the progenitor's radius ($R_0$) might have
a non-negligible contribution to the SN radius in the first
observations (obtained a few days after shock breakout).
A week after explosion $R_0$ becomes negligible, even for the largest
red supergiants known ($R_0$=300$R_\odot$, van Belle et al. 1999), so its effect
in the determination of the distance can be ignored. Hence,
we proceed now to fit the data with a straight line using equation \ref{diseqn},
and in the next section we review the validity of this approximation.

To compute the distance it suffices, in principle, to perform a 
least-squares fit to the ($\theta_i/v_i$,$t_i$) points.
To perform such fit it is necessary to know
the uncertainty in each of the $\theta_i/v_i$ points. In our case
this is rendered difficult by our lack of knowledge
of the systematic errors in $\zeta_S$ which is needed
to obtain the $\theta/v$ parameter.
By weighting the fits by the errors, the fits are going
to be biased significantly to the earlier data which carry
a lot of weight. Equal weighting seems to be a more reasonable
way to derive an ``average'' distance. To estimate the
uncertainties in the distance and explosion time
we employ the bootstrap technique described by Press et al. (1992)
which makes use of the data themselves as an estimator of the
underlying probability distribution (which we do not know).
The method consists in randomly drawing from the parent
population a synthetic dataset of the same size as the parent.
By drawing points with replacement
we end up with a randomly modified dataset, from which
we perform a uniform-weight least-squares fit in order to solve
for the time of explosion $t_0$ and the distance $D$.
From a large number (10,000) of simulations we
obtain average parameters and 
estimates of the uncertainties in $t_0$ and $D$
from the dispersion among the many bootstrap realizations.

The ridge lines in Figure \ref{sn99em.epm1.fig} correspond to
the solutions determined from the bootstrap method and the results of
the fits are given in Table 11.
The range in distance (6.9-7.8 Mpc) derived from observing through
these four filter subsets is quite small ($\pm$6\%).
This shows the good performance of the dilution
over the broad wavelength range encompassed by the $BVIZ$ filters.
The nominal error yielded by the bootstrap method for the individual subsets
ranges between 1-2\%, which proves significantly smaller than
the 6\% distance range encompassed by the four subsets.
Most likely, this is a symptom of systematic errors in the dilution factors.

\begin{figure}
\epsscale{1.0}
\plotone{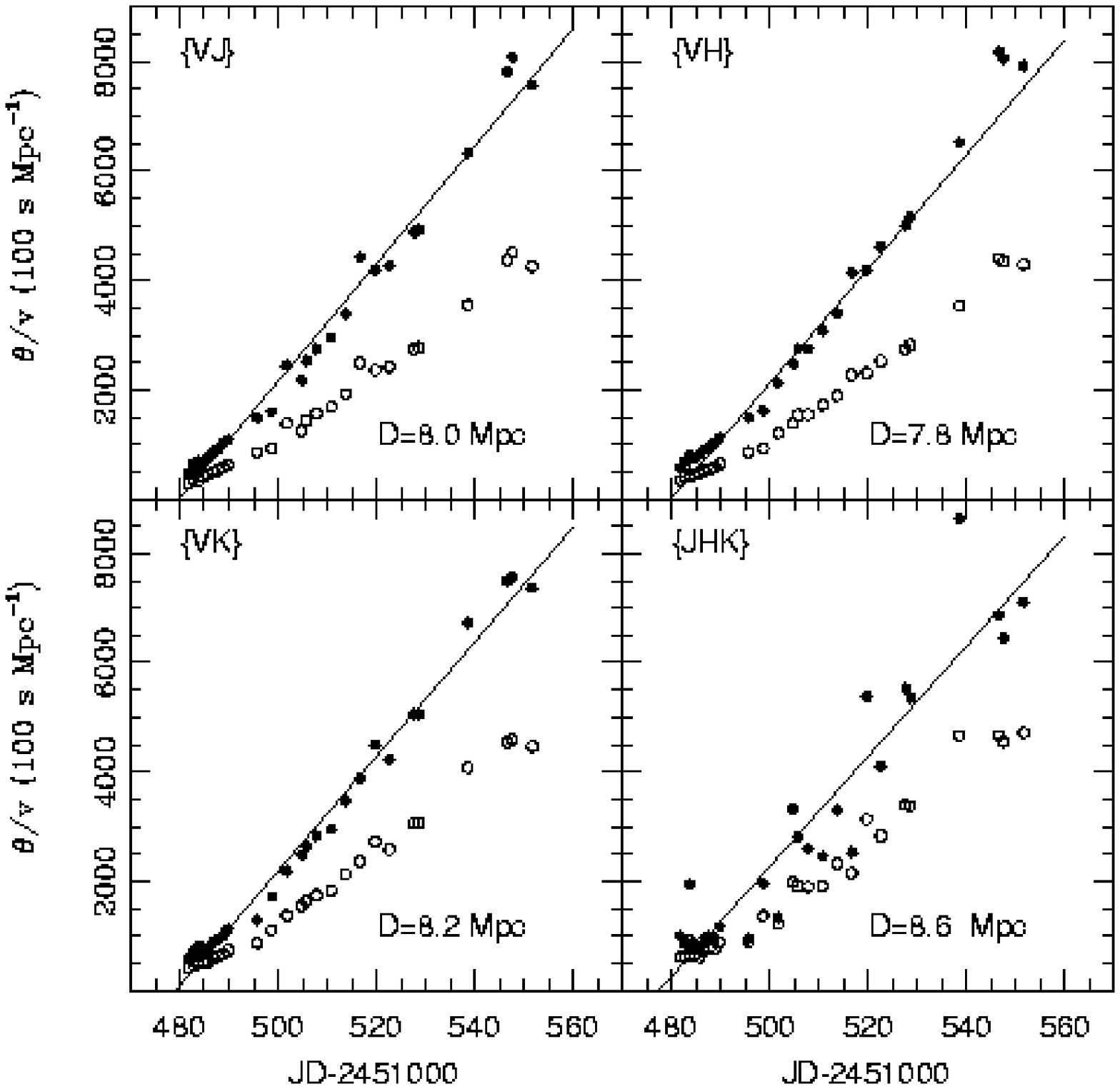}
\figcaption[figure14.ps]{Same as Figure \ref{sn99em.epm1.fig}, but for
filter subsets $\{VJ, VH, VK, JHK\}$. \label{sn99em.epm2.fig}}
\end{figure}

Figure \ref{sn99em.epm2.fig} shows $\theta$$/v$ as a function of time from
filter subsets including IR wavelengths. The scatter from
the ridge line is somewhat higher than that obtained from optical wavelengths
due to the relatively larger photometric errors 
in the IR. This problem becomes even stronger for the $\{JHK\}$ subset
because the errors in the derived color temperatures increase as the Rayleigh-Jeans limit is approached.
Despite the larger scatter in the distances derived from IR observations, this plot confirms 
the internal consistency of the dilution factors computed by E96
from IR wavelengths.  The fits to the data are listed in Table 11
which shows that the resulting distances are quite consistent with those
derived from optical observations, although the differences are somewhat
larger than the formal errors (more below).

We include also in Table 11 a solution derived from a
simultaneous least-squares fit to all of the $(\theta/v)_S$ values in Tables 7-10.
Not surprisingly, the resulting distance of 7.54 Mpc is close to the value of 7.75 Mpc
obtained from taking a straight average of the eight individual distances. We prefer the former
estimate because it weighs the distances of the individual subsets according to
their time sampling of the $(\theta/v)_S$ parameter. Also given in Table 11
is the distance of 7.82 Mpc determined from subset $\{BVIJHK\}$. In this case
we must restrict the sample of $(\theta/v)_{BVIJHK}$ values to those epochs in which we
obtained simultaneous observations with all six filters. Thus, the solution is not
weighted by the higher frequency of the optical observations so
the result is very close to the value of 7.75 Mpc obtained by averaging the individual
distances yielded by the eight subsets. In what follows we adopt the
value of 7.54 Mpc which takes into account the better sampling
of the optical light curves.

The values derived for the explosion time are listed in Table 11.
The average from the eight filter subsets is JD 2451478.8 ($\pm$1 day).
This confirms that the SN was caught at a very
young stage and that our first photometric and spectroscopic observation was obtained
at an age of only three days after shock breakout.
This is far earlier than any other SN, except for SN~1987A.

\section{DISCUSSION}

One of the challenges of EPM involves the determination of 
photospheric velocities. The spectra of SN~1999em 
show that the technique of cross-correlating the SN spectra with
the models of E96 can produce expansion velocities with an average
uncertainty of 11\%. This is significantly lower than the 20\%
precision yielded by the method of measuring velocities 
from the minimum of weak absorption features (Table 6). The cross-correlation
technique is not only more precise, but also more accurate, since
the velocity derived from individual lines tend to correspond
to the value of the thermalization surface, which expands
more slowly than the photosphere (the last scattering surface).
A clear advantage of the CC method is that it can be used during the initial hot 
phases of SN evolution when no weak lines are available.
The other advantage of the CC technique
is that it permits one to estimate velocities from lower
S/N spectra, thus extending the potential of EPM to high redshifts 
where high-quality spectra are difficult to obtain.

The effects of dust extinction can seriously hamper
the determination of distances when using ``standard candle'' techniques.
EPM, on the other hand, is quite robust to the effects of dust absorption
as pointed out by E96 and by Schmidt et al. (1992, 1994).
While extinction reduces the observed flux, it
also makes the photosphere appear cooler and hence less luminous,
so that these two effects cancel to a significant degree.
The data of SN~1999em provide empirical confirmation that the EPM distance
to this SN is not very sensitive to the adopted absorption.
We find that while the EPM distances derived from optical colors
decrease with increasing $A_{Host}(V)$, this trend reverses when
using IR filters (Figure \ref{dist_Av.fig}).
This analysis reveals that, even though IR photometry is less
affected by dust, the EPM distances derived from filter subsets
including one or more IR filters are not less sensitive to dust
than those determined from optical filters alone.
This result challenges the suggestion of Schmidt et al. (1992), namely,
that one of the advantages of using IR for EPM is that
``the uncertainty in a distance due to extinction
is less than half that incurred when optical photometry is used''.
Our analysis shows that the \{VI\} and \{VZ\} subsets have
the least sensitivity to the effects of dust. In particular,
the distances derived with the \{VI\} filters vary by a mere
7\% when the adopted absorption is varied in a wide range of
values between 0-0.45 mag. The other interesting result is
that, despite the weak sensitivity of EPM to dust,
multi-color observations can be useful for constraining $A_{Host}(V)$.

\begin{figure}
\epsscale{1.0}
\plotone{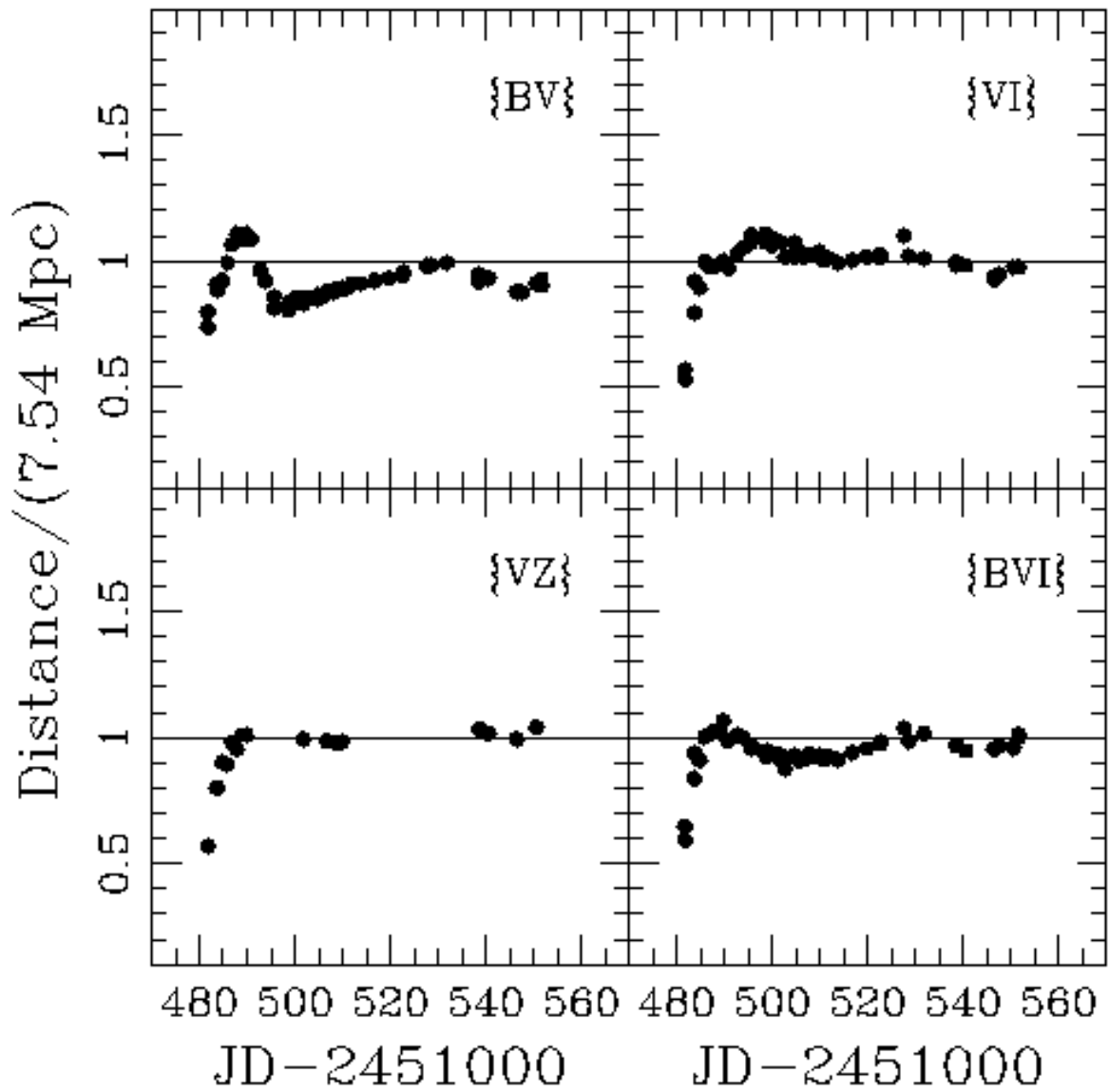}
\figcaption[figure15.ps]{EPM distance as a function of time,
for filter subsets $\{BV, VI, VZ, BVI\}$. \label{sn99em.dist.res1.fig}}
\end{figure}

\begin{figure}
\epsscale{1.0}
\plotone{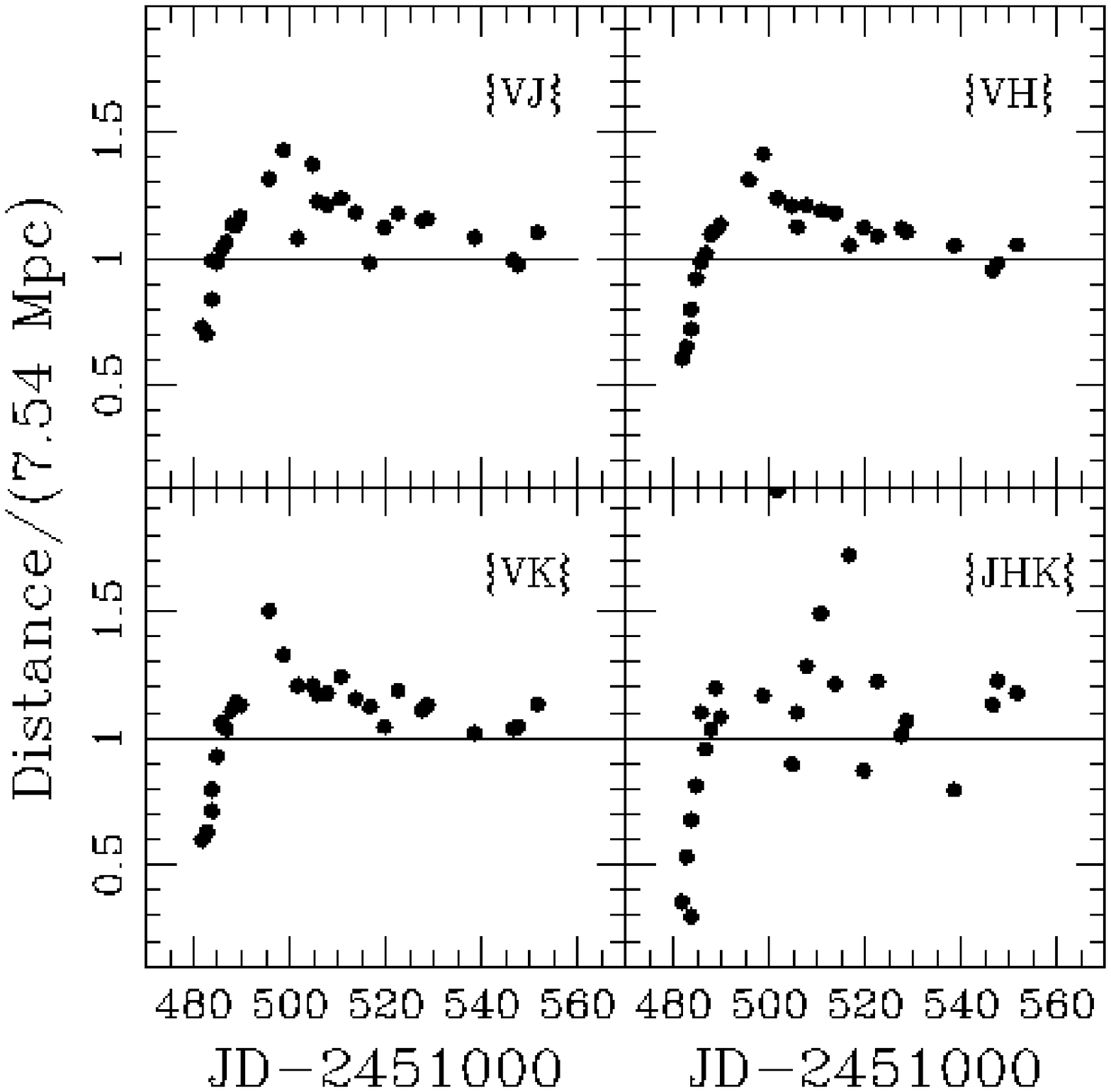}
\figcaption[figure16.ps]{Same as Figure \ref{sn99em.dist.res1.fig}, but for
filter subsets $\{VJ, VH, VK, JHK\}$. \label{sn99em.dist.res2.fig}}
\end{figure}

\begin{figure}
\epsscale{1.0}
\plotone{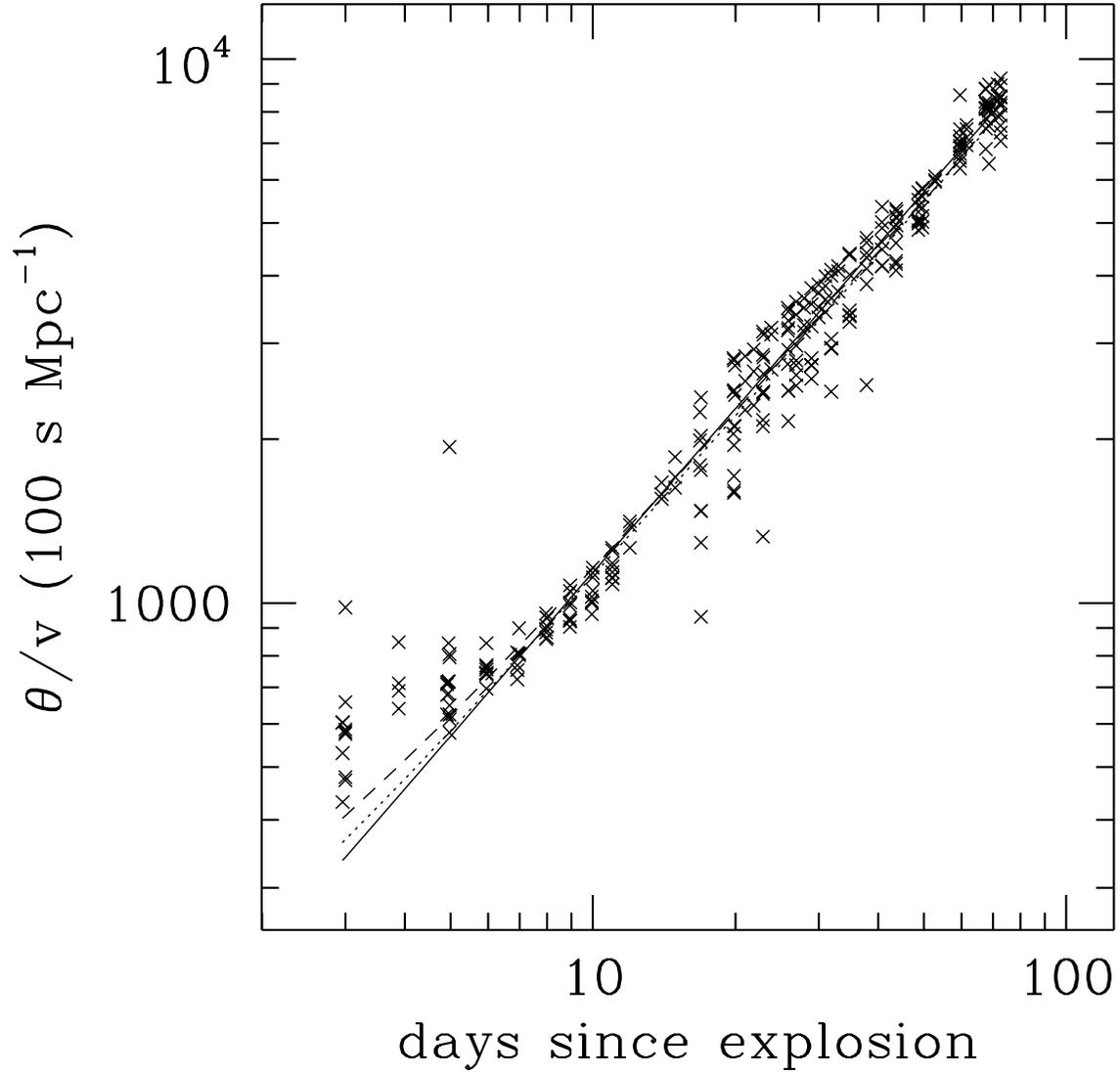}
\figcaption[figure17.ps]{$\theta$/$v$ as a function of time since explosion (JD 2451478.8)
for filter subsets $\{BV, VI, VZ, BVI, VJ, VH, VK, JHK\}$.
The solid line shows the regression line for $R_0$=0, while the dotted line
corresponds to the fit obtained with $R_0$= 5$\times$10$^{13}$cm (714 R$_\odot$). \label{all.fig}}
\end{figure}

EPM has the great advantage that observations at different epochs
are essentially independent distance measurements. The exceptional
data obtained for SN~1999em afford a unique opportunity
to perform this valuable internal check and test the dilution factors of E96
over a wide range in temperature and wavelength. Figures \ref{sn99em.dist.res1.fig} 
and \ref{sn99em.dist.res2.fig} show that the EPM distance (in units of the average
value of 7.54 Mpc obtained from the eight filter subsets) varies systematically over time
(the error of one point is highly correlated to the error in the next point).
This problem is particularly pronounced over the first week since explosion (JD 2451478.8),
in which the distances prove even 50\% lower than the average owing to the large
photometric angular radius relative to the physical size of the SN. There are
different possible causes for this discrepancy which we proceed to examine. 
The high degree of correlation in the errors derived from the different
filter subsets suggests that the small linear radius might be due to
an underestimate of the photospheric velocity. If this was the case the expansion velocity in the first
observation epoch (JD 2451481.79) would have to be 50\% higher than the
adopted value. This is well beyond the $\sim$5\% uncertainty in the CC technique, so we
can rule out errors in the photospheric velocities as the source of this discrepancy.
Another possible cause for the large initial distance residuals is the
neglect of the initial radius $R_0$ in the linear size of the SN.
To examine this issue in detail, Figure \ref{all.fig} shows how $\theta/v$
changes with time in a log-log scale, for the eight filter subsets.
Overplotted are the lines corresponding to homologous expansion
(equation \ref{gendiseqn}) for the case of a progenitor with $R_0$=0 (solid line) and 
$R_0$=5$\times$10$^{13}$cm (dotted line), which is 2$\times$ larger
than the largest supergiant known \cite{vanbelle99}. It is evident
that the two models are almost identical at later times (when the
effects of $R_0$ in the SN radius become negligible), and that
they both fit the data very well after day 7. The earliest points, on
the other hand, all fall on the high side of the lines of homologous expansion.
This plot shows that, while the initial radius of the SN progenitor can account
for some of the high $\theta/v$ values observed at the earliest epochs,
there must be other reasons to explain the relatively large angular size of the SN.
We cannot rule out, of course, that the large initial residuals are due to
incorrect dilution corrections which would act to increase the derived values
of $\theta$. For this to be true, the dilution corrections for all filter subsets
should be increased by 2$\times$. It is conceivable that such error could be caused by
circumstellar material which could lead to the formation of the photosphere at a much larger radius.
This is an interesting possibility that could be clarified with an expanded
set of atmosphere models for SNe~II.

Leaving aside the origin of the high initial distance residuals it is interesting to ask what
is their effect in the derived distance to SN~1999em.
For such purpose we employ equation \ref{gendiseqn} to solve simultaneously for
$R_0$, $t_0$, and $D$. A least-squares fit to the data, however, yields a
degenerate and non-physical solution for the three parameters, most likely
caused by the large residuals of the earliest points which demand an extremely
large $R_0$. If, instead, we fix $R_0$ to the value of a very large
progenitor, 5$\times$10$^{13}$cm, we obtain a modest increase of 4\% 
in the distance. Alternatively, we now ask what would be the distance
if we exclude the first observations. Limiting the dataset to epochs later
than JD 2451485.7 (one week after explosion), a linear fit yields
$D$=7.40$\pm$0.09 Mpc and $t_0$=2451479.7$\pm$0.5, which are indistinguishable
than the values derived from the entire dataset.

After day 7 (JD 2451485) the scatter in the EPM distances is  much lower:
9\% in $\{BV\}$, 4\% in $\{BVI\}$, 4\% in $\{VI\}$, and 4\% in $\{VZ\}$. 
Inspection of Figure \ref{sn99em.dist.res1.fig} shows that the residuals
vary systematically over time. If these were caused by the adopted velocity
the effect would be the same in the different filter subsets,
which is not the case. We believe, instead, that the problem lies in
the derived photospheric angular radii. In fact, Figure \ref{ast.fig}
reveals discrepancies of up to 10\% in the values of $\theta$ obtained
from different filter subsets. It could well be that these residuals 
are caused by the use of average dilution factors which, according
to E96, have errors between 5-10\%. Systematic errors in the
photometry could also explain the discrepancies in $\theta$. Although the nominal
uncertainties due to photon statistics are $\sim$0.015 mag, the transformation
of instrumental magnitudes to the standard system could have significant
systematic errors owing to the non-stellar nature of the SN spectrum \cite{hamuy90}.
In the IR the rms in the EPM distances after day 7 are
10\% in $\{VJ\}$, 10\% in $\{VH\}$, 9\% in $\{VK\}$, and 27\% in $\{JHK\}$.
In theses cases the observational uncertainties are partially responsible for these relatively
greater spreads, although it is evident that the EPM distances are systematically higher
than the average and that they decrease steadily with time.
We believe that this trend could be caused by systematic errors in the dilution factors
or the photometry.

The distance residuals shown in Figures \ref{sn99em.dist.res1.fig}
and \ref{sn99em.dist.res2.fig} reveal the potential problem
of applying EPM to SNe~II with small time baseline
light curves. To illustrate this point we compute distances
for SN~1999em using the data subset comprising the first days of SN evolution
(up to JD 2451490), which yields $D(BV)$=10.34, $D(VI)$=9.70,
$D(BVI)$=10.24, $D(VZ)$=9.95, $D(VJ)$=13.26, $D(VH)$=13.98,
$D(VK)$=14.46, and $D(JHK)$=20.78 Mpc. Note that these distances
are much higher than the 7.54 Mpc average value, whereas
Figures \ref{sn99em.dist.res1.fig} and \ref{sn99em.dist.res2.fig}
show that the EPM distances in the first week are on the low side from the average!
This difference turns out to be a consequence of the shallower slope displayed by the early-time
points in Figures \ref{sn99em.epm1.fig} and Figures \ref{sn99em.epm2.fig}.
A shallower slope implies a greater distance (equation \ref{diseqn}),
despite the fact that the individual
points imply smaller distances owing to their relatively larger photospheric
angular radii.  This test reveals that a ``snapshot'' distance based only
on a small time baseline light curve is clearly inappropriate.
Poorly sampled light curves, on the other hand, do not seem to be
a problem. We checked this by randomly drawing three data points from the
$\{VI\}$ dataset. From 100 realizations the computed distances fall within 10\% 
of the distance derived from the entire dataset (on average). For five data points
the distribution of distances has an rms of 6\% around the mean, while
for ten points the rms drops to only 4\%. This implies that poorly-sampled light
curves can yield precise distances, as long as the spacing in time is reasonable.

Schmidt et al. (1992) made the claim that one of the advantages of using
IR photometry for EPM is that the $JHK$ bands have fewer spectral features
(which can be checked in Figure \ref{sn99em.optir.fig}), making it easier
to derive a color temperature from broad-band photometry.
They also pointed out also that, despite the advantage of there being many fewer
lines in the IR, measuring the color temperature from IR photometry is more
difficult than in the optical since the spectrum is close to the Rayleigh-Jeans limit.
This could be particularly severe if the IR photometry has lower
precision than that at optical bands. Our IR data confirm this concern, i.e.,
the angular radii derived from filters $JHK$ have substantially higher
scatter leading thus to a much less (5$\times$) precise distance estimate
than that obtained from optical wavelengths. Filter subsets involving a
combination of one IR filter and one optical filter do not suffer from the
proximity to the Rayleigh-Jeans limit, yet the EPM distances have precisions
which are 2$\times$ lower than those including optical filters alone.
This lower precision is a symptom of the larger IR photometric errors.
Unfortunately, most of the  IR photometry of SN~1999em was obtained
with the YALO/ANDICAM camera which suffered from significant vignetting
that introduced illumination variations $\sim$50\% in all of the images (Suntzeff et al. 2001).
This made necessary to apply large photometric corrections to the SN and the field standards. 
Considering these problems, it is not surprising that the internal
precision of EPM appears lower from subsets $\{VJ, VH, VK\}$, yet it is encouraging that
the IR results are consistent with those obtained from optical wavelengths.
For a better assessment of the performance of the dilution factors in the IR
it will be necessary to obtain data with smaller observational errors.
It is worth mentioning that the initial points (up to JD 2451490) 
obtained with the LCO 1-m IR camera show a much smaller spread, comparable
to optical photometry (1-2\%), which demonstrates the potential precision
that can be reached with IR photometry. Our guess is that, for a dataset
obtained under normal circumstances, $\{VK\}$ would work the best because the
errors in temperature are very small for a given level of photometric errors
(see Figure \ref{T_color.fig}).

Table 11 shows that the distances yielded by subsets $\{BV, VI, BVI, VZ, VJ, VH, VK, JHK\}$
from the first 70 days of SN evolution lie between 6.9-8.6 Mpc. 
The lowest value corresponds to that yielded by subset
$\{BV\}$, which disagrees from the rest if we consider the formal
uncertainty of 1-2\% yielded by the bootstrap method. As discussed above,
these discrepancies are possibly caused by errors in the dilution corrections.
They could be due to metallicity effects which are expected to be relatively
stronger in the $B$ band due to line blanketing at these wavelengths.
It could well be that the metallicity of SN~1999em was lower than the solar
value adopted by E96 (as suggested by Baron et al. 2000), thus increasing 
its $B$-band observed flux relative to the models and making the distance to appear lower.
The largest value in Table 11 is due to $\{JHK\}$ which has a substantial
uncertainty of 8\% associated to it, mainly due to the lower precision in the
YALO IR data and the proximity to the Rayleigh-Jeans limit.
The distances derived from $\{VJ, VH, VK\}$ appear 9\% higher (1-3$\sigma$)
than the those derived from optical wavelengths. We investigate two possible
causes for this systematic effect. First, since most of our IR data come
from the YALO telescope, whereas the adopted filter functions for the EPM analysis
correspond to those used with the LCO IR camera (Appendix B), we repeat here
the EPM calculations using the YALO $JHK$ filter tracings. We obtain $D(VJ)$=8.27,
$D(VH)$=7.77, $D(VK)$=8.22, $D(JHK)$=7.72 Mpc, which prove insignificantly different
than the values listed in Table 11. Second, since the EPM distances shown in
Figures \ref{sn99em.dist.res1.fig} and \ref{sn99em.dist.res2.fig} change
over time, we examine now the possibility that the optical distances might
differ from those derived from $\{VJ, VH, VK\}$ due to the different sampling
of the SN evolution. By restricting the optical sample to a subset with
the same time sampling of the IR observations we get $D(BV)$=6.81, $D(BVI)$=7.39,
$D(VI)$=7.44 Mpc, which are negligibly different than the solutions obtained from
the entire optical dataset. These two tests confirm the existence of a systematic
difference between optical and IR distances.
It will be interesting to investigate whether or not these discrepancies
persist from other SNe~II with better IR data.

Altogether, Table 11 shows that the internal precision in the average distance
must lie between a minimum of 2\% (the formal statistical error from an individual
filter subset) and a maximum of 7\% (the actual scatter obtained from the eight
subsets). Adopting the average solution yielded by the eight filter subsets
our best estimate for the distance to SN~1999em is $D$$_{99em}$=7.5$\pm$0.5 Mpc. 

We cannot rule out systematic errors beyond this estimate.
Leonard et al. (2000) has recently done a detailed
multi-epoch spectropolarimetric study of SN~1999em which 
suggests a minimum asphericity of $\sim$7\% during the
plateau phase. Their lower limit could overestimate
the distance by 7\% for an edge-on view, or lead to an underestimate
of 4\% for a face-on line-of-sight. From a lower limit it
proves difficult to ascertain the actual effect on our
distance estimate. It is reassuring, on the other hand, 
the good agreement between EPM, Tully-Fisher and Cepheid distances
found from a sample of 11 galaxies \cite{schmidt94}, which suggests that the asphericity
factor is probably small among SNe~II-P. This conclusion
is further supported by Leonard et al. who used
the distance residuals in the SN~II Hubble diagram derived by
Schmidt et al. to estimate an average
asphericity for SNe~II-P of only 10\%. This value will be
constrained even more as the Hubble diagram is populated
with well-measured EPM distances.

One possibility to test the overall accuracy 
is to compare our EPM distance to other
methods. There is a  distance estimate
to NGC~1637 of 7.8$^{\rm +1.0}_{-0.9}$ Mpc based on the
brightness of red supergiants \cite{sohn98} which 
compares with our EPM distance of 
$D$$_{99em}$=7.5$\pm$0.5 Mpc. This galaxy is 
part of the 21 cm H I line profile catalog of Hanes et al. (1998)
which lists a velocity width of 180.2$\pm$1.7 km s$^{-1}$ 
that can be used to derive a Tully-Fisher distance.
Giovanelli (2001, private communication) points out that NGC~1637 has
an $I$-band extinction-corrected magnitude of 9.37 and an axial ratio of 1.62, 
which leads to an inclination-corrected velocity width of log~$W$=2.33$\pm$0.41.
The application of the Tully-Fisher template relation derived by Giovanelli et al. (1997)
yields a CMB recession velocity of 669$\pm$116 km s$^{-1}$ for NGC~1637 which,
when combined with the Cepheid-based value of H$_0$=69$\pm$5 km s$^{-1}$ Mpc$^{-1}$
derived by the same authors, leads to a distance of 9.7$\pm$1.7 Mpc.
This value is somewhat larger than that derived from our EPM analysis but,
it must be kept in mind that the H I velocity width and inclination for NGC~1637
are quite uncertain because of its lopsidedness.  Certainly, it would be very useful to
have a precise Cepheid distance to NGC~1637, in order
to further test the dilution factors of E96 and the EPM result.

The observations of SN~1999em and the EPM analysis of
this paper demonstrate that it is possible to achieve EPM distances
with {\em internal} precisions of 7\% from optical observations.
We are carrying out further tests of the {\em external} precision
and accuracy of the method from the study of the Hubble
diagram with SNe~II well in the Hubble flow. When this
study is complete we expect to have a firm assessment of
the performance of EPM.  If we confirm the results found
in this paper, the next step will be the observation of high-$z$ SNe.
As we have demonstrated, the cross-correlation technique 
will significantly extend the reach of EPM to higher redshifts, thus offering
the possibility to obtain a determination of the cosmological parameters
completely independent from the results yielded by SNe~Ia
(Riess et al. 1998, Perlmutter et al. 1999).

\section{CONCLUSIONS}

\noindent 1) We develop a technique to measure accurate
photospheric velocities by cross-correlating SN spectra
with the models of E96. The application of this technique 
to SN~1999em shows that we can reach an average uncertainty
of 11\% in velocity from an individual spectrum. 
This approach will significantly extend the reach of EPM to higher redshifts.

\noindent 2) Using the data of SN~1999em we show that EPM is
quite robust to the effects of dust. In particular,
the distances derived from the $\{VI\}$ filter
subset change by only 7\% when the adopted visual extinction in
the host galaxy is varied by 0.45 mag. Despite the
weak sensitivity of EPM to dust, our analysis 
reveals that multi-color photometry ($BVIJ$) can  be very useful
at constraining the value of A$_{Host}(V)$. In 
particular we find evidence for small ($A_{Host}(V)$$<$0.2)
dust absorption in SN~1999em, in good agreement with the independent
estimate of $A_{host}\approx$0.03-0.18 and $A_{host}$$<$0.33
from theoretical modeling of the spectra of this SN (Baron et al. 2000).
These estimates are highly inconsistent with the value
$A_{host}\sim$3.1$\pm$0.47 implied by the equivalent width
of the interstellar Na I D line measured from the SN spectrum.

\noindent 3) EPM has the advantage that observations at different epochs are
essentially independent distance measurements. The superb sampling of the
$BVIZJHK$ light-curves of SN~1999em permits us to examine in detail the internal
consistency of EPM. Our analysis shows that our first photometric and spectroscopic
observation was obtained at an age of only three days after shock breakout,
which is far earlier than any other SN, except for SN~1987A.
Our tests show that the distances computed with the dilution factors of E96
prove even 50\% lower than the average during the first week since explosion.
We cannot rule out errors in the dilution factors as the source of the problem.
It is conceivable that such error could be caused by circumstellar
material which could lead to the formation of the photosphere at a much larger radius.
Over the following 65 days, on the other hand, our analysis lends strong
credence to the models of E96, and confirms their prediction that
the use of average dilution factors can produce consistent distances
without having to craft specific models for each SN.
The $\{VI\}$ filter subset shows the greatest internal consistency
(with an average scatter of only 4\%) and the least sensitivity to the adopted
dust extinction, making it the most reliable route to cosmic distances.
Our tests show that it is necessary to obtain light curves with reasonable spacing in time
(once per week) in order to avoid systematic biases introduced by the use of average dilution factors.
A few (5-10) points properly spaced over the plateau phase
can produce distances with internal precisions of 4-6\%.

\noindent 5) When comparing distances derived from the 
first 70 days of SN evolution,
we find that those determined from $\{VJ, VH, VK\}$ appear to be 9\% higher
(1-3$\sigma$) than those yielded by the optical filter subsets
$\{BV, VI, VZ\}$. Better IR data will be required in the future
to ascertain whether this is a problem of this particular dataset
or a more general feature of the atmosphere models of E96.
The average distance obtained from filter subsets
$\{BV, VI, BVI, VZ, VJ, VH, VK, JHK\}$ is $D_{99em}$=7.5$\pm$0.5 Mpc,
where the quoted uncertainty (7\%) is a conservative estimate of the internal precision
based on the rms distance spread yielded by all these filter subsets.
This EPM distance compares with
the value 7.8$^{\rm +1.0}_{-0.9}$ Mpc derived by Sohn \& Davidge (1998)
from the brightness of red supergiants in the host galaxy of SN~1999em.
A Tully-Fisher distance of 9.7$\pm$1.7 Mpc has been derived for this galaxy,
which proves to be 1.3$\sigma$ larger than the EPM value.
A more precise Cepheid distance to the host galaxy of SN~1999em would be
very useful in order to test our results.

\acknowledgments

\noindent We are very grateful to Brian Schmidt for a thorough and
critical review of the manuscript, and to Dave Arnett and Adam Burrows
for their valuable input throughout the course of the preparation of
this paper.  MH is very grateful to Las Campanas
and Cerro Cal\'an observatories for allocating an office and providing
generous operational support to the SOIRS program during 1999-2000.
MH and JM thank the ESO, CTIO, and Las Campanas visitor support
staffs for their assistance in the course of our observing runs.
We are very grateful to R. Maiolino for providing the solar spectrum that
permitted us to perform the reductions of the IR spectroscopic data,
to J.G. Cuby for allowing us to use his high-resolution atmospheric transmission spectrum,
and to E. Baron for his help in the reductions of the SN spectrum obtained on October 30.
PAP acknowledges support from the National Science Foundation through
CAREER grant AST9501634 and from the Research Corporation though a Cottrell Scholarship.
J.M. acknowledges support from Fondo Nacional de Desarrollo
Cient\'ifico y Tecnol\'ogico, Chile, (FONDECYT), through grant
No. 1980172.
This research has made use
of the NASA/IPAC Extragalactic Database (NED), which is operated by the
Jet Propulsion Laboratory, California Institute of Technology, under
contract with the National Aeronautics and Space Administration.
This research has made use of the SIMBAD database, operated at CDS, Strasbourg, France.

\eject
\appendix

\section{THE EXPANDING PHOTOSPHERE METHOD}

The Expanding Photosphere Method involves measuring a photometric angular radius
and a spectroscopic physical radius from which a   SN distance can be derived.
Assuming that continuum radiation arises from a spherically-symmetric photosphere,
a photometric measurement of its color and magnitude determines its angular
radius $\theta$,
\begin{equation}
\theta = {R\over D} = \sqrt{{ {f_\lambda}\over{\zeta_\lambda^2 \pi B_\lambda(T) 10^{-0.4A(\lambda)}}}},
\label{angdist}
\end{equation}
where $R$ is the photospheric radius, $D$ is the distance to the supernova,
$B_\lambda(T)$ is the Planck function at the color temperature of the blackbody
radiation, $f_\lambda$ is the apparent flux density, and $A(\lambda)$ is the
dust extinction. The factor $\zeta_\lambda$ accounts for the fact that a real supernova
does not radiate like a blackbody at a unique color temperature. Its role
(as defined in Appendix C) is to convert the observed angular radius into 
the photospheric angular radius, defined as the region of total optical
depth $\tau$=2/3 or the last scattering surface. Since the
continuum opacity in the optical and NIR is dominated by electron
scattering, the opacity is grey, and the photospheric angular radius is
independent of wavelength (E96).
A measurement of the photospheric radius $R$ can then convert this angular
radius to the distance to the supernova. Because supernovae are
strong point explosions, they rapidly attain a state of homologous
expansion in which the radius at a time $t$ is given
by
\begin{equation}
R = R_0 + v(t-t_0),
\label{veleqn}
\end{equation}
where $v$ is the photospheric velocity measured from spectral lines,
$t_0$ is the time of explosion, and $R_0$ is the initial radius of the shell.
Combining these equations we get

\begin{equation}
\theta_i = \frac {R_0 + v_i(t_i-t_0)} {D},
\label{gendiseqn}
\end{equation}

\noindent where $\theta_i$ and $v_i$ are the observed quantities measured at time $t_i$.
Because the expansion is so rapid (typically $\sim10^9$~cm~s$^{-1}$), 
$R_0$ rapidly becomes insignificant. Even for a large progenitor
with $R_0$=5$\times$10$^{13}$cm (2$\times$ larger than the 
largest luminosity class I star known, van Belle et al. 1999),
the initial radius is only 10\% of the SN radius at an age of five days (and less at later times),
so it is safe to use the following approximation for all but the first days,

\begin{equation}
\frac {\theta_i} {v_i} \approx \frac {(t_i-t_0)} {D}.
\label{diseqn}
\end{equation}

\noindent This equation shows that photometric and spectroscopic data at
two or more epochs are needed to solve for $D$ and $t_0$.

Clearly, the determination of distances relies on our
knowledge of $\zeta_\lambda$.
The SN atmosphere has a large ratio of scattering to absorptive opacity,
a ratio which varies with wavelength due to line blanketing and
varying continuous absorption. The result is that the photosphere, which lies at a larger
radius than the thermalization depth where the color temperature is set,
radiates less strongly than a blackbody at that temperature, and the
color temperature itself depends upon the photometric bands employed to
measure it. $\zeta_\lambda$ is known as the ``flux dilution correction'',
though it takes into account departures from a blackbody SN for all effects.

EPM was first applied to SNe~II  by Kirshner \& Kwan (1974),
assuming that SNe~II emitted like perfect blackbodies ($\zeta_\lambda$=1).
Schmidt et al. (1992) corrected this situation by computing 
dilution factors from SNe~II atmosphere models and optical distance
correction factors derived empirically from SN~1987A. Using this
approach they computed distances to nine nearby SNe, from which they 
derived a value of the Hubble constant of 60 km s$^{-1}$ Mpc$^{-1}$.
In a subsequent paper Schmidt et al. (1994) used preliminary values of the dilution
factors computed by E96 (see below) and high-quality data
obtained at CTIO, in order to extend the Hubble diagram to $z$=0.05.
From 16 SNe they obtained a value of $H_0$=73$\pm$6
km s$^{-1}$ Mpc$^{-1}$, in good agreement with the Tully-Fisher method.

A major step forward in the knowledge of the dilution factors was
achieved by E96 from detailed NLTE models of SNe~II-P
encompassing a wide range in luminosity, density structure,
velocity, and composition. They found that the most important
variable determining $\zeta_\lambda$ was the effective temperature; for a given
temperature, $\zeta_\lambda$ changed by only 5-10\% over a very large variation in the
other parameters.

One great advantage of distances determined by EPM is that they are
independent of the ``cosmic distance ladder.'' Observations at two epochs
and a physical model for the supernova atmosphere lead directly to a
distance. Moreover, additional observations of the same supernova are
essentially independent distance measurements as the properties of the
photosphere change over time. This provides a valuable {\em internal}
consistency check.

\section{THE COMPUTATION OF SYNTHETIC MAGNITUDES}

The implementation of EPM requires fitting the observed SN magnitudes
to those of a blackbody, from which the color temperature and the angular radius
of the SN can be obtained. This process involves synthesizing broad-band
magnitudes from Planck spectra. It is crucial, therefore, to place the
synthetic magnitudes on the same photometric system employed in the
observations of the SN.

\begin{figure}
\epsscale{1.0}
\plotone{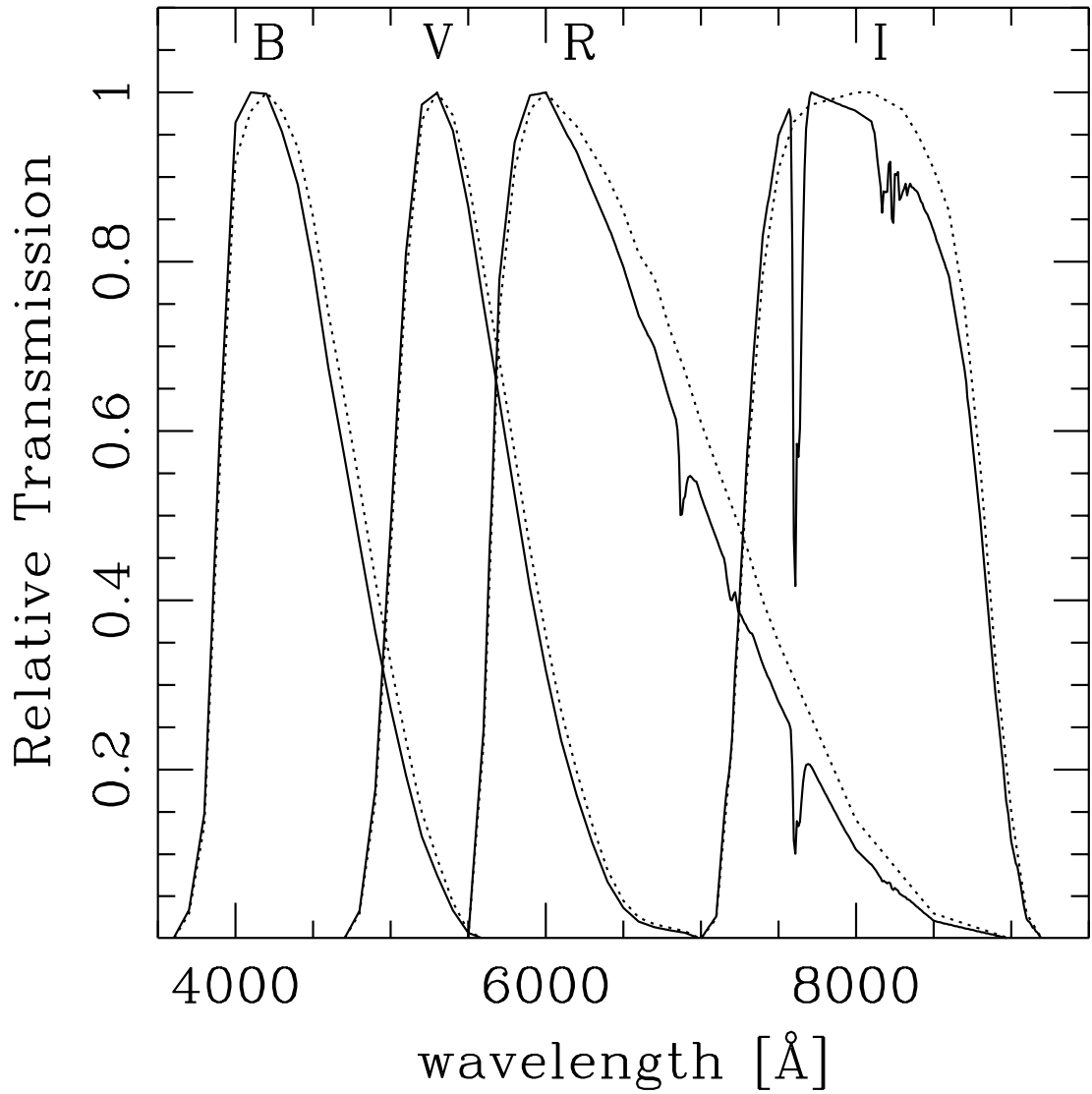}
\figcaption[figure18.ps]{$BVRI$ filters functions of Bessell (1990) 
meant for use with energy distributions (dotted curves).
With solid lines are shown the curves modified for
use with photon distributions, to which we added the telluric lines. \label{filters_bvri.fig}}
\end{figure}

Since the SN magnitudes are measured with photon detectors,
a synthetic magnitude is the convolution
of the object's photon number distribution [$N_\lambda=F_{\lambda}~\lambda/hc$]
with the filter transmission function [S($\lambda$)], i.e.,

\begin{equation}
mag = -2.5~log_{10}~\int N_{\lambda}~A(\lambda)~S(\lambda)~d\lambda ~+~ZP,
\label{mageqn}
\end{equation}

\noindent where ZP is the zero-point for the magnitude scale and
$A(\lambda$) is the factor that accounts for the attenuation of the 
stellar flux due to interstellar dust absorption [in this paper
we adopt the extinction law of Cardelli, Clayton, \& Mathis (1989) for $R_V$=3.1].
For an adequate use of equation \ref{mageqn}, S($\lambda$) must include
the transparency of the Earth's atmosphere,
the filter transmission, and the detector quantum efficiency (QE).
For $BVRI$ we adopt the filter functions
$B_{90}$, $V_{90}$, $R_{90}$, $I_{90}$
published by Bessell (1990). However, since these curves
are meant for use with energy and not photon distributions (see Appendix in Bessell 1983),
we must divide them by $\lambda$ before employing them in equation \ref{mageqn}. Also,
since these filters do not include the atmospheric telluric
lines, we add these features to the $R$ and $I$ filters (in $B$ and $V$ there
are no telluric features) using our own atmospheric transmission spectrum.
Figure \ref{filters_bvri.fig} shows the resulting curves.
For the $Z$ filter we use the transmission
curve of filter 611 and the QE of 
CCD TEK36 of the NTT/EMMI instrument. We include the
telluric lines, but we ignore continuum atmospheric
opacity which is very small at these wavelengths.
For $JHK$ we use the $J_S$, $H$, and $K_S$ filter transmissions
tabulated by Persson et al. (1998), a nominal NICMOS2 QE, 
and the IR atmospheric transmission spectrum
(kindly provided to us by J.G. Cuby).  Figure \ref{filters_zjhk.fig} shows the
resulting $ZJHK$ filter functions, along with the corresponding detector QEs.

The ZP in equation \ref{mageqn} must be determined
by forcing the synthetic magnitude of a star 
to match its observed magnitude. We
use the spectrophotometric calibration
of Vega published by Hayes (1985) in the range 3300-10405~\AA~and
the $V$ magnitude of 0.03 mag measured by Johnson et al. (1966),
from which we solve for the ZP in the $V$ band.
In principle, we can   use the same procedure for $BRI$, but
Vega's photometry in these bands is not very reliable as
it was obtained in the old Johnson standard system.
To avoid these problems we employ ten stars with excellent 
spectrophotometry \cite{hamuy94} and photometry in the modern Kron-Cousins
system (Cousins 1971, 1980, 1984). Before using these standards we
remove the telluric lines from the spectra since the filter
functions already include these features. With this
approach we obtain an average and more reliable zero-point
for the synthetic magnitude scale with rms uncertainty of
$\sim$0.01 mag. With these ZPs we find that the synthetic magnitudes of
Vega are brighter than the observed magnitudes \cite{johnson66}
by 0.016 mag in $B$, 0.025 in $R$, and 0.023 in $I$ (Table 12), which
is not so disappointing considering that this comparison requires
transforming the Johnson $RI$ magnitudes to the Kron-Cousins
system \cite{taylor86}.

In our $Z$ photometric system Vega
has a magnitude of 0.03. Note that this value is  not the
result of a measurement but, instead, of defining the zero-point
for the $Z$ photometric system to give $(V-Z)$=0 for Vega (Appendix D).

At longer wavelengths, where no continuous spectrophotometric
calibration is available for Vega (or any other star), we adopt
the solar model of Kurucz with the following parameters:
$T_{eff}$=9,400 K, log $g$=3.9, [Fe/H]=-0.5, $V_{microturb}$=0
(see Cohen et al. 1992 for a detailed description of the model and
Gray \& Corbally 1994 for the calibration of the MK spectral system).
After flux scaling this model and bringing it into agreement with the $V$=0.03 magnitude of Vega,
the model matches the Hayes calibration at the level of 1\% or better over
the $BVRI$ range, lending credence to the calibration assumed for longer wavelengths.
Figure \ref{vega.fig} shows the adopted spectrophotometric calibration for Vega
in the optical and IR.  To calculate  the zero-points in $JHK$,
we adopt the magnitude of Vega in the CIT photometric system
\cite{elias82}, namely, 0.00 mag at all wavelengths. The original
CIT system comprises stars of 4-7$^{th}$ magnitude. It has
been recently extended by Persson et al. (1998) to fainter standards
which are the stars we used for the calibration of the $JHK$ light-curves of SN~1999em.

Table 12 summarizes the zero-points computed with equation \ref{mageqn},
and the corresponding magnitudes for Vega in such system. 
For the proper use of these ZPs it is necessary to express $F_\lambda$ in 
(erg~sec$^{-1}$~cm$^{-2}$~\AA$^{-1}$), $\lambda$ in (\AA), and the
physical constants $c$ and $h$ in cgs units.
From the ten secondary standards we estimate that the uncertainty 
in the zero-points is $\sim$0.01 mag in $BVRI$.
At longer wavelengths the zero-points are more uncertain since
they come from the adopted model energy distribution of Vega, which is 
probably accurate to better than 5\%.

Following E96 we proceed to compute $b_{\overline{\lambda}}(T)$
--~the magnitude of $\pi B_\lambda(T)$ for a filter with
central wavelength $\overline{\lambda}$~--
to which we fit a polynomial of the form 

\begin{equation}
b_{\overline{\lambda}}(T) = \sum_{i=0}^{5} c_i(\lambda)\left\{\frac{10^4 K}{T}\right\}^i
\label{Tcoloreqn}
\end{equation}

\noindent in the range 4,000 K $<T<$ 25,000 K. We choose this high order
so that the residuals in magnitude are always below 0.01 mag. Table 13
gives the resulting coefficients $c_i(\lambda)$. From these
fits it is straightforward to compute the color temperature from any combination
of magnitudes. Figure \ref{T_color.fig} (top) illustrates some of these temperature-color
curves. Note that, as expected, all these curves intersect at $T\sim$10,000 K, which
corresponds to the temperature of Vega which has nearly zero
colors at all wavelengths.
The bottom panel shows the temperature difference between
our calibration and that of E96. This comparison
reveals that significant differences in color temperatures can be obtained
depending on the photometric system adopted. The disagreement is
particularly large at high temperatures where a small
difference in color translates into a large variation in temperature.

\section{THE DILUTION FACTORS IN OUR PHOTOMETRIC SYSTEM}

Figure \ref{T_color.fig} reveals that, in order to use our photometric system,
it is not possible to use the dilution factors published by E96
as these were computed in a different photometric system.
We proceed now to recompute $\zeta_\lambda$ in our photometric system from 
the model spectra of E96.
The calculation consists in fitting the models with blackbody curves $B_\lambda(T_S)$ and solving
for color temperature $T_S$ and dilution correction factor $\zeta_S$, by minimizing the quantity

\begin{equation}
\sum_{\overline{\lambda}~\epsilon~S}~[M_{\overline{\lambda}} + 5~log~(R_{ph}/10~pc)~+~5~log~\zeta_S~-~b_{\overline{\lambda}}(T_S)]^2,
\label{fiteqn}
\end{equation}

\noindent where $M_{\overline{\lambda}}$ is the broad-band absolute magnitude of the atmosphere model
for a filter with central wavelength $\overline{\lambda}$,
$R_{ph}$ is the photospheric radius,
$b_{\overline{\lambda}}(T_S)$ is the synthetic magnitude of $\pi B_\lambda(T_S)$, and
$S$ is the filter combination used to fit the
atmosphere models with blackbody curves, i.e., $S=\{BV\}=\{VI\}=\{BVI\}$, ...
As explained in Appendix A, the term ``photosphere'' corresponds 
to the last scattering surface which is independent of wavelength
in the optical and NIR. 

Figure \ref{zeta.fig} shows the resulting $\zeta_S$ factors for eight
filter combinations. The differences between the new dilution factors and those of E96
are less than 5\% and, not surprisingly, we recover the result that $\zeta_S$ is primarily
determined by temperature. For convenience, therefore, we perform polynomial fits to
$\zeta_S(T_S) \approx z(T_S)$, where

\begin{equation}
z(T_S) = \sum_{i=0}^{2} a_{S,i}\left\{\frac{10^4 K}{T_S}\right\}^i.
\label{zetaeqn}
\end{equation}

\noindent Table 14 gives the resulting coefficients $a_{S,i}$ for 
the nine  filter subsets and in Figure \ref{zeta.fig} we compare
these fits to the individual $\zeta_S$ factors. Following E96, we
remove the peculiar models s15.5.1, s25.5.1, and h10.30.1 (shown with crosses),
as they are not appropriate models for SNe~II-P.

\section{THE $Z$-BAND PHOTOMETRIC SYSTEM}

\begin{figure}
\epsscale{1.0}
\plotone{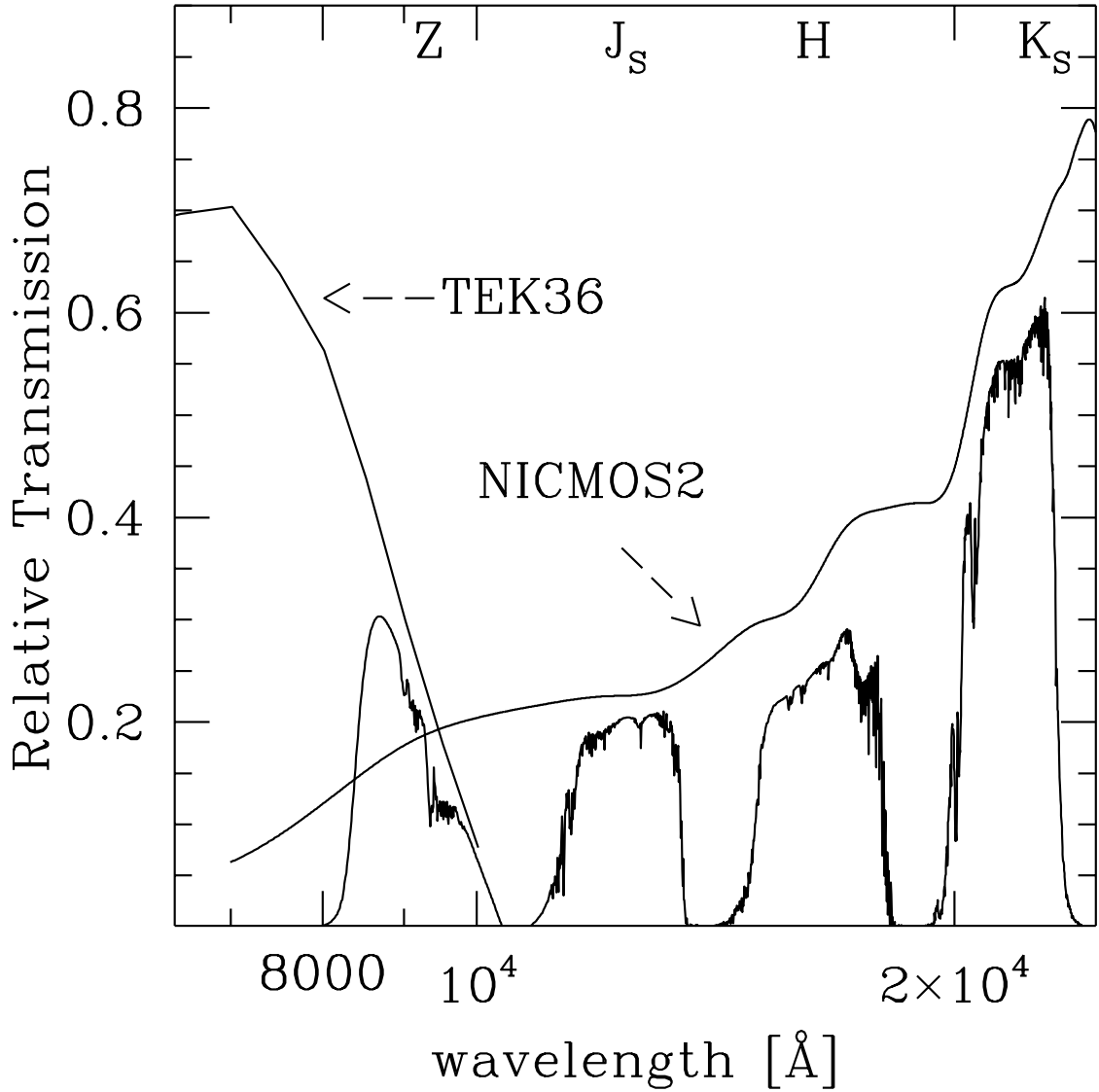}
\figcaption[figure19.ps]{$Z$, $J_S$, $H$, $K_S$ filters functions.
Also shown are the QE of TEK36 and NICMOS2 that we employed to construct
these functions. \label{filters_zjhk.fig}}
\end{figure}

\begin{figure}
\epsscale{1.0}
\plotone{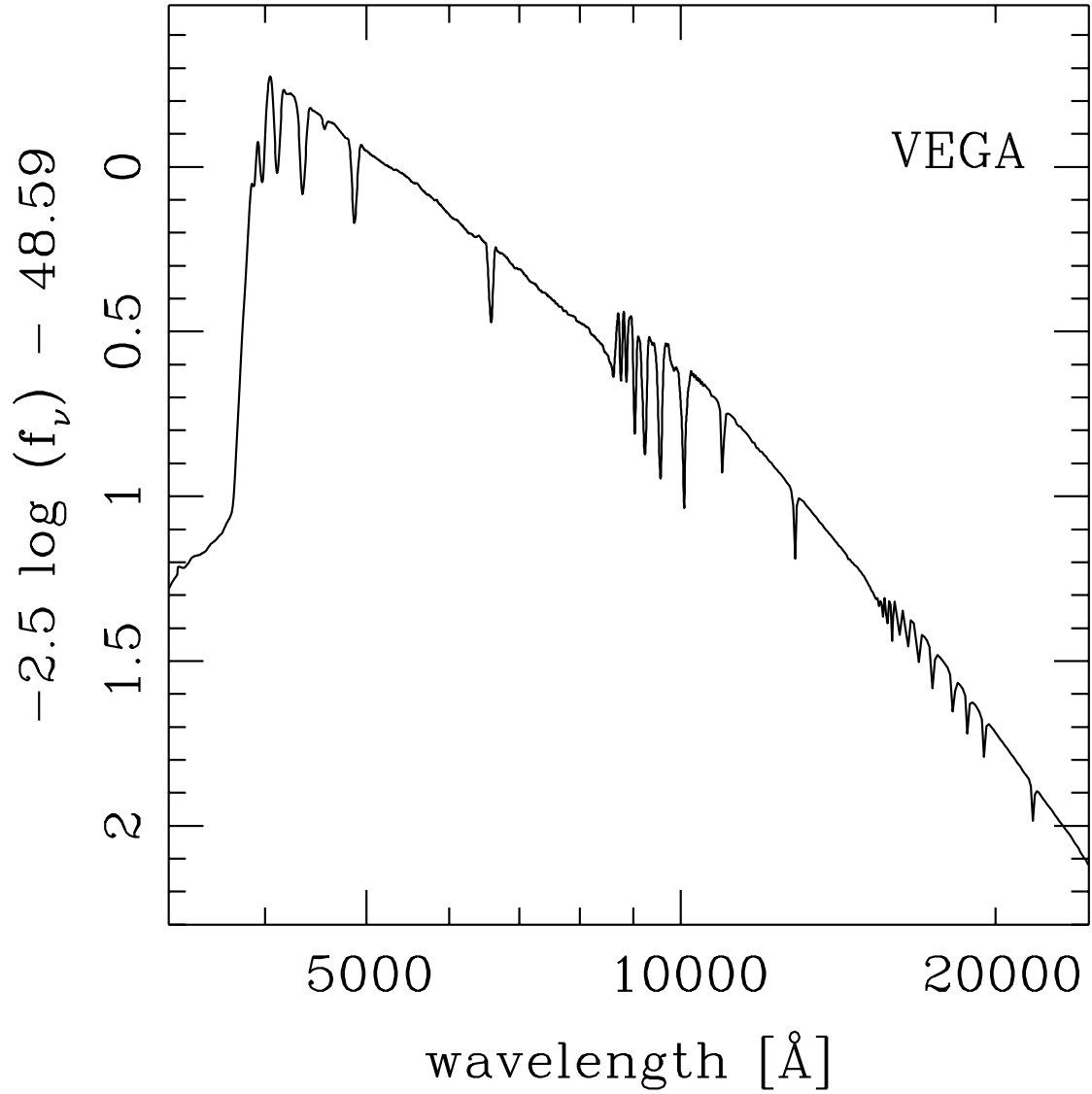}
\figcaption[figure20.ps]{Adopted spectrophotometric calibration for Vega.
In the optical ($\lambda$ $\leq$ 10,500 \AA) the calibration is from Hayes (1985),
and at longer wavelengths we adopted the Kurucz spectrum with parameters
$T_{eff}$=9,400 K, log $g$=3.9, [Fe/H]=-0.5, $V_{microturb}$=0. \label{vega.fig}}
\end{figure}

\begin{figure}
\epsscale{1.0}
\plotone{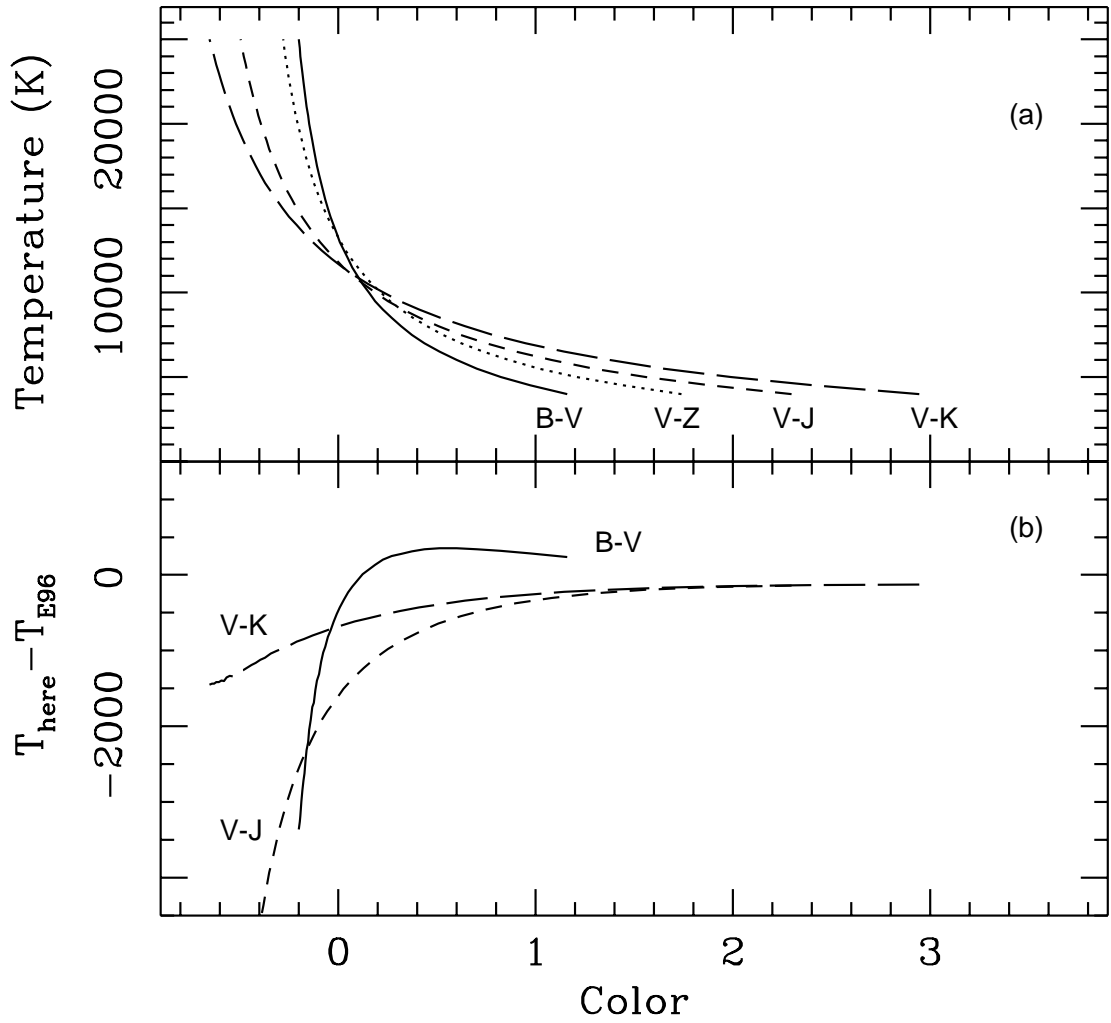}
\figcaption[figure21.ps]{(a) Relation between color temperature and four
different colors in our photometric system. (b) Difference
in color temperature between our calibration and that of
E96. \label{T_color.fig}}
\end{figure}

\begin{figure}
\epsscale{1.0}
\plotone{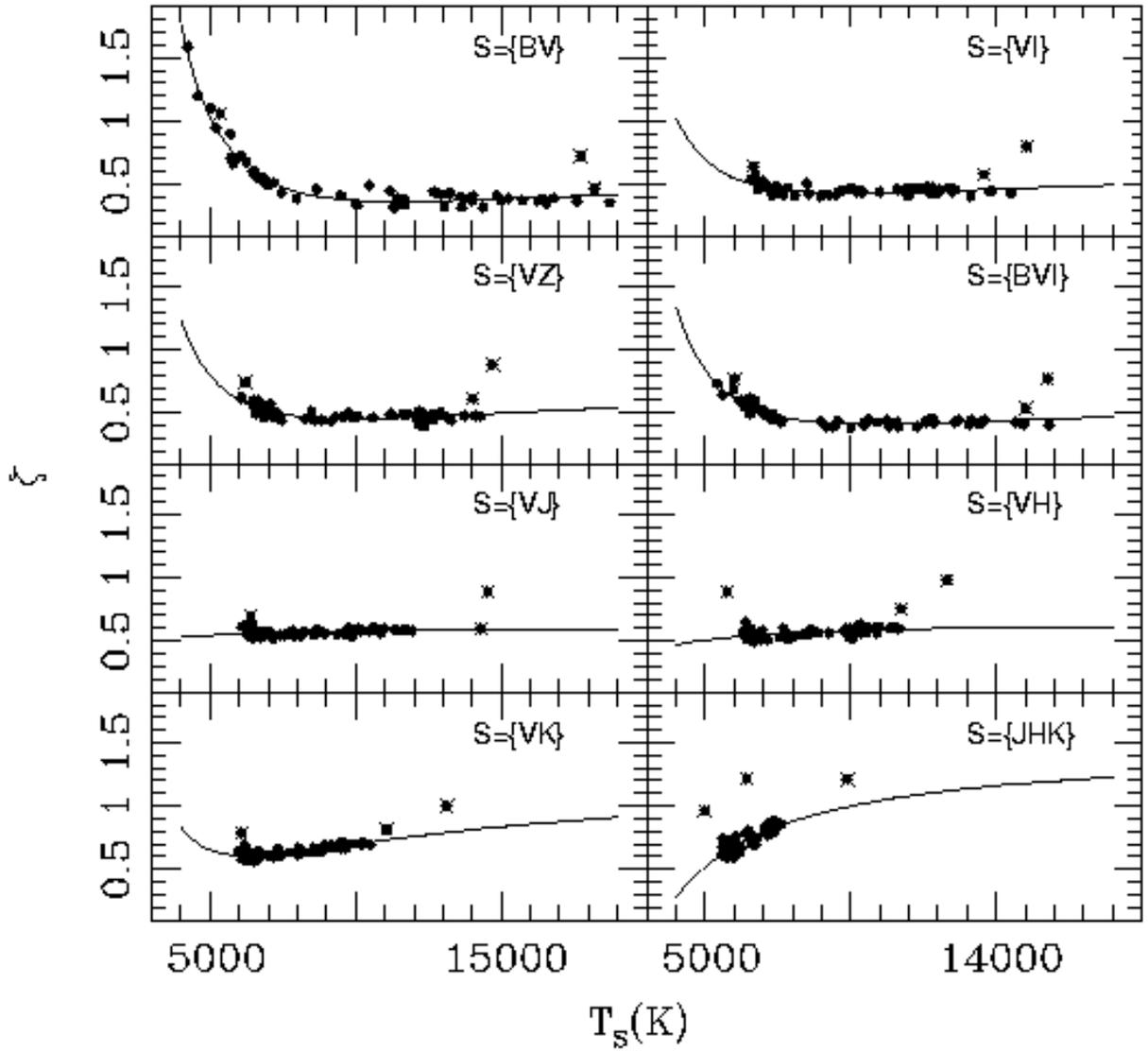}
\figcaption[figure22.ps]{Dilution factors $\zeta$ computed from
E96 atmosphere models vs. color temperature derived from
eight different filter subsets. The solid lines correspond to a polynomial
fit to $\zeta(T_S)$, from which three deviant models (shown with
crosses) are removed. \label{zeta.fig}}
\end{figure}

We define the $Z$-band as the product of the transmission of the $Z$ filter (number 611)
and the QE of CCD TEK36 of the NTT/EMMI instrument. The resulting
bandpass also includes telluric lines (see Figure \ref{filters_zjhk.fig}).
We employ this filter function to compute synthetic magnitudes from the
tertiary spectrophotometric standards published by Hamuy et al. (1994).
However, since these spectra contain telluric lines it is necessary 
first to remove these features. 
Then we use equation \ref{mageqn} and an {\em adopted} ZP of 32.724
that yields a magnitude of 0.03 for Vega. We choose this ZP
so that $(V-Z)$=0 for Vega. The resulting synthetic magnitudes for the
tertiary standards are listed in Table 15 and this is the system relative to which
we calibrated a photometric sequence around SN~1999em.
In Hamuy et al. (1994) we showed that synthetic magnitudes in the $I$ band
had  typical uncertainties of 0.018 mag. Therefore, we believe that
the $Z$ magnitudes in Table 15 have errors of 0.02 mag.

\clearpage

\begin{deluxetable} {lccccccc}
\tabletypesize{\scriptsize}
\tablecolumns{8} 
\tablenum{1}
\tablewidth{0pc}
\tablecaption{$UBVRIZ$ Photometry of SN~1999em}
\tablehead{
\colhead{JD} & 
\colhead{$U$} & 
\colhead{$B$} & 
\colhead{$V$} & 
\colhead{$R$} & 
\colhead{$I$} & 
\colhead{$Z$} & 
\colhead{Telescope} \\
\colhead{-2451000} &  
\colhead{} & 
\colhead{} & 
\colhead{} & 
\colhead{} & 
\colhead{} & 
\colhead{} & 
\colhead{}  }
\startdata
481.76 & \nodata     & \nodata     & \nodata     & \nodata     & \nodata     & 13.703(015) & CTIO 0.9-m \\
483.72 & \nodata     & \nodata     & \nodata     & \nodata     & \nodata     & 13.607(015) & CTIO 0.9-m \\
484.76 & \nodata     & \nodata     & \nodata     & \nodata     & \nodata     & 13.570(015) & CTIO 0.9-m \\
485.79 & \nodata     & \nodata     & \nodata     & \nodata     & \nodata     & 13.542(015) & CTIO 0.9-m \\
486.80 & \nodata     & \nodata     & \nodata     & \nodata     & \nodata     & 13.549(015) & CTIO 0.9-m \\
487.76 & \nodata     & \nodata     & \nodata     & \nodata     & \nodata     & 13.512(015) & CTIO 0.9-m \\
488.80 & \nodata     & \nodata     & \nodata     & \nodata     & \nodata     & 13.514(015) & CTIO 0.9-m \\
489.81 & \nodata     & \nodata     & \nodata     & \nodata     & \nodata     & 13.489(015) & CTIO 0.9-m \\
498.78 & 14.420(017) & 14.357(015) & 13.869(015) & 13.600(015) & 13.483(015) & \nodata     & CTIO 1.5-m \\
499.81 & 14.522(017) & 14.414(015) & 13.894(015) & 13.599(015) & 13.482(015) & \nodata     & CTIO 1.5-m \\
501.75 & \nodata     & \nodata     & \nodata     & \nodata     & \nodata     & 13.296(015) & CTIO 0.9-m \\
506.77 & \nodata     & \nodata     & \nodata     & \nodata     & \nodata     & 13.265(015) & CTIO 0.9-m \\
508.82 & \nodata     & \nodata     & \nodata     & \nodata     & \nodata     & 13.241(015) & CTIO 0.9-m \\
509.86 & \nodata     & \nodata     & \nodata     & \nodata     & \nodata     & 13.234(015) & CTIO 0.9-m \\
511.85 & \nodata     & 14.833(015) & 13.968(015) & 13.633(015) & 13.414(015) & \nodata     & SO 2.3-m   \\
522.70 & 16.043(015) & 15.142(015) & 14.016(015) & 13.618(015) & 13.351(015) & \nodata     & SO 1.5-m   \\
531.76 & \nodata     & 15.185(015) & 14.011(015) & 13.577(015) & 13.284(015) & \nodata     & SO 2.3-m   \\
538.56 & \nodata     & \nodata     & \nodata     & \nodata     & \nodata     & 13.090(015) & CTIO 0.9-m \\
540.55 & \nodata     & \nodata     & \nodata     & \nodata     & \nodata     & 13.102(015) & CTIO 0.9-m \\
546.55 & \nodata     & \nodata     & \nodata     & \nodata     & \nodata     & 13.112(015) & CTIO 0.9-m \\
550.55 & \nodata     & \nodata     & \nodata     & \nodata     & \nodata     & 13.108(015) & CTIO 0.9-m \\
578.63 & \nodata     & 16.060(015) & 14.409(015) & 13.828(015) & 13.492(015) & \nodata     & SO 1.5-m   \\
607.64 & \nodata     & 18.388(040) & 16.467(015) & 15.541(015) & 14.956(015) & \nodata     & SO 1.5-m   \\
636.04 & 21.300(400) & 18.370(030) & 16.658(015) & 15.721(015) & 15.239(015) & 14.924(015) & ESO NTT    \\
665.97 & \nodata     & 18.600(035) & 16.968(015) & 15.999(015) & 15.503(025) & \nodata     & ESO NTT    \\
\enddata
\end{deluxetable}

\begin{deluxetable} {lccccccc}
\tabletypesize{\scriptsize}
\tablecolumns{8}
\tablenum{2}
\tablewidth{0pc}
\tablecaption{Journal of the Spectroscopic Observations}
\tablehead{
\colhead{UT Date} & 
\colhead{Julian Date} &
\colhead{Observatory} &
\colhead{Telescope} &
\colhead{Wavelength} &
\colhead{Dispersion} &
\colhead{Weather} &
\colhead{Observer(s)} \\
\colhead{} &
\colhead{-2451000} &
\colhead{} &
\colhead{} &
\colhead{($\mu$)} &
\colhead{(\AA/pix)} & 
\colhead{} &
\colhead{}}
\startdata

1999 Oct 30 & 481.79 & Tololo   &      1.5-m    & 0.33-0.97 & 5.7     & \nodata & Smith               \\
1999 Nov 2  & 484.64 & Paranal  &     VLT/Antu  & 0.98-2.50 & 2.9-7.1 & Clear   & Hamuy,Lidman,Petr   \\ 
1999 Nov 3  & 485.67 & La Silla &     NTT       & 0.33-1.01 & 2.7-3.5 & Clear?  & Maza                \\ 
1999 Nov 9  & 491.67 & La Silla &     NTT       & 0.33-1.00 & 2.7-3.5 & Clear   & Hamuy,Brillant      \\ 
1999 Nov 14 & 496.67 & La Silla &     NTT       & 0.33-1.01 & 2.7-3.5 & Clear   & Hamuy,Doublier      \\ 
1999 Nov 18 & 500.64 & Paranal  &     VLT/Antu  & 0.98-2.54 & 2.9-7.1 & Clear   & Service Observing   \\ 
1999 Nov 19 & 501.66 & La Silla &     NTT       & 0.33-1.01 & 2.7-3.5 & Clear   & Hamuy,Doublier      \\ 
1999 Nov 28 & 510.63 & Paranal  &     VLT/Antu  & 0.98-2.53 & 2.9-7.1 & Clear   & Hamuy,Lidman,Chadid \\             
1999 Dec 16 & 528.76 & Steward  &    2.3-m      & 0.49-0.93 & 3.6     & Clear   & Corbally,Omizzolo   \\        
1999 Dec 31 & 543.76 & Steward  &    2.3-m      & 0.33-0.71 & 3.6     & \nodata & Burstein,Li         \\   

\enddata
\end{deluxetable}

\begin{deluxetable} {ccccccccc}
\tabletypesize{\scriptsize}
\tablecolumns{9}
\tablenum{3}
\tablewidth{0pc}
\tablecaption{Optical Line Identifications and Observed Wavelengths of Absorption Features of SN~1999em}
\tablehead{
\colhead{Identif.} & 
\colhead{$\lambda_{rest}$} &
\colhead{Oct 30} &
\colhead{Nov 3} &
\colhead{Nov 9} &
\colhead{Nov 14} &
\colhead{Nov 19} &
\colhead{Dec 16} &
\colhead{Dec 31} \\
\colhead{} & 
\colhead{} &
\colhead{481.79} &
\colhead{485.67} &
\colhead{491.67} &
\colhead{496.67} &
\colhead{501.66} &
\colhead{528.76} &
\colhead{543.76} }
\startdata

\nodata                 &  \nodata       &\nodata& \nodata &  3671    &   3671    &   3675     &  \nodata  &  3683 \\
Ca II K                 &   3934         &\nodata& \nodata &  3846    &   3847    &   3852     &  \nodata  &  3868 \\
Ca II H                 &   3968         &\nodata& \nodata &  3846    &   3847    &   3852     &  \nodata  &  3868 \\
H$\delta$               &   4102         &\nodata& 3996    &  4016    &   4021    &   4029     &  \nodata  &  4025 \\
\nodata                 &  \nodata       &\nodata& \nodata &  \nodata &   4095    &   4107     &  \nodata  &  \nodata \\
Sr II + Ca I + Fe II    &   4225         &\nodata& \nodata &  \nodata &   \nodata &   4169     &  \nodata  &  4178 \\
H$\gamma$               &   4340         & 4214  & 4224    &  4234    &   4240    &   4246     &  \nodata  &  4260 \\
Ti II                   &   4395         &\nodata& \nodata &  \nodata &   \nodata &   4318     &  \nodata  &  4342 \\
\nodata                 &  \nodata       &\nodata& \nodata &  \nodata &   \nodata &   \nodata  &  \nodata  &  4412 \\
Ba II + Ti II           &   4552         &\nodata& 4370    &  4425    &   4458    &   4473     &  \nodata  &  4500 \\
N II                    &   4623         &\nodata& \nodata &  \nodata &   \nodata &   \nodata  &  \nodata  &  4588 \\
\nodata                 &  \nodata       &\nodata& \nodata &  \nodata &   \nodata &   \nodata  &  \nodata  &  4626 \\
\nodata                 &  \nodata       &\nodata& \nodata &  \nodata &   \nodata &   \nodata  &  \nodata  &  4693 \\
H$\beta$                &   4861         & 4694  & 4721    &  4735    &   4746    &   4761     &  \nodata  &  4763 \\
\nodata                 &  \nodata       &\nodata& \nodata &  \nodata &   \nodata &   4836     &  \nodata  &  4817 \\
Fe II + Ba II           &   4929         &\nodata& \nodata &  \nodata &   \nodata &   \nodata  &  \nodata  &  4874 \\
Fe II                   &   5018         &\nodata& \nodata &  4911    &   4920    &   4934     &  4965     &  4969 \\
Fe II + Mg I +Ti II     &   5185         &\nodata& \nodata &  5047    &   5065    &   5081     &  5113     &  5117 \\
\nodata                 &  \nodata       &\nodata& \nodata &  \nodata &   \nodata &   \nodata  &  5178     &  5182 \\
Fe I                    &   5270         &\nodata& \nodata &  5178    &   5208    &   5229     &  5221     &  5229 \\
\nodata                 &  \nodata       &\nodata& \nodata &  \nodata &   \nodata &   \nodata  &  5263     &  5270 \\
\nodata                 &  \nodata       &\nodata& \nodata &  \nodata &   \nodata &   \nodata  &  5373     &  5376 \\
Sc II + Fe II           &   5531         &\nodata& \nodata &  \nodata &   \nodata &   5437     &  5475     &  5481 \\
Na I + Sc II            &   5685         &\nodata& \nodata &  \nodata &   \nodata &   \nodata  &  5607     &  5612 \\
He I                    &   5876         & 5665  & 5718    &  \nodata &   \nodata &   \nodata  &  \nodata  &  \nodata \\
Na I                    &   5893         &\nodata& \nodata &  \nodata &   \nodata &   5789     &  5829     &  5831 \\
Fe II + Ba II           &   6146         &\nodata& \nodata &  \nodata &   \nodata &   6042     &  6086     &  6093 \\
Fe II + Sc II           &   6244         &\nodata& \nodata &  \nodata &   \nodata &   6144     &  6184     &  6194 \\
Si II                   &   6355         &\nodata& \nodata &  \nodata &   \nodata &   \nodata  &  6316     &  6319 \\
H$\alpha$               &   6563         & 6292  & 6337    &  6362    &   6379    &   6392     &  6455     &  \nodata \\
Ca II                   & 8498+8542+8662 &\nodata& \nodata &  \nodata &   \nodata &   8344     &  8380     &  \nodata \\
\enddata
\tablecomments{All wavelengths are measured in \AA~in the SN rest frame.}
\end{deluxetable}

\begin{deluxetable} {ccccc}
\tabletypesize{\scriptsize}
\tablecolumns{5}
\tablenum{4}
\tablewidth{0pc}
\tablecaption{IR Line Identifications and Observed Wavelengths of Absorption Features of SN~1999em}
\tablehead{
\colhead{Identif.} & 
\colhead{$\lambda_{rest}$} &
\colhead{Nov 2} &
\colhead{Nov 18} &
\colhead{Nov 28} \\
\colhead{} &
\colhead{} &
\colhead{484.64} &
\colhead{500.64} &
\colhead{510.63} }
\startdata
P$\delta$  &    10049 &  \nodata    &   9901        &  9909    \\
\nodata    &   \nodata&  \nodata    &   \nodata     & 10180    \\
\nodata    &   \nodata&  \nodata    &   10418       & 10418    \\
\nodata    &   \nodata&  \nodata    &   \nodata     & 10549    \\
He I       &    10830 & 10503       &   \nodata     &  \nodata \\
P$\gamma$  &    10938 &  \nodata    &  10741        & 10788    \\
P$\beta$   &    12818 & 12473       &  12573        & 12627    \\
B$\theta$  &    16412 &  \nodata    &   \nodata     & 16219    \\
B$\eta$    &    16811 &  \nodata    &  16566        & 16612    \\
B$\zeta$   &    17367 &  \nodata    &  17108        & 17150    \\
B$\epsilon$&    18179 &  \nodata    &  17884        & 17949    \\
P$\alpha$  &    18751 &  \nodata    &   \nodata     &  \nodata \\
B$\delta$  &    19445 &  \nodata    &   \nodata     & 19210    \\
B$\gamma$  &    21656 &  \nodata    &  21289        & 21368    \\

\enddata
\tablecomments{All wavelengths are measured in \AA~in the SN rest frame.}
\end{deluxetable}

\begin{deluxetable} {cccccc}
\tabletypesize{\scriptsize}
\tablecolumns{6}
\tablenum{5}
\tablewidth{0pc}
\tablecaption{ Expansion Velocities for SN~1999em from Cross-Correlation}
\tablehead{
\colhead{JD} & 
\colhead{model} & 
\colhead{$v_{model}$} & 
\colhead{$v_{CC}$} & 
\colhead{$v_{model}-v_{SN}$} & 
\colhead{$v_{SN}$} \\
\colhead{-2451000} &  
\colhead{} & 
\colhead{(km s$^{-1}$)} & 
\colhead{(km s$^{-1}$)} & 
\colhead{(km s$^{-1}$)} & 
\colhead{(km s$^{-1}$)}  }
\startdata

481.79 & p12.10.5 & 10992 & -614 &   -728 &  11720\\
481.79 & p12.10.6 & 11002 & -591 &   -700 &  11702\\
481.79 & p12.10.9 & 11001 & -744 &   -881 &  11882\\
481.79 & s15.12.4 & 11403 &  720 &    847 &  10556\\
481.79 & s15.12.5 & 11251 &  401 &    470 &  10781\\
 & & & & & \\

484.64 & p9.10.1  & 11594 & -206 &   -287 &  11881\\
484.64 & p9.10.2  & 11564 &  274 &    375 &  11189\\
484.64 & p9.10.3  & 11329 &  477 &    655 &  10674\\
484.64 & p9.10.4  & 11375 & -366 &   -508 &  11883\\
484.64 & p12.10.3 &  8481 &-1656 &  -2288 &  10769\\
484.64 & p12.10.5 & 10992 &  -84 &   -119 &  11111\\
484.64 & p12.10.8 & 10984 &  575 &    791 &  10194\\
484.64 & p12.10.11& 10619 &  714 &    982 &   9637\\
484.64 & s15.12.1 & 11494 & 1329 &   1831 &   9663\\
484.64 & s15.12.2 & 11256 &  262 &    359 &  10897\\
484.64 & s15.12.4 & 11403 &  647 &    890 &  10513\\
484.64 & s15.12.5 & 11251 &  253 &    346 &  10905\\
 & & & & & \\

485.67 & p9.10.1  & 11594 &  467 &   548  &  11046\\
485.67 & p9.10.2  & 11564 &  454 &   533  &  11031\\
485.67 & p9.10.3  & 11329 &  240 &   280  &  11049\\
485.67 & p9.10.4  & 11375 &  114 &   132  &  11243\\
485.67 & p12.10.1 & 11141 &  875 &  1030  &  10112\\
485.67 & p12.10.2 & 11103 &  952 &  1120  &   9983\\
485.67 & p12.10.3 &  8481 & -538 &  -638  &   9119\\
485.67 & p12.10.8 & 10984 &  574 &   674  &  10310\\
485.67 & p12.10.10& 10922 &  686 &   806  &  10116\\
485.67 & p12.10.11& 10619 &  645 &   758  &   9861\\
485.67 & s15.12.1 & 11494 & 2218 &  2614  &   8880\\
485.67 & s15.12.2 & 11256 & 1866 &  2199  &   9057\\
485.67 & s15.12.3 & 11461 & 2344 &  2763  &   8698\\
485.67 & s15.19.6 &  9590 & -190 &  -227  &   9817\\
 & & & & & \\

491.67 & p6.10.1  & 12653 & 2059 &  2427  &  10226\\
491.67 & p12.10.4 & 11058 & 2061 &  2429  &   8629\\
491.67 & s15.19.1 & 10804 & 2451 &  2889  &   7915\\
491.67 & s15.19.2 & 10644 & 2382 &  2808  &   7836\\
491.67 & s15.19.3 & 10276 & 2449 &  2887  &   7389\\
491.67 & s15.19.4 & 10275 & 2479 &  2922  &   7353\\
491.67 & s15.20.1 & 10520 & 2532 &  2985  &   7535\\
491.67 & s15.28.2 &  9654 & 1665 &  1962  &   7692\\
491.67 & s15.28.4 &  8990 &  559 &   657  &   8333\\
491.67 & s25.30.1 &  5816 &  477 &   560  &   5256\\
 & & & & & \\

496.67 & s15.28.1 & 9438  & 2484 &  2928  &   6510\\
496.67 & s15.28.2 & 9654  & 2471 &  2913  &   6741\\
496.67 & s15.28.3 &10230  & 2599 &  3064  &   7166\\
496.67 & s15.28.5 & 9170  & 1436 &  1691  &   7479\\
496.67 & s25.30.1 & 5816  &  384 &   450  &   5366\\
 & & & & & \\

500.64 & p6.40.2  &  8624 & 1866 &  2572  &   6052\\
500.64 & p6.60.1  &  5409 &  236 &   323  &   5086\\
500.64 & p6.60.2  &  5170 &   95 &   128  &   5042\\
500.64 & p6.60.3  &  5384 &  101 &   136  &   5248\\
500.64 & p6.60.4  &  5263 &  158 &   215  &   5048\\
500.64 & p6.60.6  &  5409 &  253 &   346  &   5063\\
500.64 & s15.40.1 &  7040 & 1834 &  2528  &   4512\\
500.64 & s15.43.2 &  7241 & 1563 &  2154  &   5087\\
500.64 & s15.46.1 &  6453 & 1220 &  1681  &   4772\\
500.64 & s15.46.2 &  6707 & 1404 &  1935  &   4772\\
500.64 & s15.60.4 &  5215 &  318 &   436  &   4779\\
500.64 & s25.60.1 &  4667 &  310 &   425  &   4242\\
 & & & & & \\

501.66 & p6.40.2  &  8624 & 2398 &  2827  &   5797\\
501.66 & p6.60.1  &  5409 &  200 &   233  &   5176\\
501.66 & p6.60.2  &  5170 &  154 &   179  &   4991\\
501.66 & p6.60.3  &  5384 &  464 &   545  &   4839\\
501.66 & p6.60.4  &  5263 &  293 &   343  &   4920\\
501.66 & p6.60.6  &  5409 &  185 &   215  &   5194\\
501.66 & p9.60.1  &  4709 &  -13 &   -18  &   4727\\
501.66 & s15.40.1 &  7040 & 1830 &  2156  &   4884\\
501.66 & s15.43.2 &  7241 & 1608 &  1894  &   5347\\
501.66 & s15.43.3 &  6300 & 1707 &  2011  &   4289\\
501.66 & s15.46.1 &  6453 & 1169 &  1376  &   5077\\
501.66 & s15.46.2 &  6707 & 1745 &  2056  &   4651\\
501.66 & s15.60.4 &  5215 &  709 &   834  &   4381\\
501.66 & s25.60.1 &  4667 & -607 &  -719  &   5386\\
 & & & & & \\

510.63 & p6.40.3  &  7686 & 1860 &  2564  &   5122\\
510.63 & p6.60.1  &  5409 &  475 &   653  &   4757\\
510.63 & p6.60.2  &  5170 &  504 &   693  &   4477\\
510.63 & p6.60.4  &  5263 &  575 &   791  &   4473\\
510.63 & p6.60.5  &  4872 &  573 &   788  &   4084\\
510.63 & p6.60.6  &  5409 &  492 &   676  &   4733\\
510.63 & p9.60.1  &  4709 &  336 &   461  &   4248\\
510.63 & p12.60.1 &  4418 & -106 &  -149  &   4567\\
510.63 & s15.40.1 &  7040 & 2223 &  3065  &   3975\\
510.63 & s15.43.1 &  6476 & 1494 &  2059  &   4417\\
510.63 & s15.43.2 &  7241 & 1908 &  2630  &   4611\\
510.63 & s15.43.3 &  6300 & 1709 &  2355  &   3945\\
510.63 & s15.46.1 &  6453 & 1621 &  2234  &   4219\\
510.63 & s15.46.2 &  6707 & 1799 &  2480  &   4227\\
510.63 & s15.60.1 &  3438 & -463 &  -642  &   4080\\
510.63 & s15.60.2 &  5794 &  863 &  1188  &   4606\\
510.63 & s15.60.3 &  3452 & -261 &  -363  &   3815\\
510.63 & s15.60.4 &  5215 &  698 &   960  &   4255\\
510.63 & s25.60.1 &  4667 &  750 &  1032  &   3635\\
510.63 & h10.30.1 &  3864 & -668 &  -925  &   4789\\
510.63 & h10.60.1 &  2786 & -453 &  -628  &   3414\\
 & & & & & \\

528.76 & p6.40.1  &  7340 & 3290 &  3879  &   3461 \\
528.76 & p6.40.3  &  7686 & 3481 &  4105  &   3581 \\
528.76 & p6.60.5  &  4872 & 1595 &  1879  &   2993 \\
528.76 & s15.60.1 &  3438 & 1003 &  1181  &   2257 \\
528.76 & s15.60.3 &  3452 &  837 &   985  &   2467 \\
528.76 & h10.30.1 &  3864 &  711 &   836  &   3028 \\
528.76 & h10.60.1 &  2786 &  140 &   162  &   2624 \\
 & & & & & \\

543.76 & p6.40.1  &  7340 & 3861 &  4553  &   2787\\
543.76 & s15.60.1 &  3439 & 1430 &  1684  &   1755 \\
543.76 & s15.60.3 &  3452 & 1297 &  1527  &   1925\\
543.76 & h10.30.1 &  3864 & 1472 &  1734  &   2130 \\
\enddata
\end{deluxetable}

\begin{deluxetable} {ccccc}
\tabletypesize{\scriptsize}
\tablecolumns{3}
\tablenum{6}
\tablewidth{0pc}
\tablecaption{ Expansion Velocities for SN~1999em}
\tablehead{
\colhead{JD} & 
\colhead{$v_{lines}$} & 
\colhead{$v_{Fe}$} & 
\colhead{$v_{Sc}$} & 
\colhead{$v_{CC}$} \\
\colhead{-2451000} &  
\colhead{(km s$^{-1}$)} & 
\colhead{(km s$^{-1}$)} & 
\colhead{(km s$^{-1}$)} & 
\colhead{(km s$^{-1}$)}  }
\startdata

481.79  &  \nodata    & \nodata   & \nodata   & 11328(611)  \\
484.64  &  \nodata    & \nodata   & \nodata   & 10776(721)  \\
485.67  &  8067       & \nodata   & \nodata   & 10023(859)  \\
491.67  &  6997(1451) & 6739(483) & \nodata   &  7817(1238) \\
496.67  &  5631(1473) & 5947(125) & \nodata   &  6652(811)  \\
500.64  &  \nodata    & \nodata   & \nodata   &  4975(442)  \\
501.66  &  4808(1004) & 5065(60)  & 4831      &  4976(402)  \\
510.63  &  \nodata    & \nodata   & \nodata   &  4307(415)  \\
528.76  &  3132(702)  & 3209(58)  & 2736(46)  &  2916(496)  \\
543.76  &  2972(661)  & 2973(63)  & 2441(3)   &  2149(452)  \\
\enddata
\end{deluxetable}

\begin{deluxetable} {cccccccccc}
\tabletypesize{\scriptsize}
\tablecolumns{10}
\tablenum{7}
\tablewidth{0pc}
\tablecaption{ EPM Quantities Derived for SN~1999em from Subsets \{BV\} and \{VI\}}
\tablehead{
\colhead{JD-} & 
\colhead{$v_{ph}$} &  
\colhead{$T_{BV}$} & 
\colhead{$\theta\zeta_{BV}$} &
\colhead{$\zeta_{BV}$} &
\colhead{($\theta$/$v$)$_{BV}$} &
\colhead{$T_{VI}$} & 
\colhead{$\theta\zeta_{VI}$} &
\colhead{$\zeta_{VI}$} &
\colhead{($\theta$/$v$)$_{VI}$} \\
\colhead{2451000} &
\colhead{(km s$^{-1}$)} &
\colhead{(K)} &
\colhead{(10$^{11}$cm Mpc$^{-1}$)} &
\colhead{} &
\colhead{(100 s Mpc$^{-1}$)} &
\colhead{(K)} &
\colhead{(10$^{11}$cm Mpc$^{-1}$)} &
\colhead{} &
\colhead{(100 s Mpc$^{-1}$)} }
\startdata
481.76 & 11763 & 19490(1471) & 220(16) & 0.431 & 433(39)   & 13392(589) & 323(15) & 0.453 & 606(41) \\
481.80 & 11741 & 18387(1292) & 234(17) & 0.420 & 475(41)   & 12683(487) & 347(14) & 0.448 & 661(43) \\
483.72 & 10752 & 16362(990) & 270(18) & 0.400 & 627(52)    & 12966(512) & 348(15) & 0.450 & 719(47) \\
483.78 & 10722 & 16087(1613) & 278(30) & 0.397 & 652(77)   & 14501(1685) & 310(38) & 0.461 & 627(84) \\
484.76 & 10255 & 15122(830) & 297(19) & 0.388 & 747(60)    & 12870(504) & 356(15) & 0.449 & 773(50) \\
485.79 & 9788 & 14524(759) & 305(19) & 0.383 & 813(65)     & 12683(487) & 356(15) & 0.448 & 811(53) \\
486.80 & 9354 & 14278(730) & 309(19) & 0.381 & 869(69)     & 11717(408) & 391(15) & 0.441 & 946(60) \\
487.76 & 8964 & 13878(685) & 316(19) & 0.377 & 934(73)     & 11086(359) & 415(16) & 0.438 & 1057(66) \\
488.80 & 8562 & 13064(598) & 338(20) & 0.372 & 1060(82)    & 10649(328) & 437(16) & 0.437 & 1168(73) \\
489.81 & 8194 & 12709(1119) & 347(38) & 0.370 & 1146(139)  & 10320(709) & 454(41) & 0.436 & 1271(132) \\
490.79 & 7857 & 12059(728) & 367(29) & 0.367 & 1271(118)   & 9766(270) & 486(17) & 0.436 & 1421(87) \\
492.79 & 7224 & 10467(366) & 449(24) & 0.371 & 1675(122)   & 9804(272) & 492(18) & 0.436 & 1563(96) \\
493.74 & 6949 & 9869(322) & 490(26) & 0.378 & 1867(135)    & 9791(271) & 496(18) & 0.436 & 1638(101) \\
495.67 & 6437 & 8659(295) & 596(35) & 0.410 & 2258(174)    & 9640(360) & 506(26) & 0.436 & 1802(129) \\
495.74 & 6419 & 8086(209) & 676(33) & 0.438 & 2401(168)    & 9882(277) & 494(18) & 0.436 & 1766(109) \\
498.65 & 5758 & 7096(182) & 849(45) & 0.521 & 2833(207)    & 9531(291) & 517(21) & 0.436 & 2059(134) \\
498.68 & 5752 & 7081(159) & 844(40) & 0.522 & 2810(195)    & 9251(239) & 535(19) & 0.437 & 2128(129) \\
498.78 & 5731 & 7385(173) & 773(37) & 0.491 & 2749(191)    & 9229(237) & 533(18) & 0.437 & 2126(129) \\
499.81 & 5527 & 7133(161) & 816(39) & 0.516 & 2859(198)    & 8948(221) & 552(19) & 0.439 & 2276(138) \\
500.70 & 5362 & 6906(233) & 858(60) & 0.543 & 2944(254)    & 8937(341) & 547(31) & 0.439 & 2321(174) \\
501.71 & 5187 & 6388(128) & 1023(48) & 0.622 & 3173(218)   & 8906(219) & 560(19) & 0.439 & 2456(149) \\
501.75 & 5180 & 6537(135) & 967(46) & 0.596 & 3131(216)    & 8855(216) & 560(19) & 0.440 & 2459(149) \\
502.67 & 5031 & 6486(210) & 982(70) & 0.605 & 3229(281)    & 8449(391) & 604(44) & 0.444 & 2702(239) \\
504.67 & 4738 & 5917(183) & 1197(87) & 0.720 & 3507(310)   & 8676(259) & 578(26) & 0.442 & 2764(185) \\
504.74 & 4728 & 6093(151) & 1109(65) & 0.680 & 3452(265)   & 8252(185) & 621(20) & 0.447 & 2937(176) \\
505.72 & 4600 & 5687(102) & 1294(61) & 0.781 & 3601(247)   & 8296(187) & 614(20) & 0.446 & 2991(179) \\
506.77 & 4472 & 5815(106) & 1219(57) & 0.746 & 3652(250)   & 8065(176) & 638(21) & 0.450 & 3169(189) \\
507.80 & 4356 & 5488(138) & 1399(90) & 0.843 & 3811(311)   & 8090(177) & 636(21) & 0.450 & 3246(194) \\
508.82 & 4249 & 5579(98) & 1338(63) & 0.814 & 3870(265)    & 7983(172) & 648(21) & 0.452 & 3371(201) \\
509.86 & 4148 & 5323(95) & 1498(74) & 0.901 & 4009(282)    & 7983(172) & 647(21) & 0.452 & 3448(205) \\
510.75 & 4068 & 5252(87) & 1556(73) & 0.928 & 4123(282)    & 7779(162) & 680(22) & 0.457 & 3657(217) \\
511.85 & 3976 & 5224(86) & 1562(73) & 0.939 & 4185(287)    & 7696(158) & 686(22) & 0.459 & 3760(223) \\
513.72 & 3835 & 5021(80) & 1739(81) & 1.026 & 4417(303)    & 7522(161) & 718(25) & 0.464 & 4038(245) \\
516.71 & 3645 & 4770(75) & 1985(96) & 1.155 & 4718(327)    & 7369(144) & 742(23) & 0.469 & 4342(256) \\
519.72 & 3485 & 4571(67) & 2243(105) & 1.276 & 5043(346)   & 7263(139) & 764(24) & 0.473 & 4635(273) \\
522.59 & 3353 & 4449(63) & 2433(114) & 1.361 & 5333(366)   & 7191(136) & 780(24) & 0.475 & 4891(288) \\
522.70 & 3348 & 4338(60) & 2555(120) & 1.445 & 5281(363)   & 6956(126) & 809(25) & 0.486 & 4977(292) \\
527.63 & 3139 & 4360(256) & 2562(461) & 1.427 & 5720(1067) & 7289(349) & 754(64) & 0.472 & 5095(501) \\
528.59 & 3098 & 4307(82) & 2650(163) & 1.470 & 5819(461)   & 6799(151) & 859(35) & 0.494 & 5617(361) \\
531.76 & 2956 & 4205(57) & 2816(133) & 1.556 & 6122(421)   & 6607(113) & 899(27) & 0.505 & 6023(352) \\
538.56 & 2584 & 3981(53) & 3328(162) & 1.776 & 7250(506)   & 6507(129) & 920(34) & 0.511 & 6962(433) \\
538.60 & 2582 & 3894(85) & 3618(286) & 1.875 & 7475(699)   & 6618(196) & 897(49) & 0.504 & 6895(513) \\
540.55 & 2451 & 3941(71) & 3394(216) & 1.821 & 7605(615)   & 6523(160) & 902(43) & 0.510 & 7216(497) \\
546.55 & 2176 & 3792(47) & 3851(183) & 1.999 & 8850(611)   & 6332(103) & 956(29) & 0.524 & 8392(489) \\
546.61 & 2176 & 3823(54) & 3781(200) & 1.960 & 8865(646)   & 6436(107) & 933(28) & 0.516 & 8311(485) \\
547.60 & 2172 & 3768(53) & 3977(211) & 2.031 & 9018(658)   & 6426(106) & 937(28) & 0.517 & 8347(487) \\
550.55 & 2159 & 3733(75) & 4058(299) & 2.077 & 9051(806)   & 6337(161) & 953(48) & 0.523 & 8437(600) \\
551.61 & 2154 & 3752(46) & 3970(189) & 2.051 & 8988(621)   & 6260(100) & 974(29) & 0.529 & 8547(497) \\
551.66 & 2154 & 3693(45) & 4258(203) & 2.133 & 9271(641)   & 6308(102) & 972(29) & 0.525 & 8589(500) \\
\enddata
\end{deluxetable}

\begin{deluxetable} {cccccccccc}
\tabletypesize{\scriptsize}
\tablecolumns{10}
\tablenum{8}
\tablewidth{0pc}
\tablecaption{ EPM Quantities Derived for SN~1999em from Subsets \{BVI\} and \{VZ\}}
\tablehead{
\colhead{JD-} & 
\colhead{$v_{ph}$} &  
\colhead{$T_{BVI}$} & 
\colhead{$\theta\zeta_{BVI}$} &
\colhead{$\zeta_{BVI}$} &
\colhead{($\theta$/$v$)$_{BVI}$} &
\colhead{$T_{VZ}$} & 
\colhead{$\theta\zeta_{VZ}$} &
\colhead{$\zeta_{VZ}$} &
\colhead{($\theta$/$v$)$_{VZ}$} \\
\colhead{2451000} &
\colhead{(km s$^{-1}$)} &
\colhead{(K)} &
\colhead{(10$^{11}$cm Mpc$^{-1}$)} &
\colhead{} &
\colhead{(100 s Mpc$^{-1}$)} &
\colhead{(K)} &
\colhead{(10$^{11}$cm Mpc$^{-1}$)} &
\colhead{} &
\colhead{(100 s Mpc$^{-1}$)} }
\startdata

481.76 & 11763 & 15674(448) & 278(8) & 0.443 & 533(31)    & 12827(409) & 339(11) & 0.475 & 607(36) \\
481.80 & 11741 & 14707(364) & 301(8) & 0.435 & 589(33)   & \nodata  & \nodata & \nodata & \nodata \\
483.72 & 10752 & 14295(341) & 316(8) & 0.431 & 682(39)    & 12526(388) & 362(12) & 0.472 & 713(43) \\
483.78 & 10722 & 15434(1071) & 291(22) & 0.441 & 615(55)   & \nodata  & \nodata & \nodata & \nodata \\
484.76 & 10255 & 13794(315) & 332(9) & 0.427 & 759(43)    & 12454(382) & 370(12) & 0.472 & 766(46) \\
485.79 & 9788 & 13451(297) & 335(9) & 0.424 & 807(45)     & 11335(309) & 409(13) & 0.462 & 904(53) \\
486.80 & 9354 & 12741(263) & 358(9) & 0.419 & 913(51)     & 11335(309) & 407(13) & 0.462 & 943(55) \\
487.76 & 8964 & 12181(238) & 375(9) & 0.416 & 1007(56)    & 10543(262) & 444(13) & 0.457 & 1084(63) \\
488.80 & 8562 & 11608(213) & 397(9) & 0.413 & 1122(62)    & 10469(258) & 447(13) & 0.457 & 1142(66) \\
489.81 & 8194 & 11386(545) & 402(25) & 0.412 & 1190(96)   & 10073(297) & 470(16) & 0.456 & 1257(76) \\
490.79 & 7857 & 10378(219) & 453(12) & 0.412 & 1398(80)   & \nodata  & \nodata & \nodata & \nodata \\
492.79 & 7224 & 10091(156) & 475(10) & 0.413 & 1593(87)  & \nodata  & \nodata & \nodata & \nodata \\
493.74 & 6949 & 9826(147) & 494(11) & 0.415 & 1714(93)   & \nodata  & \nodata & \nodata & \nodata \\
495.67 & 6437 & 9135(192) & 543(18) & 0.422 & 2001(120)   & \nodata  & \nodata & \nodata & \nodata \\
495.74 & 6419 & 8981(120) & 556(12) & 0.424 & 2041(111)   & \nodata  & \nodata & \nodata & \nodata \\
498.65 & 5758 & 8267(127) & 626(16) & 0.441 & 2463(139)   & \nodata  & \nodata & \nodata & \nodata \\
498.68 & 5752 & 8132(97) & 635(13) & 0.446 & 2476(134)   & \nodata  & \nodata & \nodata & \nodata \\
498.78 & 5731 & 8299(101) & 612(12) & 0.440 & 2427(131)   & \nodata  & \nodata & \nodata & \nodata \\
499.81 & 5527 & 8033(94) & 638(13) & 0.449 & 2572(139)   & \nodata  & \nodata & \nodata & \nodata \\
500.70 & 5362 & 7919(171) & 652(24) & 0.453 & 2680(167)   & \nodata  & \nodata & \nodata & \nodata \\
501.71 & 5187 & 7572(83) & 700(14) & 0.469 & 2876(155)   & \nodata  & \nodata & \nodata & \nodata \\
501.75 & 5180 & 7644(85) & 686(14) & 0.466 & 2843(153)    & 8088(144) & 650(17) & 0.473 & 2654(149) \\
502.67 & 5031 & 7288(173) & 766(35) & 0.485 & 3136(212)   & \nodata  & \nodata & \nodata & \nodata \\
504.67 & 4738 & 7527(141) & 713(23) & 0.472 & 3191(191)   & \nodata  & \nodata & \nodata & \nodata \\
504.74 & 4728 & 7340(95) & 734(17) & 0.482 & 3222(177)   & \nodata  & \nodata & \nodata & \nodata \\
505.72 & 4600 & 6897(68) & 805(16) & 0.513 & 3411(183)   & \nodata  & \nodata & \nodata & \nodata \\
506.77 & 4472 & 6889(68) & 807(16) & 0.514 & 3510(188)    & 7587(125) & 710(18) & 0.487 & 3259(183) \\
507.80 & 4356 & 7063(94) & 777(18) & 0.501 & 3562(197)   & \nodata  & \nodata & \nodata & \nodata \\
508.82 & 4249 & 6710(64) & 841(16) & 0.530 & 3738(200)    & 7409(119) & 739(18) & 0.495 & 3519(197) \\
509.86 & 4148 & 6609(66) & 861(17) & 0.540 & 3848(207)    & 7354(117) & 749(19) & 0.497 & 3632(203) \\
510.75 & 4068 & 6426(58) & 914(17) & 0.559 & 4018(215)   & \nodata  & \nodata & \nodata & \nodata \\
511.85 & 3976 & 6376(57) & 920(17) & 0.565 & 4098(219)   & \nodata  & \nodata & \nodata & \nodata \\
513.72 & 3835 & 6130(57) & 1004(21) & 0.597 & 4386(237)   & \nodata  & \nodata & \nodata & \nodata \\
516.71 & 3645 & 5992(52) & 1040(20) & 0.617 & 4626(248)   & \nodata  & \nodata & \nodata & \nodata \\
519.72 & 3485 & 5791(47) & 1114(21) & 0.650 & 4915(262)   & \nodata  & \nodata & \nodata & \nodata \\
522.59 & 3353 & 5684(45) & 1160(22) & 0.670 & 5161(275)   & \nodata  & \nodata & \nodata & \nodata \\
522.70 & 3348 & 5528(43) & 1205(22) & 0.702 & 5127(273)   & \nodata  & \nodata & \nodata & \nodata \\
527.63 & 3139 & 6360(221) & 961(68) & 0.567 & 5401(468)   & \nodata  & \nodata & \nodata & \nodata \\
528.59 & 3098 & 5567(67) & 1245(35) & 0.694 & 5793(333)   & \nodata  & \nodata & \nodata & \nodata \\
531.76 & 2956 & 5316(39) & 1329(25) & 0.751 & 5984(319)   & \nodata  & \nodata & \nodata & \nodata \\
538.56 & 2584 & 5030(42) & 1519(35) & 0.832 & 7064(388)   & 6347(85) & 969(23) & 0.565 & 6632(367) \\
538.60 & 2582 & 5134(73) & 1468(54) & 0.801 & 7101(441)   & \nodata  & \nodata & \nodata & \nodata \\
540.55 & 2451 & 5120(64) & 1475(48) & 0.805 & 7479(446)   & 6272(82) & 979(23) & 0.573 & 6977(386) \\
546.55 & 2176 & 4941(34) & 1530(28) & 0.861 & 8168(435)   & 6280(83) & 973(23) & 0.572 & 7819(432) \\
546.61 & 2176 & 5123(41) & 1424(28) & 0.804 & 8139(438)   & \nodata  & \nodata & \nodata & \nodata \\
547.60 & 2172 & 5083(40) & 1448(29) & 0.816 & 8176(440)   & \nodata  & \nodata & \nodata & \nodata \\
550.55 & 2159 & 5002(69) & 1559(56) & 0.841 & 8589(530)   & 6249(84) & 982(24) & 0.575 & 7907(439) \\
551.61 & 2154 & 4890(33) & 1563(29) & 0.878 & 8262(440)   & \nodata  & \nodata & \nodata & \nodata \\
551.66 & 2154 & 4864(33) & 1597(29) & 0.887 & 8359(445)   & \nodata  & \nodata & \nodata & \nodata \\

\enddata
\end{deluxetable}

\begin{deluxetable} {cccccccccc}
\tabletypesize{\scriptsize}
\tablecolumns{10}
\tablenum{9}
\tablewidth{0pc}
\tablecaption{ EPM Quantities Derived for SN~1999em from Subsets \{VJ\} and \{VH\}}
\tablehead{
\colhead{JD-} & 
\colhead{$v_{ph}$} &  
\colhead{$T_{VJ}$} & 
\colhead{$\theta\zeta_{VJ}$} &
\colhead{$\zeta_{VJ}$} &
\colhead{($\theta$/$v$)$_{VJ}$} &
\colhead{$T_{VH}$} & 
\colhead{$\theta\zeta_{VH}$} &
\colhead{$\zeta_{VH}$} &
\colhead{($\theta$/$v$)$_{VH}$} \\
\colhead{2451000} &
\colhead{(km s$^{-1}$)} &
\colhead{(K)} &
\colhead{(10$^{11}$cm Mpc$^{-1}$)} &
\colhead{} &
\colhead{(100 s Mpc$^{-1}$)} &
\colhead{(K)} &
\colhead{(10$^{11}$cm Mpc$^{-1}$)} &
\colhead{} &
\colhead{(100 s Mpc$^{-1}$)} }
\startdata

481.80 & 11741 & 13207(318) & 331(8) & 0.586 & 481(27)  & 11318(192) & 400(7) & 0.590 & 578(31) \\
482.69 & 11271 & 10963(348) & 421(17) & 0.581 & 643(41)  & 10338(178) & 456(9) & 0.583 & 693(37) \\
483.76 & 10732 & 10993(210) & 427(9) & 0.581 & 685(37)  & 10541(163) & 452(8) & 0.585 & 720(38) \\
483.78 & 10722 & 12596(342) & 364(10) & 0.585 & 581(33)  & 9923(207) & 498(13) & 0.581 & 800(45) \\
484.76 & 10255 & 11310(224) & 418(9) & 0.582 & 700(38)  & 10706(169) & 449(8) & 0.586 & 748(40) \\
485.73 & 9815 & 10714(488) & 440(26) & 0.580 & 773(60) & 10279(284) & 466(17) & 0.583 & 815(50) \\
486.77 & 9367 & 10218(178) & 469(9) & 0.579 & 864(47) & 9909(142) & 489(8) & 0.580 & 900(47) \\
487.75 & 8968 & 10102(173) & 471(9) & 0.578 & 909(49) & 9836(140) & 490(8) & 0.580 & 942(50) \\
488.76 & 8577 & 9645(156) & 502(10) & 0.577 & 1015(54) & 9525(130) & 511(8) & 0.577 & 1033(54) \\
489.81 & 8194 & 9464(188) & 514(11) & 0.576 & 1089(59) & 9304(155) & 527(9) & 0.576 & 1118(59) \\
495.74 & 6419 & 9211(141) & 548(10) & 0.575 & 1483(79) & 9206(120) & 548(9) & 0.575 & 1486(78) \\
498.68 & 5752 & 9305(240) & 531(19) & 0.576 & 1602(99) & 9252(245) & 535(21) & 0.575 & 1618(102) \\
501.71 & 5187 & 7668(93) & 719(13) & 0.568 & 2438(129) & 8324(96) & 624(9) & 0.566 & 2125(111) \\
504.74 & 4728 & 8530(118) & 588(11) & 0.572 & 2172(116) & 7983(87) & 657(10) & 0.562 & 2469(129) \\
505.72 & 4600 & 7943(101) & 661(12) & 0.570 & 2524(134) & 7646(79) & 707(10) & 0.558 & 2752(143) \\
507.80 & 4356 & 7775(96) & 681(12) & 0.569 & 2751(146) & 7834(129) & 673(18) & 0.561 & 2754(156) \\
510.75 & 4068 & 7744(95) & 685(12) & 0.569 & 2962(157) & 7653(79) & 699(10) & 0.558 & 3079(160) \\
513.72 & 3835 & 7418(89) & 737(13) & 0.567 & 3389(180) & 7475(119) & 727(19) & 0.556 & 3406(193) \\
516.71 & 3645 & 6643(68) & 906(15) & 0.562 & 4428(233) & 6953(64) & 828(12) & 0.549 & 4140(215) \\
519.72 & 3485 & 6980(76) & 823(14) & 0.564 & 4189(221) & 7067(71) & 804(13) & 0.550 & 4190(219) \\
522.59 & 3353 & 7056(77) & 808(14) & 0.565 & 4271(225) & 6885(63) & 848(12) & 0.548 & 4616(239) \\
527.63 & 3139 & 6804(95) & 861(17) & 0.563 & 4874(261) & 6809(82) & 860(13) & 0.547 & 5011(262) \\
528.59 & 3098 & 6793(71) & 860(14) & 0.563 & 4935(260) & 6740(60) & 874(12) & 0.545 & 5172(268) \\
538.60 & 2582 & 6549(80) & 917(16) & 0.561 & 6329(336) & 6557(68) & 914(13) & 0.542 & 6529(340) \\
546.61 & 2176 & 6375(62) & 952(15) & 0.560 & 7818(411) & 6352(52) & 959(13) & 0.539 & 8182(424) \\
547.60 & 2172 & 6286(60) & 981(16) & 0.559 & 8084(425) & 6399(53) & 945(13) & 0.540 & 8063(417) \\
551.66 & 2154 & 6491(64) & 916(15) & 0.561 & 7585(399) & 6463(55) & 924(12) & 0.541 & 7933(411) \\
\enddata
\end{deluxetable}

\begin{deluxetable} {cccccccccc}
\tabletypesize{\scriptsize}
\tablecolumns{10}
\tablenum{10}
\tablewidth{0pc}
\tablecaption{ EPM Quantities Derived for SN~1999em from Subsets \{VK\} and \{JHK\}}
\tablehead{
\colhead{JD-} & 
\colhead{$v_{ph}$} &  
\colhead{$T_{VK}$} & 
\colhead{$\theta\zeta_{VK}$} &
\colhead{$\zeta_{VK}$} &
\colhead{($\theta$/$v$)$_{VK}$} &
\colhead{$T_{JHK}$} & 
\colhead{$\theta\zeta_{JHK}$} &
\colhead{$\zeta_{JHK}$} &
\colhead{($\theta$/$v$)$_{JHK}$} \\
\colhead{2451000} &
\colhead{(km s$^{-1}$)} &
\colhead{(K)} &
\colhead{(10$^{11}$cm Mpc$^{-1}$)} &
\colhead{} &
\colhead{(100 s Mpc$^{-1}$)} &
\colhead{(K)} &
\colhead{(10$^{11}$cm Mpc$^{-1}$)} &
\colhead{} &
\colhead{(100 s Mpc$^{-1}$)} }
\startdata

481.80 & 11741 & 9944(130) & 477(7) & 0.698 & 583(30)  & 5640(164) & 727(22) & 0.628 & 987(57) \\
482.69 & 11271 & 9165(209) & 543(18) & 0.673 & 716(43)  & 6337(484) & 693(50) & 0.721 & 852(75) \\
483.76 & 10732 & 9451(116) & 528(8) & 0.682 & 722(38)  & 6574(232) & 681(22) & 0.749 & 848(51) \\
483.78 & 10722 & 8970(166) & 579(15) & 0.667 & 810(45)  & 4732(167) & 970(41) & 0.465 & 1947(128) \\
484.76 & 10255 & 9607(142) & 523(10) & 0.687 & 743(40)  & 6736(278) & 666(26) & 0.766 & 848(54) \\
485.73 & 9815 & 9642(405) & 511(31) & 0.688 & 757(59) & 7532(1186) & 601(83) & 0.842 & 728(107) \\
486.77 & 9367 & 9079(106) & 557(8) & 0.670 & 888(46) & 6716(244) & 689(23) & 0.764 & 962(58) \\
487.75 & 8968 & 9014(105) & 558(8) & 0.668 & 932(48) & 6731(245) & 684(23) & 0.766 & 996(60) \\
488.76 & 8577 & 8850(107) & 572(9) & 0.663 & 1006(53) & 7118(294) & 662(25) & 0.805 & 959(60) \\
489.81 & 8194 & 8563(117) & 601(9) & 0.654 & 1121(58) & 6592(234) & 721(24) & 0.751 & 1172(70) \\
495.74 & 6419 & 9094(130) & 559(10) & 0.670 & 1298(69) & 8814(520) & 569(28) & 0.934 & 949(66) \\
498.68 & 5752 & 8276(147) & 639(18) & 0.645 & 1722(98) & 6160(435) & 789(54) & 0.700 & 1960(166) \\
501.71 & 5187 & 7705(81) & 712(11) & 0.629 & 2184(114) & 8580(458) & 635(28) & 0.919 & 1332(89) \\
504.74 & 4728 & 7520(74) & 730(10) & 0.624 & 2472(129) & 5447(160) & 939(29) & 0.598 & 3323(195) \\
505.72 & 4600 & 7379(81) & 754(13) & 0.621 & 2640(140) & 6020(220) & 880(32) & 0.682 & 2807(174) \\
507.80 & 4356 & 7298(106) & 764(19) & 0.619 & 2834(158) & 6417(329) & 827(43) & 0.731 & 2597(188) \\
510.75 & 4068 & 7388(132) & 746(23) & 0.621 & 2951(174) & 6825(363) & 776(40) & 0.776 & 2461(176) \\
513.72 & 3835 & 7024(87) & 816(17) & 0.614 & 3468(189) & 6177(286) & 890(42) & 0.702 & 3306(228) \\
516.71 & 3645 & 6809(56) & 863(11) & 0.610 & 3879(200) & 7613(323) & 782(29) & 0.848 & 2529(157) \\
519.72 & 3485 & 6495(50) & 951(11) & 0.607 & 4495(231) & 5356(146) & 1093(31) & 0.583 & 5384(311) \\
522.59 & 3353 & 6815(68) & 865(14) & 0.610 & 4228(223) & 6077(225) & 949(35) & 0.689 & 4109(256) \\
527.63 & 3139 & 6439(67) & 963(13) & 0.607 & 5056(262) & 5555(163) & 1069(32) & 0.615 & 5539(324) \\
528.59 & 3098 & 6469(53) & 950(13) & 0.607 & 5054(262) & 5659(175) & 1047(33) & 0.631 & 5360(317) \\
538.60 & 2582 & 6127(54) & 1055(13) & 0.607 & 6733(347) & 5116(132) & 1206(34) & 0.541 & 8636(494) \\
546.61 & 2176 & 6261(64) & 989(19) & 0.606 & 7495(401) & 6004(235) & 1016(40) & 0.679 & 6871(439) \\
547.60 & 2172 & 6240(49) & 997(13) & 0.606 & 7570(392) & 6207(215) & 988(33) & 0.706 & 6451(389) \\
551.66 & 2154 & 6339(48) & 962(12) & 0.606 & 7368(379) & 5892(181) & 1016(31) & 0.664 & 7104(416) \\
\enddata
\end{deluxetable}

\begin{deluxetable} {ccc}
\tabletypesize{\scriptsize}
\tablecolumns{3}
\tablenum{11}
\tablewidth{0pc}
\tablecaption{ Distance and Explosion Time for SN~1999em}
\tablehead{
\colhead{Filter} & 
\colhead{$D$} & 
\colhead{$t_0$} \\
\colhead{Subsets} &  
\colhead{(Mpc)} & 
\colhead{(JD-2451000)} }
\startdata

\{BV\}     & 6.91$\pm$0.09 & 478.3$\pm$0.4 \\
\{BVI\}    & 7.40$\pm$0.08 & 477.9$\pm$0.3 \\
\{VI\}     & 7.31$\pm$0.10 & 479.6$\pm$0.4 \\
\{VZ\}     & 7.79$\pm$0.10 & 477.6$\pm$0.3 \\
\{VJ\}     & 8.01$\pm$0.29 & 479.9$\pm$0.8 \\
\{VH\}     & 7.82$\pm$0.28 & 479.8$\pm$0.9 \\
\{VK\}     & 8.22$\pm$0.23 & 479.1$\pm$0.8 \\
\{JHK\}    & 8.56$\pm$0.69 & 477.5$\pm$2.1 \\
 & & \\
All(8)     & 7.54$\pm$0.08 & 478.8$\pm$0.5 \\
 & & \\
\{BVIJHK\} & 7.82$\pm$0.15 & 478.9$\pm$0.5 \\
\enddata
\end{deluxetable}

\begin{deluxetable} {crrrrrrrr}
\tabletypesize{\scriptsize}
\tablecolumns{9}
\tablenum{12}
\tablewidth{0pc}
\tablecaption{Photometric Zero-points and Synthetic Magnitudes for Vega and the Sun}
\tablehead{
\colhead{}  & 
\colhead{$B$} & 
\colhead{$V$} &
\colhead{$R$} &
\colhead{$I$} &
\colhead{$Z$} &
\colhead{$J_S$} &
\colhead{$H$} &
\colhead{$K_S$} }
\startdata
Zero-point & 35.287  & 34.855  & 35.060  & 34.563  & 32.724 & 32.230 & 32.098 &  32.175 \\
Vega & 0.014 & 0.030 & 0.042 & 0.052 & 0.030 & 0.00 & 0.00 & 0.00 \\
Sun  & -26.083 & -26.752 & -27.120 & -27.451 & -27.592 & -27.918 & -28.243 & -28.285 \\
\enddata
\end{deluxetable}

\begin{deluxetable} {crrrrrrrr}
\tabletypesize{\scriptsize}
\tablecolumns{9}
\tablenum{13}
\tablewidth{0pc}
\tablecaption{Fits to $b_{\overline{\lambda}}(T)$\tablenotemark{a}}
\tablehead{
\colhead{$i$}  & 
\colhead{c$_i(B)$} & 
\colhead{c$_i(V)$} &
\colhead{c$_i(R)$} &
\colhead{c$_i(I)$} &
\colhead{c$_i(Z)$} &
\colhead{c$_i(J_S)$} &
\colhead{c$_i(H)$} &
\colhead{c$_i(K_S)$} }
\startdata
0    & -45.144 & -44.766 & -44.597 & -44.345 & -44.232 & -43.913 & -43.767 & -43.638 \\
1    &   7.159 &   6.793 &   6.628 &   6.347 &   6.262 &   6.022 &   5.878 &   5.737 \\
2    &  -4.301 &  -4.523 &  -4.693 &  -4.732 &  -4.810 &  -4.859 &  -4.914 &  -4.881 \\
3    &   2.639 &   2.695 &   2.770 &   2.739 &   2.778 &   2.772 &   2.797 &   2.757 \\
4    &  -0.811 &  -0.809 &  -0.831 &  -0.811 &  -0.825 &  -0.819 &  -0.829 &  -0.813 \\
5    &   0.098 &   0.096 &   0.099 &   0.096 &   0.098 &   0.097 &   0.098 &   0.096 \\
\enddata
\tablenotetext{a}{~$b_{\overline{\lambda}}(T)$=$\sum_{i} c_i(\lambda)T_4^{-i}$, $T_4=T/10^4 K$.}
\end{deluxetable}

\begin{deluxetable} {crrrr}
\tabletypesize{\scriptsize}
\tablecolumns{5}
\tablenum{14}
\tablewidth{0pc}
\tablecaption{Fits to $\zeta(T_S)$\tablenotemark{a}}
\tablehead{
\colhead{$S$}  & 
\colhead{$a_0$} & 
\colhead{$a_1$} & 
\colhead{$a_2$} & 
\colhead{$\sigma$ \tablenotemark{b}}} 
\startdata
$\{BV\}$    & 0.7557 & -0.8997 & 0.5199 & 0.048 \\
$\{BVI\}$   & 0.7336 & -0.6942 & 0.3740 & 0.027 \\
$\{VI\}$    & 0.7013 & -0.5304 & 0.2646 & 0.029 \\
$\{VZ\}$    & 0.8185 & -0.7137 & 0.3510 & 0.031 \\
$\{VJ\}$    & 0.6104 & -0.0323 & 0.0000 & 0.025 \\
$\{VH\}$    & 0.6548 & -0.0737 & 0.0000 & 0.031 \\
$\{VK\}$    & 1.2865 & -0.8571 & 0.2700 & 0.051 \\
$\{JHK\}$   & 1.4787 & -0.4799 & 0.0000 & 0.046 \\
$\{BVIJHK\}$& 1.1551 & -0.9790 & 0.3913 & 0.024 \\
\enddata
\tablenotetext{a} {~$\zeta(T_S) = \sum_i a_{S,i}(\frac{10^4 K}{T_S})^i$.}
\tablenotetext{b} {$\sigma$ is the rms of the fit.}
\end{deluxetable}

\begin{deluxetable} {cr}
\tabletypesize{\scriptsize}
\tablecolumns{2}
\tablenum{15}
\tablewidth{0pc}
\tablecaption{Standard Stars for the $Z$-band}
\tablehead{
\colhead{$Star$}  & 
\colhead{$Z$} \\
\colhead{} &
\colhead{$\pm$0.020} }
\startdata

LTT 377         & 10.523 \\ 
LTT 1020        & 10.648 \\
EG 21           & 11.619 \\
LTT 1788        & 12.369 \\
LTT 2415        & 11.526 \\
Hiltner 600     & 10.132 \\
L745-46A        & 12.670 \\
LTT 3218        & 11.652  \\
LTT 3864        & 11.374  \\
LTT 4364        & 11.181  \\
Feige 56        & 11.154 \\
LTT 4816        & 13.792  \\
CD -32          & 10.041  \\
LTT 6248        & 10.966  \\
EG 274          & 11.359  \\
LTT 7379        &  9.376  \\
LTT 7987        & 12.437  \\
LTT 9239        & 11.144  \\
Feige 110       & 12.249 \\
LTT 9491        & 14.071 \\
\enddata
\end{deluxetable}

\clearpage

\clearpage

\clearpage

\clearpage

\clearpage

\clearpage

\clearpage

\clearpage

\clearpage

\clearpage

\clearpage

\clearpage

\clearpage

\clearpage

\clearpage

\clearpage

\clearpage

\clearpage

\clearpage

\clearpage

\clearpage

\clearpage

\end{document}